\documentclass[english]{aastex}
\usepackage[T1]{fontenc}
\usepackage{amstext}
\usepackage{amssymb}
\usepackage{graphicx}

\makeatletter

\slugcomment{}
\shorttitle{}
\shortauthors{}
\newcommand*{\x}[1]{{#1}}

\makeatother

\usepackage{babel}
\begin{document}

\title{Metallicity-Dependent Galactic Isotopic Decomposition for Nucleosynthesis }

\author{Christopher West$^{\text{1,3}}$}

\email{west0482@umn.edu}

\and{}

\author{Alexander Heger$^{\text{1,2},3}$}

\email{alexander.heger@monash.edu}

\affil{$^{\text{1}}$Minnesota Institute for Astrophysics}

\affil{School of Physics and Astronomy, University of Minnesota }

\affil{$^{\text{2}}$Monash Centre for Astrophysics }

\affil{School of Mathematical Sciences, Monash University}

\affil{$^{\text{3}}$Joint Institute for Nuclear Astrophysics}
\begin{abstract}
All stellar evolution models for nucleosynthesis require an initial
\emph{isotopic} abundance set to use as a starting point. Generally,
our knowledge of isotopic abundances of stars is fairly incomplete
except for the Sun. We present a first model for a complete average
isotopic \x{decomposition} as a function of metallicity. Our model
is based on the underlying nuclear astrophysics processes, and is
fitted to observational data, rather than traditional forward galactic
chemical evolution modeling which integrates stellar yields beginning
from big bang nucleosynthesis. We \x{first} decompose the isotopic
solar abundance pattern into contributions from astrophysical sources.
Each contribution is then assumed to \x{scale} as a function of metallicity.
The resulting total isotopic \x{abundances are} summed into elemental
\x{abundances} and fitted to available halo and disk stellar data
to constrain the model's free parameter values. This procedure allows
us to use available elemental observational data to reconstruct and
constrain both the much needed complete isotopic evolution that is
not accessible to current observations, and the underlying astrophysical
processes. As an example, our model finds a best fit for Type Ia contributing
$\simeq.7$ to the solar Fe abundance, and Type Ia onset occurring
at $[\mathrm{Fe/H}]\simeq-1.1$, in agreement with typical values. 
\end{abstract}


\section{Introduction}

Yields from stellar simulations depend on the initial isotopic composition
of the star. For example, the initial composition is important during
hydrostatic burning phases for neutron capture reactions on initial
metals, affecting odd-z nuclei abundances. In massive stars the weak
\emph{s}-process yields are constrained by both the initial CNO abundance,
which is responsible for providing a neutron source, and also the
initial Fe abundance which supplies the seeds for neutron capture
\citep{r26}. The initial Fe abundance is also important in intermediate
and low-mass asymptotic giant branch (AGB) stars for seeding the main
\emph{s}-process yields \citep{r15,r16,r18}. The detailed stellar
abundances affect the opacity of the star (e.g., the \citealp{r72}),
which in turn will affect the structure as well as mass and angular
momentum loss, which in turn changes the late stellar evolution. Knowing
the initial abundances of heavy isotopes is crucial for understanding
$\gamma$-process abundances, which use \emph{s}- and \emph{r}-process
isotopes as seeds \citep{r29,r70}. Since these \emph{p}-isotopes
are so rare, any difference in the seed abundances propagates to the
resulting $\gamma$-process yields. Finally, some fraction of the
initial stellar composition is not processed and will return to the
interstellar medium (ISM) unchanged, hence the final abundance pattern
will directly inherit any poorly estimated initial abundances, but
in many cases what is made in the stars and what was there initially
is difficult to disentangle. 

The purpose of galactic chemical evolution (GCE) is to understand
how the abundances of the elements and their isotopes evolved from
the Big Bang to today, and can be used for obtaining the isotopic
abundances at any metallicity to use as stellar simulation inputs.
Traditional GCE models typically split a model galaxy into one or
more zones that require functional forms for infall and star formation
rates \citep{r34,r48,r47,r12}. Other processes such as interstellar
medium (ISM) mixing, and galaxy mergers are often not addressed, although
there have been fairly recent efforts to incorporate mergers in a
hierarchical model \citep{r87}. The model galaxy is then usually
evolved by integrating stellar yields over time, hence these models
require nucleosynthesis yields from stellar simulations as inputs.
The difficulty with this approach is that in order to provide self-consistent
nucleosynthesis yields, the stellar simulations need a complete initial
set of isotopic abundances. 

Ideally, such a set would come directly from the GCE simulation. This
would require knowledge of the complete set of stellar yields for
the specific abundances of the GCE model at each time step and for
each zone: the complicating factor is in reality composition is \emph{not
}just a function of metallicity, but also a function of environment,
i.e., time and space. Instead the inputs for the stellar simulations
often use scaled solar abundances or the results of some other approximation.
Furthermore most GCE models usually evolve only a subset of the stable
isotopes, such as just the iron peak isotopes \citep{r12} or everything
from hydrogen up to the iron peak \citep{r34,r17}, or only elements.
An approach different than GCE modeling to describe chemical abundances
has been done recently by \citet{Ting2012}. They perform a principle
component analysis on elemental data that takes correlations among
elemental ratios and roughly delineates different processes responsible
for these correlations, as well as identify likely sites and metallicity
regimes congruent with them. These efforts offer verification of the
established paradigm concerning many astrophysical processes, and
may help constrain others whose properties remain incompletely known.

The approach taken here is complementary to traditional GCE methods,
however, one must take care not to confuse our model with a proper
GCE model. We construct an astrophysical model of all stable isotopes,
based on physical principles for the production sites and mechanisms.
Effectively, we scale isotopic abundances as a function of a chosen
model parameter. This isotopic model is then mapped to an elemental
model by summing the isotopic abundances to their respective elements.
We then fit the elemental abundances against available observational
data to obtain numerical values for the free parameters of the model.
Our completed model then gives the average \emph{isotopic} history
of the Galaxy, subject to the approximations employed. The benefit
of this approach to isotopic GCE is that it is not necessary to know
or model dynamic and galactic evolution processes employed in traditional
GCE models such as infall, ISM mixing, and galaxy mergers, whose uncertainties
are poorly constrained. 

Compared to full GCE calculations our approach is rather simplistic
and approximate, precisely because we do not integrate stellar yields
or address dynamic and galactic evolution processes employed in traditional
GCE models, and we assume a unique and \emph{typical} abundance distribution
for a given metallicity rather than allowing for a spread in distribution
as found in nature. The intention here is to improve upon the typical
standard of scaling isotopic solar abundances by a constant factor,
which effectively treats all isotopic production as primary. Our improvements
to this standard is to incorporate secondary processes and Type Ia
contributions with separate scalings, to better approximate their
relative values at desired input metallicities for nucleosynthesis
studies, and the resulting isotopic histories from our model can be
used as inputs in stellar models, and comparison to abundances from
other sources like Damped Lyman-$\alpha$ systems and dwarf galaxies.
This represents an improvement over the previous standard of guessing
\emph{ad-hoc} assumptions for the interpolation between the known
endpoints of solar and big bang nucleosynthesis (BBN), or the use
of scaled solar abundances. This model could also be seen as a first
iteration step for more sophisticated first principle GCE models,
but at a higher order approximation than just scaled solar abundances,
but not a replacement for them. Hence, whereas the approximations
and comparisons of our model are sufficient for this purpose, they
would be quite unsatisfactory for describing GCE itself.

This paper has the following outline: Section\,\ref{sec:Astrophysical-Processes-and}
introduces the astrophysical processes and sites considered by the
model, and the separation of the solar isotopic abundance pattern
into contributions from these processes is discussed. In Section\,\ref{sec:Model-Description},
the model itself is introduced and the scaling of the processes as
a function of our parametrization is explained, using their relative
solar contributions and the BBN abundance pattern as fixed boundary
conditions. This also defines the relevant free parameters for fitting
the elemental abundances with stellar data. In Section\,\ref{sec:Fitting-Scaling-Model},
the elemental model is fit to available data and best fit parameter
values are found. In Section\,\ref{sec:Results-and-Discussion},
the resulting elemental model is discussed and additional results
given by the model. The final Section\,\ref{sec:Conclusions} addresses
constraints of the model and discusses possible extensions for future
work.

\section{Astrophysical Processes and Solar Abundance Decomposition\label{sec:Astrophysical-Processes-and}}

Our model attempts to cover the essential key astrophysical processes
responsible for the production of the isotopes in the \x{Galaxy}.
We first use these processes to decompose the solar system abundances.
Here the isotopic solar abundance pattern was taken from a new data
set by Lodders et al. \citeyearpar{r20} which gives updated values
relative to their 2003 publication \citep{r19}. In the following
we organize the processes roughly by the mass range of isotopes to
which they contribute.

\subsection{Big-Bang Nucleosynthesis}

When the Universe was less than $\sim$100 seconds old, protons and
neutrons were in thermal equilibrium with each other by weak interactions
with neutrinos. Upon continuing expansion, however, the temperature
dropped sufficiently to ``freeze out'' neutrino interactions. At
this ``freeze-out'' temperature, corresponding to $\mathrm{k_{\text{B}}\mathit{\mathrm{T}}}$$\sim$0.8\,MeV,
the weak interaction rate became slower than the hubble expansion,
and the neutron-to-proton ratio (after some subsequent $\beta$ decay)
was fixed to $\mathrm{n/p\simeq1/7}$ \citep{r67}. At this point
BBN began with deuterium formation, and proceeded to produce non-negligible
abundances of $^{\text{2}}$H,$^{\text{3}}$He, $^{\text{4}}$He,
$^{\text{7}}$Li, and minute abundances of $^{\text{6}}$Li and isotopes
up through oxygen.

We use the theoretical BBN abundance pattern from \citet{r3} and
provided by \citet{r5}. Isotopic contributions from this BBN pattern
are taken for $\textsuperscript{1}$H, $\textsuperscript{2}$H, $^{3}$He,
$^{4}$He, and $^{7}$Li. The remaining negligible contributions of
$^{6}$Li and the isotopes heavier than $^{7}$Li are not used. A
constant $^{7}$Li abundance at low metallicities has been observed
that suggested a primordial abundance, but whose value was much smaller
than the predicted BBN abundance \citep{r84}. This ``Spite plateau''
remains unexplained, and for the present model we use the theoretical
BBN abundance for $^{7}$Li.

\subsection{Light Isotopes}

\subsubsection{Helium}

The remaining helium not made during BBN is a product of hydrogen
burning and is \x{scaled} as a primary process. The ``release time
scale'' of helium can vary depending on the stellar mass, and relative
to metallicity it may \x{scale} slightly slower than a true primary
process (the same is true for C and N). We do not account for this
delay in the present model.

\subsubsection{$\nu$-Process}

The ``light'' $\nu$-process involves interactions among neutrinos
and lighter nuclei in CCSNe environments \citep{r36,r40,r39}. The
neutrinos elevate nuclei to excited states, which then decay by nucleon
emission \citep{r8}. The target nuclei for these interactions that
produce light isotopes are CNO isotopes made from hydrogen and helium,
hence this production is primary. This process produces $\textsuperscript{11}$B
\citep{r9}, and some $^{7}$Li \citep{r28}. The ``heavy'' $\nu$-process
also involves neutrino interactions, but with target nuclei made from
either the \emph{s}- or \emph{r}-processes, and is responsible for
heavy nuclei production such as $^{\text{180}}$Ta and $^{\text{138}}$La.
Due to the requirement of pre-existing \emph{s}- or \emph{r}-process
metals to serve as the target nuclei, the ``heavy'' $\nu$-process
behaves like the $\gamma$-process with respect to metallicity, discussed
below in Section\,\ref{sub:P-Isotopes}. Hence we do not distinguish
between the ``heavy'' $\nu$-process and $\gamma$-process contributions.
In our model, the $\nu$-process stands for the ``light'' $\nu$-process,
and the $\gamma$-process includes the ``heavy'' $\nu$-process.

\subsubsection{Galactic Cosmic Ray Spallation}

Galactic Cosmic Ray (GCR) spallation events occur when energetic protons
or $\alpha$-particles impact on existing CNO nuclei in the ISM \citep{r37,r38}.
GCR spallation contributes to $^{6}$Li, $^{9}$Be, $^{10}$B, and
$^{11}$B \citep{r27,r28}. Since spallation occurs on pre-existing
CNO nuclei in the ISM, this process is traditionally considered secondary.
Observations, however, show a primary dependence on metallicity for
$^{9}$Be \citep{r27}, which is in conflict with the understanding
of spallation events. \citet{Prantzos2012} and \citet{r28} proposes
a solution to this problem, and states that GCRs accelerated by the
winds of rotating massive stars could be abundant in CNO isotopes.
If these GCRs then hit ISM protons or $\alpha$-particles,
this would satisfy the condition of a primary event. We adopt this
proposed solution, and assign primary GCR spallation as a mechanism
for LiBeB production, along with secondary GCR spallation \citep{r27,r28,Prantzos2012}.

\subsubsection{Classical Novae}

White dwarfs accreting material from a companion star can undergo
outbursts powered by thermonuclear runaway in the accreted layer \citep{r55,r42}.
Nucleosynthesis occurs on the accreted material that is rich in H
and He and dredge-up of primary CNO (and ONeMg for ONeMg novae) into
the envelope, hence this process is primary. Unlike most other primary
processes that immediately begin to enrich the ISM however, there
is likely some onset timescale for novae contributions similar to
Type Ia SNe. We do not consider this delay for novae in our model.
Simulations have shown differing isotopic production below the iron
peak, depending on the composition of the core and its mass \citep{r56,r14,r42}.
Hence precise abundance determinations are difficult to isolate. Many
CO novae simulations show production of $^{7}$Li, $^{13}$C, $^{15}$N,
$^{17}$O, and $^{19}$F that dominate the ejecta, whereas ONeMg novae
additionally show contributions to other metals up to $^{40}$K \citep{r14}.
We take all contributions beyond $^{7}$Li to be negligible compared
to massive star contributions. This approximation holds well for CNO
isotopes that have large contributions from massive stars, but for
isotopes such as $^{19}$F the approximation is less than ideal. In
fact it believed that $^{19}$F may also be produced in the $\nu$-process
in core-collapse supernovae (CCSNe; \citealp{r76}), during hydrostatic
nucleosynthesis in He shell of thermally pulsating asymptotic giant
branch stars (TP-AGB; \citealp{r78}), and in the He core of heavy
mass loss Wolf-Rayet stars \citep{r79}. The decomposition from these
sources is still a matter of debate (\citealp{r80}; and references
therein), and we do not address this complication in the current model.

\subsubsection{Light Isotope Decomposition}

Identifying the precise (non-BBN) Li, Be, and B contributions presents
a challenge. At present, there is no consensus for explaining the
solar abundance pattern for the isotopes of these elements using the
processes that could be responsible. Due to the difficulty in determining
what fraction each actually contributes to the light isotope solar
abundance pattern, novae, the $\nu$-process, and primary GCR spallation
are placed into a single category due to their shared primary nature.
Standard GCR spallation is treated separately since it is secondary.
The relative solar abundance decomposition between these two categories
for $^{6,7}$Li,$^{9}$Be, and $^{10,11}$B is estimated from \citet{Prantzos2012}.
Both Li isotopes are given $\approx$30\,\% secondary contributions,
and the $^{9}$Be, and $^{10,11}$B isotopes are given 25\,\% secondary
contributions. The remaining non-BBN contributions for all LiBeB isotopes
are assigned the novae/$\nu$-process/primary GCR spallation category.
Note the decomposition for these light isotopes are at best known
to within $\approx$5\,\%.

\subsection{Low and Intermediate-Mass Stars}

Stellar winds from low and intermediate-mass stars are rich in C and
N isotopes \citep{r71}, and provide significant contributions to
the solar abundances for these isotopes. Fitting the contributions
from these sources to data is problematic, since contamination from
massive stars is always present and difficult to separate out. Hence
we do not independently address stellar wind contributions in our
model and instead combine their contributions with those of massive
stars, both of which are primary processes.

\subsection{Intermediate-Mass and Iron Group Isotopes\label{sub:Intermediate-Mass-and-Iron}}

Hydrostatic burning in massive stars ($\sim$$10-100$\,$\mathrm{M_{\text{\ensuremath{\odot}}}}$)
synthesizes most isotopes from helium up to the iron peak \citep{r41,r29}.
Stellar winds can eject some of this material over the star's life
but the explosive stellar death dominates the metal yields. CCSNe
likely produces between 1/3 and 2/3 of the solar abundance iron peak
isotopes \citep{r34}. They also produce the majority of the alpha
isotopes and many of the intermediate isotopes from $^{16}$O to the
iron group. Type Ia SNe are thermonuclear explosions of accreting
white dwarfs (eg., \citealp{r25,r59,r75}), and primarily provides
the remaining iron peak solar abundances, with some enrichment of
other metals \citep{r58,r57}. In fact, simulations show Type Ia production
of minute trace contributions to the isotopes below the iron peak
\citep{r25}, with the exception of $^{40}$K. Both CCSNe and Type
Ia produce their isotopic yields explosively, which destroys much
of the initial metal composition. The evolutions of their isotopic
products are considered primary. 

Yields for Type Ia supernovae were taken from the W7 model \citep{r25}
for isotopes with mass numbers $\mathrm{12\le\mathit{A}\le56}$. The
category of ``massive star contributions'' is defined in this context
to be the collection of all primary isotopic production with mass
numbers $\mathrm{12\le\mathit{A}\le68}$ not attributed to Type Ia
SNe. This includes all isotopic enrichment to the ISM driven by massive
star stellar winds and production from CCSNe, the \emph{r}-process,
the $\nu$-process, novae yields, and stellar winds from low and intermediate-mass
stars, with production from CCSNe dominating the isotopic abundances
in this category. The solar contributions from massive stars were
taken from the yields of a massive star simulation \citep{r10} fitted
to stars in the range $-3.1\le\mathit{\mathit{\mathrm{[Fe/H]}}}\le-2.9$
from the \citet{r6} data set. We use the common definition: $\mathrm{[X]\equiv Log(X/X_{\odot})}$.
Note that the ``iron metallicity'' (relative to solar), {[}Fe/H{]},
should be distinguished from the total metallicity, {[}Z{]}: The former
is a conventional proxy for the latter. The simulation included stars
in the mass range $10-100\mathrm{\, M}_{\odot}$, with a Salpeter
initial mass function (IMF) and a low mixing of 0.02512 \citep{r60}
employed in a running boxcar method \citep{r10}. The explosion energy
of the supernovae was set to be $\mathit{E}=1.2\,\mathrm{B}$, where
$1\,\mathrm{B}=10^{51}\,\mathrm{erg}$. 

The fit of the yields to the \citet{r6} data set gave a $\chi$ value
of 2.218 (see \citealt{r10} for fitting procedure). The heavier massive
star contributions for mass numbers $\mathrm{57\le\mathit{A}\le68}$
were not taken from Heger and Woosley's massive star simulation, and
were instead calculated as residuals from the main and weak \emph{s}-processes,
discussed below in Section\,\ref{sub:Heavy-Isotopes}. 

Under the assumption that Type Ia are responsible for some fraction
\emph{$f$} of the observed solar $^{56}$Fe abundance, each W7 yield
was scaled to this fraction. The scaling factor is given as: $f\cdot\mathit{X}_{56}^{\odot}/\mathit{X}_{56}^{Ia}$,
where $\mathit{X}_{56}^{\odot}$ is the solar abundance of $^{56}$Fe,
and $\mathit{X}_{56}^{Ia}$ is the W7 yield for $^{56}$Fe. Hence
each isotopic abundance was scaled by this factor, which shifts the
entire abundance pattern until the yield for $^{56}$Fe is equal to
$f\cdot\mathit{X}_{56}^{\odot}$. The fraction \emph{$f$} represents
a free parameter in the model, which is determined by fitting the
elemental \x{scalings} against available data (in Section\,\ref{sec:Fitting-Scaling-Model}).
The massive star yields were scaled to the remaining contribution
to solar $^{56}$Fe not accounted for by Type Ia. This factor is given
as: $(1-f)\cdot\mathrm{\mathit{X}_{56}^{\odot}/\mathit{X}_{56}^{massive}}$,
where $\mathit{X}_{56}^{\mathrm{massive}}$ is the massive star yield
for $^{56}$Fe. An additional scaling of both types of SNe data in
the range $\mathrm{12\le\mathit{A}\le56}$ was then required to ensure
that the massive star and Type Ia contributions summed to the solar
abundance for every isotope, as the first scaling using \emph{$f$}
would only guarantee that $^{56}$Fe satisfied this requirement. This
additional scaling preserved the ratio of each isotopic contribution
between the W7 and massive star yields,

\begin{equation}
\mathrm{\mathit{X}_{\mathit{i},f}^{massive}=\left(\frac{\mathit{X}_{\mathit{i}}^{\odot}}{\mathit{X}_{\mathit{i},0}^{massive}+\mathit{X}_{\mathit{i},0}^{Ia}}\right)\mathit{X}_{\mathit{i},0}^{massive}}\label{eq:s_decomp_massive}
\end{equation}

\begin{equation}
\mathrm{\mathit{X}_{\mathit{i},f}^{Ia}=\left(\frac{\mathit{X}_{\mathit{i}}^{\odot}}{\mathit{X}_{\mathit{i},0}^{massive}+\mathit{X}_{\mathit{i},0}^{Ia}}\right)\mathit{X}_{\mathit{i},0}^{Ia}}
\end{equation}

where $\mathit{X}_{\mathit{i},0}$, $\mathit{X}_{\mathit{i},\mathrm{f}}$
are the original and (scaled) fitted abundances of isotope \emph{i},
for either the massive or Type Ia contributions (denoted as superscripts),
and $\mathit{X}_{\mathit{i}}^{\odot}$ is the solar abundance of isotope
\emph{i}. Note that for clarity this procedure has been explained
using two successive scalings, but in practice this can be done with
a single scaling that achieves both. This second scaling preserves
the isotopic ratios \emph{across} each model, but the ratios \emph{within}
each model undergo some distortion. That is, the overall abundance
patterns of the W7 and massive star simulations are altered. Nevertheless,
the final abundance patterns we use still show Type Ia contributing
mostly to the Fe peak, and massive stars contributing to CNO and $\alpha$-elements
up to the Fe peak. 

The use of the solar metallicity W7 model is approximate. In our Galaxy,
large contributions to the SN Ia yields may come from sub-solar progenitors.
The exact nature and properties of these sources, however, are still
uncertain (see, for e.g., \citealp{Bours2013,r81}), and given these
uncertainties the inclusion of the W7 yields, while not a complete
description, is also not unreasonable. To investigate the impact of
this approximation, we also computed the solar abundance decomposition
for the massive stars and Type Ia using a composition between that
of the W7 and W70 models to estimate a sub-solar composition. We then
compared the ratios of the isotopes between \textbf{$^{12}$}C and
\textbf{$^{56}$}Fe from our original decomposition and this new one.
Of the 53 isotopes in question, 45 have ratios of the new abundance
(using sub-solar Type Ia) over the old abundance (solar Type Ia) that
are within 2.0. Of the 8 that remain, the largest differences are
\textbf{$^{40}$}K, which now has a nonzero abundance, and \textbf{$^{15}$}N,
\textbf{$^{41}$}K, \textbf{$^{43}$}Ca, and \textbf{$^{47}$}Ti,
which have ratios of \textasciitilde{} 300, 15, 40, and 20, respectively
(the others have ratios within 5). This does present non-trivial corrections
to the solar abundance decomposition, but except for a few isotopes,
the changes are quite minor (within a factor of 2). It is unlikely
that these changes would noticeably impact the fittings to data we
perform later.

\subsection{Heavy Isotopes\label{sub:Heavy-Isotopes}}

\subsubsection{Weak S-Process}

The \emph{s}-process is one of the four trans-iron processes for making
the heavier nuclides distinguished here. It synthesizes isotopes via
slow neutron capture (relative to the beta decay rate; \citealp{r41}).
It is responsible for approximately half the heavy isotopes beyond
iron \citep{r26,r50}. Since this process is characterized by neutron
capture rates that are slow compared to the beta decay rate of the
target nucleus, production proceeds along the path of isotopic stability,
with $^{56}$Fe playing the role of the seed nucleus. In practice
many metals could seed this process, but $^{56}$Fe will stand as
the sole target nucleus due to its large abundance and neutron capture
cross section relative to other potential seeds. The weak component
of the \emph{s}-process occurs in massive stars, during convective
core He burning and shell C burning \citep{r26}. The neutron source
in core He burning is from the reaction $^{22}$Ne($\alpha$,n)$^{25}$Mg.
The $^{22}$Ne nuclei are produced from the burning of $^{14}$N made
previously in the CNO cycle. Subsequent C shell burning can produce
neutrons by itself and provide $\alpha$-particles to reignite the
$^{22}$Ne($\alpha$,n)$^{25}$Mg neutron source. These neutrons are
captured on initial $^{56}$Fe present in the star. 

The weak \emph{s}-process component can only synthesize isotopes along
the line of stability up to a mass number of $A\mathrm{\approx100}$
\citep{r51}, due its smaller neutron exposures relative to the main
component. Both components of the\emph{ s}-process rely on the existence
of metals that formed the initial composition of the star, and therefore
are considered secondary processes. The \emph{neutron source} for
the main\emph{ s}-process, however, is primary (as well as some of
the shell C burning contributions to the weak \emph{s}-process), which
could result in a behavior between that of a primary and secondary
process.

For both the main and the weak components of the \emph{s}-process,
bottlenecks exist at closed neutron shells, allowing the abundances
at these corresponding mass numbers to accumulate into peaks. Three
such peaks exist at approximate mass numbers 88, 138, and 208. Elements
of interest at these peaks for representing the \x{scaling} of \emph{s}-process
elements include strontium, barium, and lead. Good recent compilations
of the main \emph{s}-process can be found in \citet{r18}, and a good
review of the \emph{s}-process can be found in \citet{r16}. 

Modeling the weak \emph{s}-process has been less successful than models
for the main component \citep{r26}. Since flow equilibrium is not
reached during the neutron exposure, uncertainties in the neutron
capture cross-sections affect the yields of all subsequent isotopes
\citep{r50}. We need to reproduce the solar system abundances for
the \emph{s}-only isotopes for both the main and weak components.
For the weak \emph{s}-process, the \emph{s}-only isotopes $^{70}$Ge,
$^{76}$Se, $^{80}$Kr, $^{82}$Kr, $^{86}$Sr, and $^{87}$Sr owe
their solar abundances to \emph{both} the main and weak components
\citep{r51}, and the relative contributions from each represent a
poorly known \emph{a priori} constraint. Additionally, recent stellar
model calculations for the weak \emph{s}-process have difficulties
in both producing sufficient \emph{s}-only isotopic yields without
at the same time overproducing many other isotopes beyond their solar
abundances \citep{r26}. 

For the purposes of making a reasonable assessment of the weak \emph{s}-process
contributions to the solar system abundances, we performed a calculation
using the updated online MACs compilation from the KADoNIS project
\emph{www.kadonis.org} \citep{r4}. The branching points addressed
in our calculation that lie along the weak \emph{s}-process path are
the unstable nuclei $^{64}$Cu and $^{80}$Br, with corresponding
$\beta$$^{+}$ and $\beta$$^{-}$ thermal branching ratios calculated
from \citet{r31}. The branching point at $^{79}$Se was ignored,
and was taken via $\beta$$^{-}$ decay to $^{79}$Br prior to any
neutron capture (which would give a $^{80}$Se population). This approximation
is due to the high temperature dependence of the $\beta$$^{-}$ decay
rate of $^{79}$Se \citep{r11,r21,r35}. $^{93}$Zr was treated as
stable for this calculation, since the thermal $\beta$$^{-}$ decay
branching ratio is small relative to neutron capture rate \citep{r31}.
After the neutron exposure has ended, the remaining $^{93}$Zr abundance
is taken to $^{93}$Nb. The $^{85}$Rb branching was omitted and it
was assumed that it entirely $\beta$$^{-}$ decays to $^{86}$Sr,
again because the neutron capture branching ratio is negligible \citep{r31}.

The differential equations for the weak \emph{s}-process abundances
under the classical approximation \citep{r16} were solved numerically.
We have the following system of linear differential equations:

\begin{equation}
\mathrm{\frac{d\mathbf{\mathit{\mathbf{X}}}}{d\mathit{\tau}}=A\cdot\mathbf{\mathit{\mathbf{X}}\mathrm{,}}}
\end{equation}

where $\mathit{\mathbf{\mathit{\mathbf{X}}}}$ is a vector of the
isotopic abundances, $\mathbf{\mathbf{\mathbf{\mathit{\mathbf{X}}}}\mathrm{=(\mathit{X}_{Fe56},\mathit{X}_{Fe57}},}\ldots\mathrm{,\mathit{X}_{Ru101})}$,
$\tau$ is the neutron exposure, and $\mathrm{A}$ is the matrix of
Maxwellian averaged neutron capture cross-sections and branching ratios
for beta decay, with components $\mathrm{A_{\mathit{\mathrm{ij}}}=-\sigma_{\mathit{i}}}$
for $\mathrm{\mathrm{\mathrm{\mathit{j}=\mathit{i}}}}$, with $\mathrm{A_{\mathit{\mathrm{i,j}}-1}=+\sigma_{\mathit{i}}}$,
$\mathrm{A_{\mathit{i}+1,j}=\sigma_{\mathit{i}}\beta_{\mathit{i}}}$,
and $\mathrm{A_{\mathrm{\mathit{i}+2,\mathrm{\mathit{j}}}}=\sigma_{i}(1-\beta_{\mathrm{\mathit{i}}})}$,
where $\beta_{\mathit{i}}$ is the beta decay branching ratio for
isotope \emph{i} at branching points (where applicable). Note $\mathrm{(1-\beta_{\mathit{i}})}$
is the $\beta^{+}$ branching ratio, and all other elements in the
sparse matrix are zero.

In solving this system, we did not assume a continuous neutron exposure
distribution. Instead, different values for single exposures were
used. A linear combination of the yields from different single neutron
exposures was determined that best fits the abundances of the \emph{s}-only
isotopes in the weak \emph{s}-process range. In addition, to ensure
that the sum of \emph{s}-only isotopes for the weak and main components
were not overproduced with respect to solar, the abundances to which
the \emph{s}-only isotopes were fit were taken as the residuals of
the main s-process yields subtracted from the solar abundances. In
addition, 15\,\% of the solar $^{80}$Kr and 3\,\% of the solar
$^{82}$Kr were attributed to the $\nu$\emph{p}-process \citep{r15,r26}
and likewise subtracted. 

We found a linear combination of five neutron exposures that reproduces
the desired \emph{s}-only isotopic abundance residuals for $^{70}$Ge,
$^{80}$Kr, and $^{86}$Sr Fig.\,\ref{fig1}. The remaining \emph{s}-only
isotopes, $^{76}$Se, $^{82}$Kr, and $^{87}$Sr were under-produced.
The only over-productions that resulted were for $^{65}$Cu and $^{89}$Y.
We found the neutron exposures to be: 0.2, 0.25, 0.3, 0.35, and 0.4\,$\mathrm{mb^{-1}}$.
The respective coefficients that the abundances from each of these
exposures were weighted were: 1.5, 0.2, 0.4, 0.6, and 0.7. We attempted
to fit as many \emph{s}-only isotopes to their residuals without generating
over-productions, and it was not possible to fit all \emph{s}-only
isotopes without causing additional isotopic over-abundances. The
over-productions of $^{65}$Cu and $^{89}$Y as well as the under-productions
of $^{76}$Se, $^{82}$Kr, and $^{87}$Sr were scaled with the main
\emph{s}-process abundances, to fit with the solar abundances. This
scaling was equivalent to the scaling done for massive stars and Type
Ia yields: the isotopic ratios were preserved. 

\includegraphics[scale=0.5,angle=90]{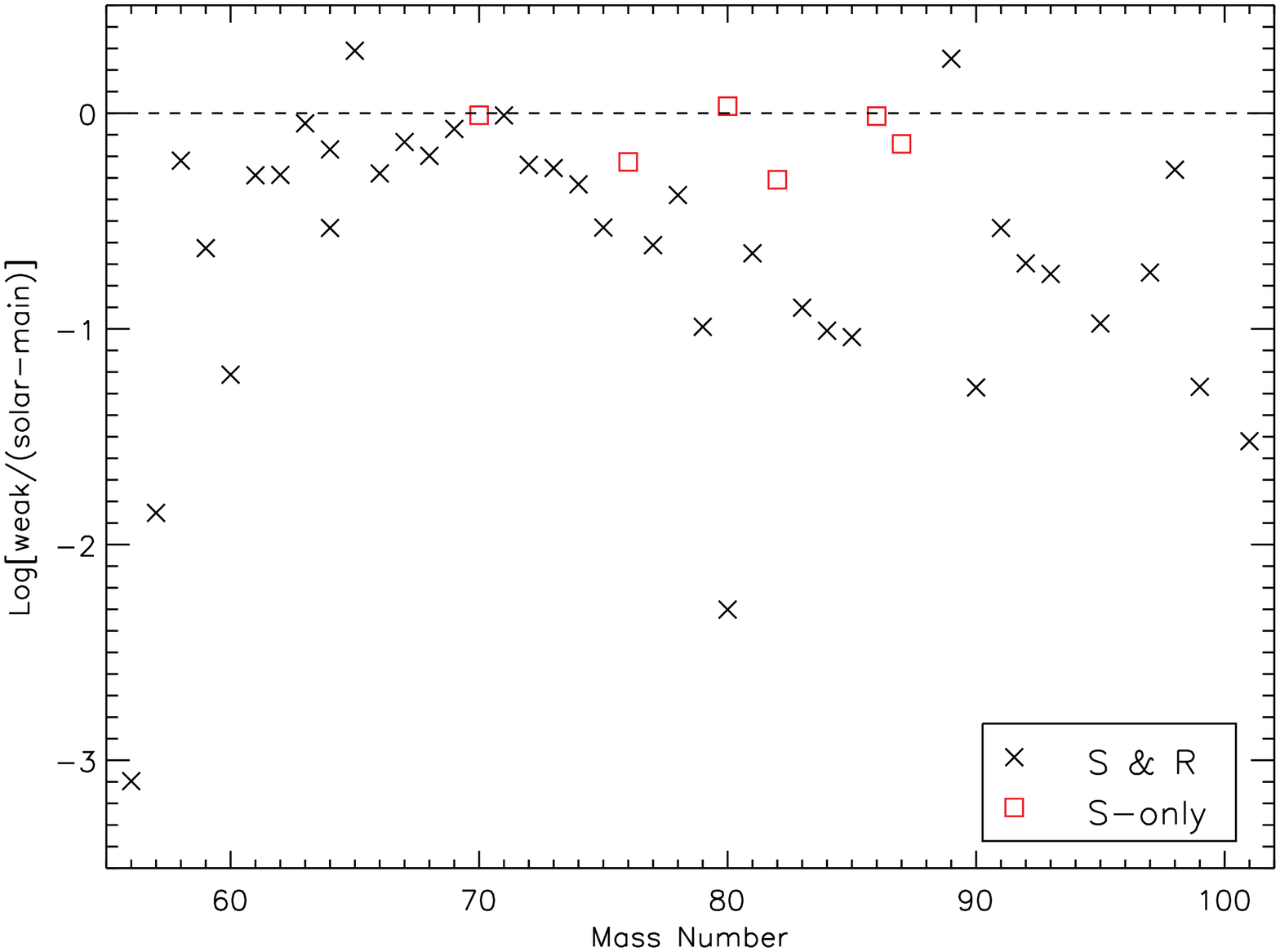}

\figcaption{\label{fig1}The calculated weak \emph{s}-process contributions to
the solar abundance pattern. \emph{\textsl{Red boxes}}: \emph{s}-only
isotopes. \textsl{Black x's}: isotopes with contributions from the
\emph{r}- and \emph{s}-processes. The yields for the \emph{s}-only
isotopes $^{70}$Ge, $^{80}$Kr, and $^{86}$Sr are reproduced well,
with underproductions for $^{76}$Se, $^{82}$Kr, and $^{87}$Sr.
The only over-productions were for the non \emph{s}-only isotopes
$^{65}$Cu and $^{89}$Y. }

In addition to producing isotopes along the path of stability, there
is indication that the weak \emph{s}-process component in massive
stars is responsible for seeding a non-negligible \emph{p}-isotopic
production component. The massive star yields from \citet{r29} show
significant isotopic productions for $^{74}$Se, $^{78}$Kr, and $^{84}$Sr,
which cannot be accounted for by the $\nu p$-process (since their
stellar models include weak \emph{s}-process and $\gamma$-process
reactions, but do not include the $\nu p$-process). A mechanism for
producing these \emph{p}-isotopes are $\gamma$-process reactions
on weak \emph{s}-process seeds, where the weak \emph{s}-process would
enhance the abundances of unstable proton-rich isotopes which would
then undergo photo-disintegration events to give stable \emph{p}-isotopes.
The \x{scaling} of these \emph{p}-isotopic yields should thus track
the weak \emph{s}-process, rather than the traditional $\gamma$-process
(as far as its dependence on metallicity), since the abundances are
made in-situ from existing weak \emph{s}-process seeds. It is unclear
exactly how much this weak \emph{s}-process enhanced $p$-process
(WSEP) should contribute to the solar $^{74}$Se, $^{78}$Kr, and
$^{84}$Sr abundances. We decided to attribute half of the solar abundances
to each the WSEP isotopes and the $\nu$\emph{p}-process for these
three \emph{p}-isotopes. 

The uncertainties in our calculation are constrained by the errors
of the MACs given in the compilation from the KADoNIS project \emph{www.kadonis.org}
\citep{r4}, and vary by isotope. We do not propagate the uncertainties
through our equations, since our treatment of the weak \emph{s}-process
is only approximate.

\subsubsection{Lighter Element Primary Process}

Indication for the need of an additional primary process distinct
from the \emph{r}-process appears to be implied by ultra-metal poor
(UMP) stellar abundances, and was first implemented by \citet{r68}
in a two component phenomenological model. This process, sometimes
referred to as the weak \emph{r}-process \citep{r69} or charged-particle
reaction process \citep{r44}, was named in more general terms by
\citet{r52} as the lighter element primary process (LEPP), and is
needed to explain an observed excess of some lower mass ($A<130$)
elements, notably Sr, Y, and Zr, that can not be accounted for by
neutron capture processes, photo-disintegration, or CCSNe. Investigation
of the triple-$\alpha$ and $^{12}$C($\mathrm{\alpha}$,n)$^{16}$O
rates indicates that their present $\mathrm{2\sigma}$ uncertainty
can not account for the necessary production of Sr, Y, and Zr by the
weak \emph{s}-process in massive stars \citep{r65}. A more recent
nucleosynthesis calculation by \citet{r66} suggests the interesting
possibility of the needed abundances being produced in the neutrino-driven
winds of ultra-metal poor (UMP) CCSNe, although their yields suffer
over-productions of additional isotopes in order to provide the necessary
Sr, Y, and Zr. In our model we do not separate out the yet unknown
LEPP process, but instead the abundances produced by this mechanism
are absorbed into the massive star category.

\subsubsection{Main S-Process}

The main component of the \emph{s-}process occurs in the thermally
pulsating AGB stellar phase for stars with $\mathrm{M\lesssim1.5M_{\text{\ensuremath{\odot}}}}$.
During hydrogen burning, protons are thought to mix downward into
the helium layer, which can then be captured on synthesized $^{12}$C
to form a $^{13}$C pocket \citep{r49}. After the subsequent helium
flash, $\alpha$-particles are convectively dredged up through this
pocket initiating a neutron source via $^{13}$C($\alpha$,n)$^{16}$O.
This neutron source drives the main \emph{s}-process by capture on
pre-existing metals contained in the star throughout several helium
flash cycles. Due to the longer neutron exposures operative in the
main \emph{s}-process (relative to the weak component of the \emph{s}-process),
the main isotopic contributions are isotopes with mass numbers $\mathrm{\mathit{A}\geq88}$
\citep{r82,r83}. 

The main \emph{s}-process contributions to the solar abundance pattern
were taken from \citet{r1}, and these yields were re-normalized to
the Lodders et al. \citeyearpar{r20} abundances. Many of the \emph{s}-isotopes
with mass numbers $A\geq88$ were consistent with their solar values,
however, some were under-produced. For the under-produced \emph{s}-only
isotopes that fell within $\mathrm{0.1\, dex}$ below their solar
abundance, we assumed them to be their nominal solar values. This
cutoff at $\mathrm{0.1\, dex}$ is admittedly somewhat arbitrary.
The uncertainties in neutron capture cross-sections and modeling of
the AGB stars introduces error, yet remarkably so many \emph{s}-only
isotopes are reproduced by \citet{r1} to within $25\%$ ($\simeq\mathrm{0.1\, dex}$)
of their solar values. Our choice of this cutoff value acknowledges
the existence of errors that are often difficult to enumerate, while
assuming that since so many \emph{s}-only isotopes are reproduced
close to their solar values, the model is likely reliable.

The resulting abundance pattern is given in Fig.\,\ref{fig2}. Below
a mass number of $\mathrm{\mathit{A}}=88$, the \emph{s}-only isotopes
show an explicit drop in their production, and the weak \emph{s}-process
has to contribute in this region to reproduce the needed solar abundances.
For mass numbers above $\mathit{\mathrm{\mathit{A}}}=88$, three \emph{s}-only
isotopes are under-produced (by more than $\mathrm{0.1\, dex}$) by
\citet{r1}: $^{152}$Gd, $^{187}$Os, and $^{192}$Pt. We attributed
the residual abundances of these three isotopes to the $\gamma$-process.
This treatment of the residuals is very approximate, and while $^{152}$Gd
and $^{192}$Pt lie on the proton-rich side of the stability line,
$^{187}$Os is preceded by the more proton rich $^{184}$Os and $^{186}$Os,
both of which would be more likely candidates for $\gamma$-process
enrichment. The contribution of $^{152}$Gd to elemental Gd is only
$\sim$$0.2\,\%$, hence relatively little $\gamma$-process enrichment
is needed to populate the residual abundance. The isotope $^{192}$Pt
contributes $\sim$$0.79\,\%$ to elemental Pt, however, the \emph{p}-isotope
$^{190}$Pt contributes even less at $\sim$$0.01\,\%$, so it is
unlikely that a residual abundance from the $\gamma$-process would
populate the needed $\sim$$0.7\,\%$ $^{192}$Pt without also raising
the abundance of the more rare $^{190}$Pt.

\includegraphics[scale=0.5,angle=90]{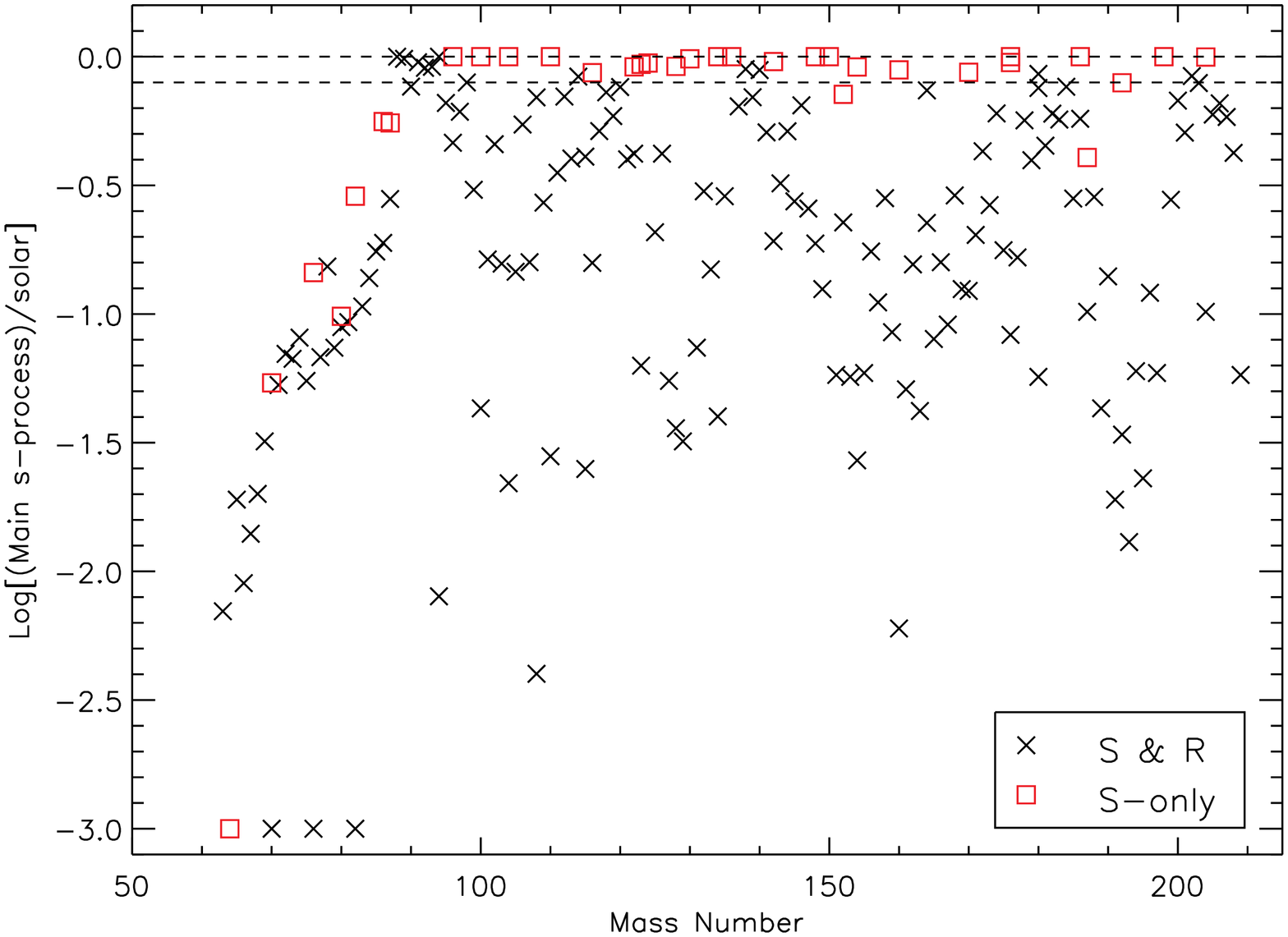}

\figcaption{\label{fig2} Adopted main \emph{s}-process abundances from Bisterzo
(2011), relative to the solar abundance pattern. \emph{\textsl{Red Boxes}}:
\emph{s}-only isotopes. \emph{\textsl{Black x's}}: isotopes with
contributions from the \emph{r}- and \emph{s}-processes. Under-produced
\emph{s}-only isotopes that fell within $\mathrm{0.1\, dex}$ (dashed
line) below their solar abundance were assumed to be their nominal
solar values.}

The main \emph{s}-process produces three characteristic peaks at closed
neutron shells, and we sub-divide the main \emph{s}-process abundances
into \emph{ls}, \emph{hs}, and ``strong'' components. The \emph{ls}
component is taken to be all abundances from Fig.\,3 up to and including
Sr, the \emph{hs} component is taken to be all abundances after Sr
and up to and including Ba. The strong component is assigned to Pb
and Bi isotopes, and is discussed in more detail below. Using the
solar main \emph{s}-process abundances given in Fig.\,3 as a starting
point, we \x{scale} the \emph{ls}, \emph{hs}, and ``strong'' components
separately due to the different neutron exposures required for their
production.

\subsubsection{Strong S-Process}

Originally, a strong component of the \emph{s}-process was introduced
to address the underproduction of Pb not accounted for by the main
component at solar metallicities \citep{r54}, and it was believed
that a third type of neutron exposure was needed to generate the remaining
Pb. Sufficient production was later found to be present in the low
metallicity regime of the main component, and the strong \emph{s}-process
component has since been re-interpreted as a low metallicity effect
of the main component \citep{r53}. As the metallicity (and hence
Fe) decreases the neutron-to-seed ratio increases, providing a sufficient
neutron exposure to make the heavy Pb and Bi isotopes. Simulations
of low metallicity AGB stars show production of ``strong'' \emph{s}-process
isotopes at $\mathrm{[Fe/H]=-2.6}$ that exceed production at solar
metallicities by several dex, depending on $^{13}$C pocket efficiencies
\citep{r86}. This implies nearly all of the solar ``strong'' \emph{s}-process
abundances are made at low metallicities.

Here we continue to refer to Pb and Bi \emph{s}-process contributions
as coming from the ``strong'' component, since they are \x{scaled}
distinctly from the \emph{hs} and \emph{ls} parts of the main \emph{s}-process,
but it should be noted this is a convention introduced for clarity,
and the strong \emph{s}-process is indeed the low metallicity regime
of the main \emph{s}-process, having a distinctly stronger neutron-to-seed
ratio.

\subsubsection{R-Process}

The \emph{r}-process synthesizes isotopes beyond the iron peak using
rapid neutron capture (relative to the beta decay rate; \citealp{r41}).
The location of this process has not been universally accepted, but
historically was first thought to occur in CCSNe environments \citep{r43}.
More recently it has been postulated to occur in shocked surface layers
of O-Mg-Ne proto-neutron stars \citep{r24,r44}, and also simulations
have shown success in reproducing \emph{r}-process signatures from
$\nu$-driven nucleosynthesis in the He-shell during CCSNe in the
low metallicity regime ($\mathrm{Z<-3.0}$; \citealp{r64}). A recent
principle component analysis has linked r-process elemental abundances
with alpha-elements, further suggesting CCSNe as a possible site for
this process \citep{Ting2012}.

The \emph{r}-process proceeds far beyond the neutron-rich side of
stability and (as with the \emph{s}-process) also bottlenecks at closed
neutron shells. Since the closed neutron shells are encountered on
the neutron rich side of stability, the proton number is lower at
these bottlenecks for the \emph{r}-process than it is for the \emph{s}-process,
hence after decay to stability the \emph{r}-process peaks always occur
at lower mass numbers relative to the \emph{s}-process peaks, at approximately
80, 130, and 195. Elements of interest for representing the \x{scaling}
of \emph{r}-process elements that have \emph{r}-only isotopes include
germanium, europium, and platinum. The explosive environment of the
\emph{r}-process does not \emph{directly} depend on the initial metallicity
of the star, and therefore is primary. Note, however, that the stellar
populations that may be responsible for the \emph{r}-process could
depend on Z, e.g., through metallicity-dependence of evolution and
stellar mass loss.

The \emph{r}-process solar abundance contributions were determined
using the residual method, in the mass range $\mathrm{69\leq\mathit{A}\leq238}$.
The \emph{r}-process contributions in the range $\mathrm{56\leq\mathit{A}\leq68}$
were \emph{not} differentiated from the ``massive star category.''
This choice was made to facilitate the decomposition of the solar
abundance pattern in this range, where both CCSNe yields (from $\alpha$-rich
freeze-out, the $\alpha$-process, and previous nuclear burning) and
\emph{r}-process yields contribute to abundances as primary processes,
and are difficult to separate.

\subsubsection{P-Isotopes\label{sub:P-Isotopes}}

The $\nu$p-process occurs in CCSNe environments where high neutrino
fluxes create proton-rich ejecta via weak reactions with neutrons
and protons \citep{r7}. Additional neutrino interactions with the
left-over protons after $\alpha$-rich freeze-out produces a neutron
abundance, which then undergo $\mathrm{(n,p)}$ strong reactions.
The resulting protons can then be captured allowing the synthesis
of proton rich isotopes up to mass number of $\mathrm{\mathit{A}\simeq100}$
\citep{r7,r22}. Since the evolution of this process depends upon
$\nu$-interactions with free nucleons, it is independent of the initial
star's metallicity, and is considered a primary process.

Proton-rich isotopes beyond mass number $\mathrm{\mathit{A}\simeq100}$
are believed to be created by the $\gamma$-process, where successive
photo-disintegration events on pre-existing metals occur in the interior
layers of supernovae \citep{r45,r46}. The target nuclei for such
events are previously synthesized metals from the \emph{r}- or \emph{s}-processes.
This process generally begins with $\mathrm{(\gamma,n)}$ reactions
moving the target nuclei to the proton rich side of the stability
line, where the rate of $\mathrm{(\gamma,p)}$ and $\mathrm{(\gamma,\alpha)}$
start to dominate. Whereas all \emph{p}-isotope abundances are relatively
small and do not dominate their respective elemental abundances, they
cannot be accounted for by \emph{s}- or \emph{r}-process production,
and so both of the \emph{p}-isotope production processes discussed
are required to provide a complete galactic-chemical history.

Similar to the main and weak components of the \emph{s}-process, we
expect to have a transition region between the $\nu$\emph{p}-process
and $\gamma$-process. A calculation of $\nu$\emph{p}-process yields
for a 15M$_{\odot}$ star was performed by \citet{r33}. Their results
show sufficient production of Mo and Ru isotopes relative to their
solar abundances, which is much needed due to well known deficiencies
in the $\gamma$-process production of these isotopes. From their
results, the $\nu$\emph{p}-process yields are shown to decrease quite
rapidly beyond mass number $\mathrm{\mathit{A}=100}$. This suggests
that the region of overlap is fairly abrupt, with only $^{102}$Pd
beyond a peak at $^{98}$Ru owing non-negligible fractions of their
abundances to the $\gamma$-process. 

A more complete analysis, perhaps using a grid of stellar masses in
order to determine the distribution in the transition region, would
likely offer only minor corrections to the final elemental \x{scalings}.
Instead, new free parameters from the stellar and nucleosynthesis
modeling would be introduced. The complete isotopic decomposition
for the solar abundance pattern is shown in Section\,\ref{sub:Solar-Abundance-Decomposition},
and can be used for future isotopic reference.

\section{Model Description\label{sec:Model-Description}}

Our simple model tries to describe a typical average galactic composition,
where the initial state is a homogeneous BBN composition. The final
state is taken as a homogeneous composition equivalent to the isotopic
solar abundance pattern. Each process responsible for isotopic production
(and depletion for the case of D and $^{3}$He) is assumed to enrich
(deplete) our model galaxy under the instantaneous mixing approximation,
to preserve homogeneity across the entire metallicity range considered.
Each isotope from our decomposition of the solar abundance pattern
is \x{scaled} as a function of a chosen normalized dimensionless
parameter \emph{$\xi$}. At \emph{$\xi$}=0 the galaxy is in the BBN
composition, and at \emph{$\xi$}=1 the galaxy has the solar composition.
The benefit of this choice of parametrization is that its range is
congruent with the total metallicity (relative to solar), Z/Z$_{\odot}$.
A comparison between \emph{$\xi$} and Z/Z$_{\odot}$ is shown in
Section\,\ref{sec:Results-and-Discussion}. This choice of parameter
motivates choosing functional forms for the \x{scalings} of each
process by addressing the predicted dependence on total metallicity
these processes would obey. Note that this comparison should not be
interpreted as a physical condition imposed upon this parameter. Indeed
the parameter $\xi$ is not a physical quantity,\textbf{ }rather it
is a technical parameter that takes continuous values in the range
{[}0,1{]}, and is chosen to closely approximate Z for ease of discussion.
Hence, we assume the relation between \textbf{\emph{$\xi$}}\textbf{
}and Z/Z$_{\odot}$ to be log($\xi$)={[}Z{]}. This ansatz is checked
later in Section\,\ref{sec:Results-and-Discussion}.

The parametrization of our model is an attempt to scale the isotopic
abundances characterized by typical trends in our Galaxy. For consistency
we do not directly use metallicity as the argument of the scaling
functions; indeed metallicity is an output of the model, and hence
cannot be used as an input. In practice, normalized metallicity Z/Z$_{\odot}$
tracks $\xi$\emph{ }very closely, see Fig.\,\ref{fig10}, and the
deviation is typically less than 2\%. Whereas our model is not a traditional
GCE model that uses scaled stellar yields, our parametrization allows
us to improve upon the standard of scaling solar abundances by a constant
factor, for use in nucleosynthesis studies.

In the following we address the functional form for each process described
in Section\,\ref{sec:Astrophysical-Processes-and}.

\subsection{Massive Stars}

The results of massive star simulations for Population III stars (discussed
previously in Section\,\ref{sub:Intermediate-Mass-and-Iron}), in
addition to being used to compute the solar abundance decomposition,
are also used to model the isotopic abundance pattern for $\mathrm{12\le\mathit{A}\le68}$
at a low metallicity of $\mathrm{[Fe/H]=-3}$. The abundances from
this simulation were normalized to the $^{56}\mathrm{Fe}$ abundance
at $\mathrm{[Fe/H]=-3}$, under assumption that Type Ia contributions
to $^{56}\mathrm{Fe}$ are negligible at this metallicity, hence each
abundance was multiplied by the factor, $^{56}\mathrm{Fe}_{\odot}/X_{i}^{\mathrm{sim}}$,
where $X_{i}^{\mathrm{sim}}$ is the abundance of isotope \emph{i}
from the PopIII simulation. These normalized abundances represent
a third fixed point for the massive star isotope abundances (in addition
to the solar abundances and BBN abundances). Then the massive abundances
were interpolated linearly in log-space between their solar values
(found in Section\,2.4) and their respective abundances given by
the normalized PopIII simulation. That is, for each isotope the abundances
were scaled according to,

\begin{equation}
\mathrm{log}\left(X_{i}^{*}\left(\xi\right)\right)=m_{i}\left(\mathrm{log}\left(\xi\right)-\mathrm{log}\left(\xi_{low}\right)\right)+\mathrm{log}\left(X_{i}^{\mathrm{sim}}\right)\label{eq:massive_interpolation}
\end{equation}
where $X_{i}^{*}\left(\xi\right)$ is the massive abundance of isotope
\emph{i} as a function of the model parameter, and $\mathrm{log}\left(\xi_{low}\right)=-2.5$
(corresponding to a metallicity of $\mathrm{[Fe/H]=-3}$). The slope
is defined as, $m_{i}\equiv\left(\mathrm{log}\left(X_{i,f}^{\mathrm{massive}}\right)-\mathrm{log}\left(X_{i}^{\mathrm{sim}}\right)\right)/\left(\mathrm{log}\left(\xi_{\odot}\right)-\mathrm{log}\left(\xi_{low}\right)\right)$,
where $X_{i,f}^{\mathrm{massive}}$ is the massive star contribution
to the solar abundance (found in Equation\,\ref{eq:s_decomp_massive}),
and obviously $\mathrm{log}\left(\xi_{\odot}\right)=0$. The interpolation
given in Equation\,\ref{eq:massive_interpolation} is extrapolated
in both directions from $\mathrm{[Fe/H]=-3}$ to the BBN abundances
(zero for these isotopes) and from $\mathrm{[Fe/H]=0}$ to super-solar
values, which gives massive abundances for these isotopes across the
entire metallicity range.

As mentioned above, the parameter $\xi$ takes a value of $\mathrm{log(\xi)=-2.5}$
at $\mathrm{[Fe/H]=-3}$. The reasoning behind this choice is as follows.
At solar metallicity massive stars are responsible for roughly 30\,\%
of the total iron, but nearly 100\,\% of the alpha isotopes. So then
$\mathrm{[Fe]_{massive}=-0.5}$, and $[\alpha]_{\mathrm{massive}}=0$.
At lowest metallicities, alpha isotopes effectively comprise the aggregate
of metals by mass, since secondary sources and Type Ia SNe do not
provide enrichment until later. Hence $[Z]\backsimeq[\alpha]_{\mathrm{massive}}$
holds in low metallicity regimes and will trail $\mathrm{[Fe/H]}$
by a nearly constant $0.5\,\mathrm{dex}$ until Type Ia onset. We
then use our model parameter \emph{$\xi$} in place of Z/Z$_{\odot}$.
This comparison between $\xi$ and $\mathrm{[Fe/H]}$ is approximate,
and relies on the normalized metallicity Z/Z$_{\odot}$ tracking $\xi$\emph{
}very closely, see Fig.\,\ref{fig10}.

\subsection{Type Ia SNe}

There are three constraints for choosing a parametrization for Type
Ia isotopes as a function of $\xi$. First, Type Ia contributions
experience a delay before they begin to enrich the ISM. This is due
to the time necessary for, in the accreting white dwarf model, an
intermediate to low mass star to evolve through its main sequences
and then accrete sufficient material to surpass the Chandrasekhar
mass limit. In the meanwhile, massive stars continue to enrich the
ISM, making the galaxy more metal rich. The metallicity at which Type
Ia are able to begin contributing to ISM enrichment is typically constrained
to the interval $\mathrm{-2<[Fe/H]<-1}$ depending on environment,
galaxy size, etc., but is usually favored towards the upper bound.
Hence, Type Ia contributions should be negligible below some value
of $\xi$. Second, upon crossing the Type Ia onset value, contributions
should gradually rise to their solar values. Third, at high $\xi$
the scalings should behave essentially like a primary process. Note
that this last constraint is an assumption of our model and could
be true only for a flat star formation rate (SFR). In fact investigation
into Type Ia progenitors by \citet{r85} predicts that the ratio of
CCSNe to Type Ia rates to be increasing with redshift, a conclusion
reached by posing a two component delay time distribution (DTD) progenitor
model. If indeed there exists a Type Ia progenitor that operates at
low redshifts and experiences a longer delay between the formation
of the WD and the later Type Ia explosion (named 'tardy' by \citealp{r85}),
a larger number of CCSNe relative to number of CCSNe at the birth
of the WD contaminate the ISM, thus increasing the metallicity at
a faster rate than it was increasing at the birth of the WD. Hence
once the Type Ia SNe do explode, their products would have a metallicity
dependence different than linear, due to this increasing rate ratio
of CCSNe/SNe Type Ia. We are careful to note that it is unclear exactly
how a non flat SFR would impact our scaling for Type Ia SNe, since
the above discussion relies on time, and a relation between our model
parameter and time cannot be formed. We simply note that there may
indeed be an effect. The analogy presented above with respect to flat
SFRs serves to construct a rough behavior of Type Ia abundances, and
does not mean to attach the physical quantity of time to our model
parameter. Furthermore, Type Ia onset may not occur at one unique
metallicity, there may be a spread - in how far the different environments
have evolved in terms of Z (and O as its main tracer) - when Type
Ia sets in. Hence, the corresponding parameter value $\xi$ for Ia
onset is not unique for all constituents, but rather just a typical
average that fits the different environments that we sample as as
function of it.

One specific form which satisfies all three constraints discussed
is given, e.g., by a scaled and shifted hyperbolic tangent base function,

\begin{equation}
\mathrm{\mathit{X}{}_{\mathit{i}}^{Ia}(\xi)=\mathit{X}{}_{\mathit{i},\odot}^{Ia}\cdot\mathit{\xi}\cdot[\tanh(\mathit{\mathit{a}\cdot\mathit{\xi-b}})+\tanh(\mathit{b})]/[\tanh(\mathit{a-b})+\tanh(\mathit{b})]}.
\end{equation}
The specific value for Type Ia onset is determined by fitting the
free parameters, \emph{$a$} and \emph{$b$}, against available data.
Note the hyperbolic tangent function is tempered with a linear factor
of $\xi$ to ensure the behavior is linear near solar, where otherwise
the $\mathrm{\tanh(x)}$ function would asymptote. Note further that
whereas this function is phenomenologically motivated it is not unique.
The $\mathrm{\arctan(x)}$ and $\mathrm{erf(x)}$ also satisfy the
above constraints, however, the $\mathrm{erf(x)}$ is more computationally
expensive than the $\mathrm{arctan(x)}$ and $\mathrm{tanh(x)}$ functions.

\subsection{Neutron Capture and P-Isotopes}

The\emph{ }main \emph{s}-process (\emph{ls, hs, }and ``strong''
component), weak \emph{s}-process, \emph{r}-process, $\nu$p-process,
and $\gamma$-process contributions are parametrized as power laws,

\begin{equation}
\mathrm{\mathit{\mathit{X}{}_{i}^{{\normalcolor {\normalcolor \mathrm{strong}}}}(\xi)}=\mathit{c\left[1-\frac{\tanh\left(d\cdot\xi+g\right)}{\tanh\left(d+g\right)}\right]+X{}_{i,\odot}^{\mathrm{\mathit{\mathrm{strong}}}}}}
\end{equation}

\begin{equation}
\mathrm{\mathit{\mathit{X}{}_{i}^{\mathrm{{\normalcolor {\normalcolor l\mathrm{s}}}}}(\xi)}=\mathit{X{}_{i,\odot}^{\mathrm{\mathit{\mathrm{ls}}}}}\cdot\mathit{\xi}^{\mathit{l}}}
\end{equation}

\begin{equation}
\mathrm{\mathit{\mathit{X}{}_{i}^{\mathrm{{\normalcolor {\normalcolor h\mathrm{s}}}}}(\xi)}=\mathit{X{}_{i,\odot}^{\mathrm{\mathit{\mathrm{hs}}}}}\cdot\mathit{\xi}^{\mathit{h}}}
\end{equation}

\begin{equation}
\mathrm{\mathit{X}_{\mathit{\mathrm{\mathrm{\mathit{i}}}}}^{ws}(\mathit{\xi})=\mathit{X}{}_{\mathit{i},\odot}^{ws}\cdot\mathit{\xi}^{\mathit{w}}}
\end{equation}

\begin{equation}
\mathrm{\mathit{X}{}_{\mathit{i}}^{r}(\xi)=\mathit{X}{}_{\mathit{i},\odot}^{r}\cdot\mathit{\xi^{p}}}
\end{equation}

\begin{equation}
\mathrm{\mathit{X}{}_{\mathit{i}}^{\nu p}(\mathit{\xi})=\mathit{X}{}_{\mathit{i},\odot}^{\nu p}\cdot\mathit{\xi^{p}}}
\end{equation}

\begin{equation}
\mathrm{\mathit{X}{}_{\mathit{i}}^{\gamma}(\xi)=\mathit{\mathit{X}}{}_{\mathit{i},\odot}^{\gamma}\cdot\mathit{\xi}^{\frac{\mathit{h}+\mathit{p}}{2}+1}}
\end{equation}
The free parameters \emph{$p$} and \emph{$h$} denotes the power
of the $\mathit{\xi}$-dependence for primary and secondary (\emph{hs})
processes, respectively. The weak component of the\emph{ s}-process
was given its own parameter \emph{$w$}. The motivation for choosing
a power law dependence on these parameters lies in the relations between
abundance and metallicity for primary and secondary events, namely
that primary events produce abundances linear in metallicity, and
secondary produce abundances quadratic in metallicity. The chosen
model parameter $\mathit{\xi}$ takes the place of total metallicity
(as discussed in the beginning of this Section), and the exponents
are parametrized for fitting with data. The metallicity dependence
for the $\gamma$-process is less clear. It is possible for the target
nuclei for this process to be either secondary or primary in origin,
and produced in a previous astrophysical environment via the \emph{s}-
or \emph{r}-process. Hence photo-disintegration events can potentially
be a tertiary process. All $\gamma$-process isotopes have low (<1-10\,\%)
contributions to their respective elemental abundances, hence we chose
not to assign a separate free parameter for this process, since the
fit to data would likely be poorly constrained. Instead a compromise
was struck between the possible \emph{s}- or \emph{r}-process origins
and the chosen exponent for the $\gamma$-process represents an equal
primary and secondary seed; about half the metals which might be target
nuclei for this process are created by the \emph{s}-process, and the
other half from the \emph{r}-process. Note that the proposed $\gamma$-process
abundances enhanced from weak \emph{s}-process seeds in the WSEP process
(discussed at the end of section 2.4.) are scaled the same as the
weak \emph{s}-process, not as $\gamma$-process yields. 

The choice for the ``strong'' \emph{s}-process function is motivated
phenomenologically. At low ($\mathrm{[Fe/H]\lesssim-2.0}$) metallicities
essentially all of the solar abundances for the ``strong'' \emph{s}-process
have been made. Hence the abundances should be close to constant between
$-2.0\lesssim\mathrm{[Fe/H]\lesssim0}$. Additionally, since AGB stars
experience a time delay in their contributions, below $\mathrm{[Fe/H]\lesssim-2.0}$
the abundances drop smoothly, hitting zero at some unknown metallicity.
The free parameters \emph{$c$, $d$, }and\emph{ $g$ }are used to
adjust the shape of the hyperbolic tangent function, and constrain
the peak abundance achieved at low metallicity, the metallicity at
which the abundances begin to drop to zero, and the rate at which
the abundances drop to zero.

\subsection{Hydrogen Burning, Classical Novae, $\nu$-Process, and Galactic Cosmic
Ray Spallation}

The \x{scaling} of standard GCR spallation abundances has the same
functional form as the \emph{s}-process \x{scaling}. Novae, primary
GCR spallation, $\nu$\emph{p}-process, deuterium, and helium (from
hydrogen burning) abundances were \x{scaled} the same as the \emph{r}-process,
but with an offset added to reflect BBN abundances at $\xi=0$. The
remaining isotope of hydrogen,$^{1}\mathrm{H}$, was \x{scaled} according
to,

\begin{equation}
X{}_{i}^{\mathrm{H}}(\xi)=X{}_{i,\odot}^{\mathrm{H}}\cdot[1.0-\xi\cdot Z_{\odot}-Y(\xi)-D(\xi)],
\end{equation}
where \emph{$\mathit{Y}(\xi)$} is the mass fraction of the helium
isotopes, $D(\xi)$ is the mass fraction of deuterium, and Z$_{\odot}$
is the solar value of total metallicity, given as Z$_{\odot}$=0.0153
\citep{r20}. Effectively, this \x{scaling} is simply a restatement
of the sum of mass fractions, $\mathrm{X+Y+Z=1}$.

\section{Fitting Scaling Model to Observational Data\label{sec:Fitting-Scaling-Model}}

The isotopic scaling functions were summed into elemental scaling
functions for the purpose of fitting the free parameters. An example
of this algorithm is given in the Section\,\ref{sub:Comparison-to-Linear}.
Two compilations of stellar abundance data were used for the fitting.
The \citet{r6} low metallicity data set contains over 1000 stars
from the Milky Way (MW) halo and dwarf galaxies. Note that this set
is a compilation from various sources and there is hence an unknown
source of varying systematic errors. Nevertheless, the spread of the
averages of this data (described below) is much bigger than the provided
errors and likely spans these systematic errors, though may be subject
to systematic offsets. The dwarf galaxy abundances were removed from
the set, since it is likely that dwarf galaxies exhibit a different
GCE than the MW. In fact a future extension of the current model could
be applied to dwarf galaxies to give insight into their GCE relative
to the MW. Additionally, \citet{r6} data from stars in binary systems
was removed from the {[}Ba/Fe{]} and {[}Sr/Fe{]} data, since binary
systems experience more enriched s-process abundances due to accretion.
Binary stars were identified from the online CHARA catalog http://www.chara.gsu.edu/\textasciitilde{}taylor/catalogpub/catalogpub.html
\citep{r32}, and were removed from the data set. In addition to removing
the known binaries identified by the CHARA catalog, stars that simultaneously
satisfied the following criteria: $\mathrm{[Fe/H]<-2.0}$, $\mathrm{[Eu/Fe]>0.5}$,
and $\mathrm{[Ba/Fe]>0.0}$ were also removed, since they are likely
also candidates for binary systems not covered by the CHARA catalog.
The other compilation used was the \citet{r30} data set, which contains,
in part, 725 stars with magnesium abundances. The \citet{r30} data
is needed for high metallicity abundances which fill a paucity of
Mg data in the range $\mathrm{-1<[Fe/H]<0}$ offered by \citet{r6}.

All data is given in units of {[}X/Fe{]} as a function of {[}Fe/H{]}.
The {[}Fe/H{]} axis was split into 300 bins in the range $\mathrm{-5<[Fe/H]<1}$,
and the {[}X/Fe{]} axis was split into 1500 bins in the range of the
elemental data. Each data point was then assigned a Gaussian distribution
using nominal values for the errors in the observations, 

\begin{equation}
\mathrm{f{}_{i}=\exp\left\{ -0.5\cdot\left([(x_{i}-x_{0})/\sigma_{x}]^{2}+[(y_{i}-y_{0})/\sigma_{y}]^{2}\right)\right\} ,}\label{eq:gaussian}
\end{equation}

where f$\ensuremath{_{\mathrm{i}}}$ is the value of the distribution
at bin $\mathrm{x_{\text{i}}}$ on the {[}Fe/H{]} axis and bin $\mathrm{y_{\text{i}}}$
on the {[}X/Fe{]} axis, $\mathrm{x_{\text{0}}}$ and $\mathrm{y_{0}}$
are the {[}Fe/H{]} and {[}X/Fe{]} values of each data point, and $\mathrm{\sigma}_{x}$
and $\mathrm{\sigma}_{y}$ are the nominal errors for each $\mathrm{x_{\text{0}}}$
and $\mathrm{y_{0}}$. A nominal error of $\mathrm{0.1\, dex}$ was
assumed for {[}Fe/H{]} and {[}Mg/Fe{]}, and a nominal error of $\mathrm{0.15\, dex}$
was assumed for {[}Ba/Fe{]}, {[}Sr/Fe{]}, {[}Eu/Fe{]}, and {[}O/Fe{]}.
The Gaussian contributions were summed for each bin, and assigned
an average and standard deviation for each bin. These binned averages
and standard deviations were used for sampling the parameter spaces.
We note that all fitting is done in the logarithmic space as is common
practice. One could argue that abundances and yields are ``additive,''
hence the linear space should be used. This is left for future work.
Indeed fitting in the linear space would be impractical for large
dynamic ranges on the one hand; for ratios and small ranges the logarithmic
space should suffice.

\subsection{Type Ia Parameters}

For fitting the Type Ia parameters $a$, $b$, and the fraction of
the solar $^{56}$Fe abundance attributed to Type Ia SNe, $f$, {[}Mg/Fe{]}
data was chosen due the large number of data points that exist for
this element as well as it owing its \x{abundance} to both massive
stars and Type Ia SNe. The parameter spaces for $a$, $b$, and $f$
were chosen to minimize $\mathrm{\chi}_{\mathrm{r}}^{2}$ between
the {[}Mg/Fe{]} \x{scaling} model and the averages of the binned
data. The results are given in Fig.\,\ref{fig3}.

\includegraphics[scale=0.5,angle=90]{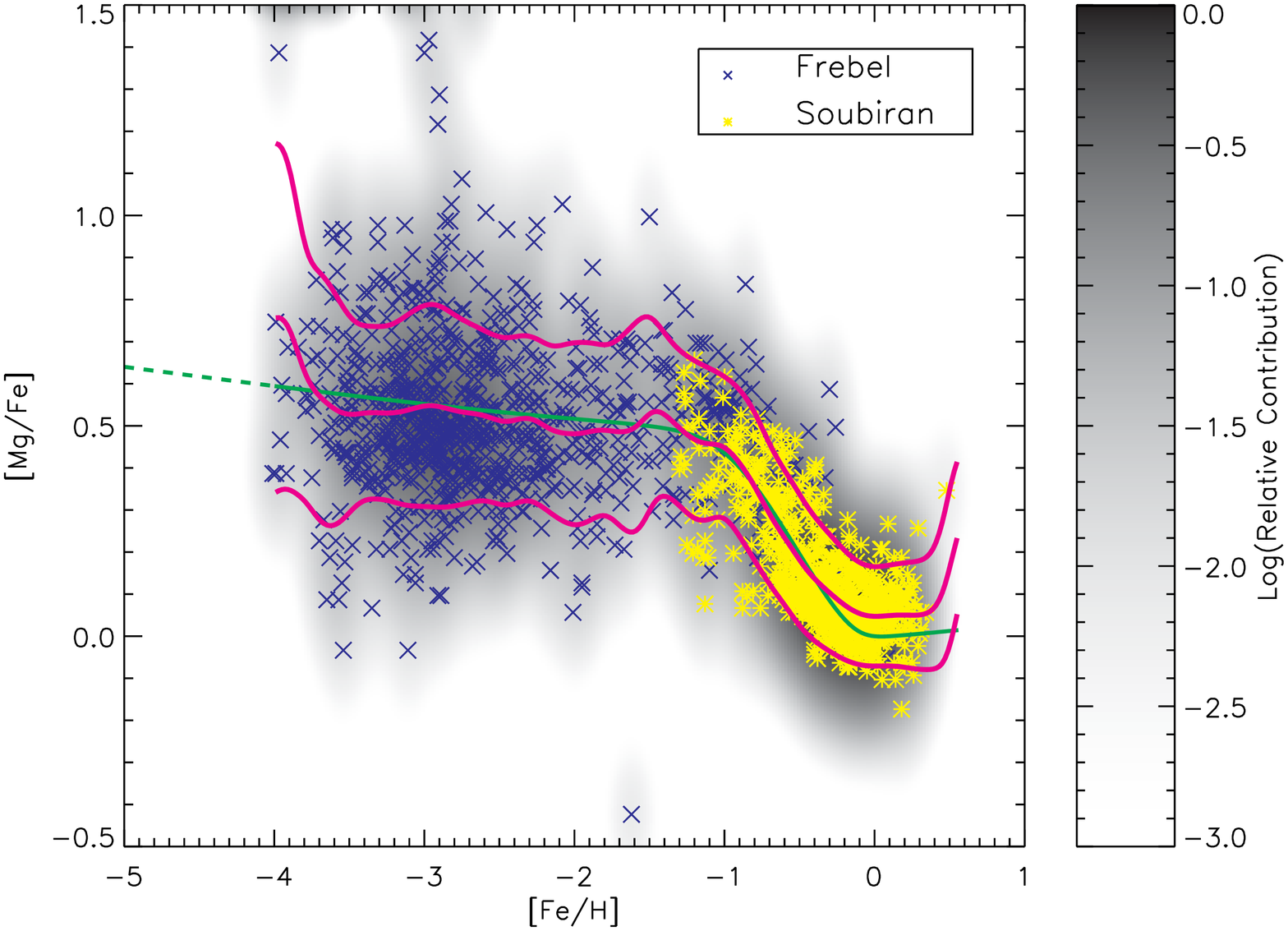}

\figcaption{\label{fig3}The resulting \x{model for} {[}Mg/Fe{]} found by parameter
fitting. \emph{\textsl{Blue x's}}: \citet{r6} data points. \emph{\textsl{Yellow asteriks}}:
\citet{r30} data. The dark shadow background shows the errors in
the data, depicted by a Gaussian distribution about the data points.
The averages and standard deviations for the Gaussian contributions
per bin are given by the central and exterior solid thick (pink) lines,
and the model itself is shown as a solid thin (green) line. There
are a couple outlying data points on the graph whose Gaussian contributions
can be seen at the upper limit of the {[}Mg/Fe{]} axis.}

As shown in Fig.\,\ref{fig3}, the magnesium \x{scaling} falls well
within the standard deviation of the data, and tracks the average
well. The determined best fit parameter values are \emph{$a$ }$\mathbf{=\mathrm{5.024}}$,
\emph{$b$ }$=2.722$, and $f$\emph{ }$\mathrm{=0.693}$, with a
resulting $\mathrm{\chi_{r}^{2}=0.0317}$. This low $\mathrm{\mathrm{\chi}}_{\mathrm{r}}^{2}$
reflects the large spread inherent in stellar abundances and the observational
uncertainties that exist, and also the fact that the bins (and data)
are not uncorrelated, as each star is ``spread'' out over many bins
due to the assigned Gaussians. The drop of the curve to solar values
from its peak at $\mathrm{[Fe/H]\simeq-1}$ is caused by Type Ia onset,
when Fe production begins to dominate over magnesium. 

The relatively flat line (only $\approx$0.1\,dex drop in $[\mathrm{Mg/Fe}]$
over 2\,dex in $[\mathrm{Fe/H}]$) depicted in the figure below $\mathrm{\mathrm{[Fe/H]\lesssim-4}}$
describes magnesium and iron abundances scaling together. This is
a consequence of our model assuming massive stars are the sole and
unique source of metals at low metallicity. Whereas it is likely that
magnesium and iron co-evolve in this range, due to their shared primary
(massive star) origin that dominates at low metallicities, it is unsubstantiated
from any data that the correlation is as exact as the model forces
it to be. Indeed, the model cannot predict \x{abundances} below $\mathrm{[Fe/H]\simeq-4}$
with any reliability, given the paucity of data at such low metallicities.
It could be argued that ``average'' \x{scaling of abundances} is
not even a well defined concept in this range, since individual astrophysical
events can have such a large stochastic effect on the metallicity
content. In this sense, our model assumes ``\x{scaling}'' is a
consequence of varying amounts of mixing with BBN from the same stellar
sources, and until sufficient stellar events can reliably produce
an average, our model is not a statistically accurate description
of this mixing.

\subsection{R-Process and \emph{hs} Parameters}

For determining the value for the\emph{ r}-process parameter $p$,
the chosen data was {[}Eu/Fe{]}. Europium is an \emph{r}-process peak
element with two isotopes, $^{151}$Eu and $^{153}$Eu, both of which
have dominant ($\sim$ 85\,\% its solar value) contributions from
the \emph{r}-process. We determined the optimized value for $p$ using
data from the Frebel set (2010). The best fitting \x{scaling} for
{[}Eu/Fe{]} is given in Fig.\,\ref{fig4}.

\includegraphics[scale=0.5,angle=90]{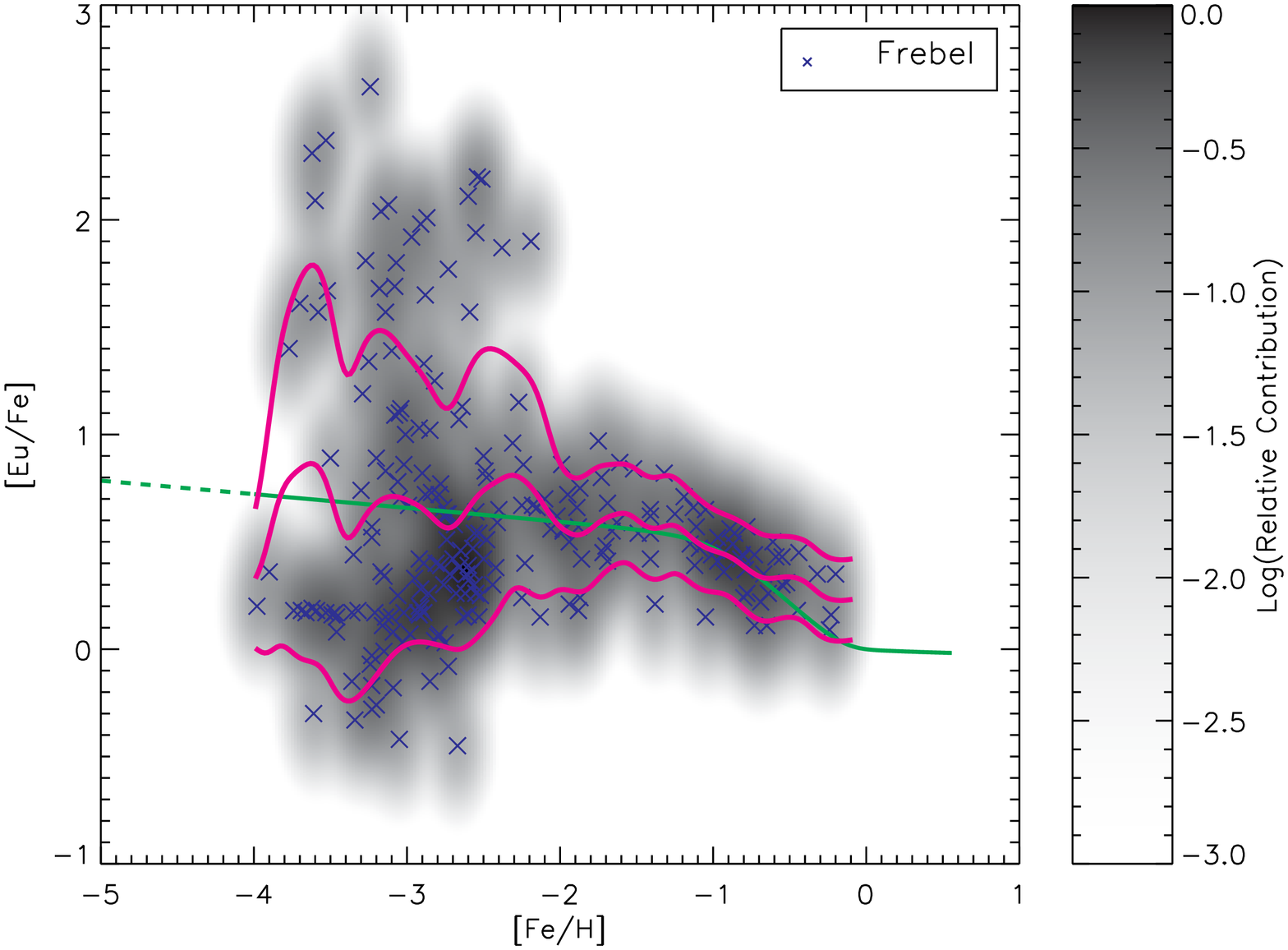}

\figcaption{\label{fig4}The resulting \x{model for} {[}Eu/Fe{]} found by parameter
fitting. The full \x{scaling} is shown in the thin solid (green)
line. \emph{\textsl{Blue x's}}: Frebel (2010) data points. The rest
of the figure follows the convention of Fig.\,3. }

The best fit parameter value was found to be $p$\emph{ }$=0.938$
with $\mathrm{\chi_{r}^{2}=0.040}$. The previously found best fit
values for $a$, $b$, and $f$ (from Section 4.1) were used for Fe.
Note that a nominal value of $h$$=1.5$ for the \emph{s}-process
parameter was assigned for the purpose of fitting the {[}Eu/Fe{]}
\x{model}. The choice of this nominal value for $h$ has a negligible
impact on the best fit value for $p$ due to the small \emph{s}-process
component of Eu, and for comparison an optimization of the parameter
space for $p$ with an $h$ value of 2 yields a best fit value of
$p$ $=0.935$ with $\mathrm{\chi_{r}^{2}=0.041}$, a difference of
0.3\,\% in $p$, and a difference of 2.5\,\% in $\chi_{\mathrm{r}}^{2}$.

For fitting the values for the heavy main \emph{s}-process parameter
$h$, the chosen data was {[}Ba/Fe{]}. This element has two \emph{s}-only
isotopes, $^{134}$Ba and $^{136}$Ba, along with three isotopes with
contributions from both the \emph{s}- and \emph{r}-processes, $^{135}$Ba,
$^{137}$Ba, and $^{138}$Ba, and a small elemental contribution ($\sim$0.2\,\%
its solar value) from two $\gamma$-process isotopes, $^{130}$Ba,
and $^{133}$Ba. In Fig.\,\ref{fig5}, we plot results for the \x{scaling}
of barium with the best fit value for $h$. The previously found best
fit values (from Section 4.1) for $a$, $b$, $f$, and $p$ were
used for Fe and the \emph{r}-process contributions to Ba.

\includegraphics[scale=0.5,angle=90]{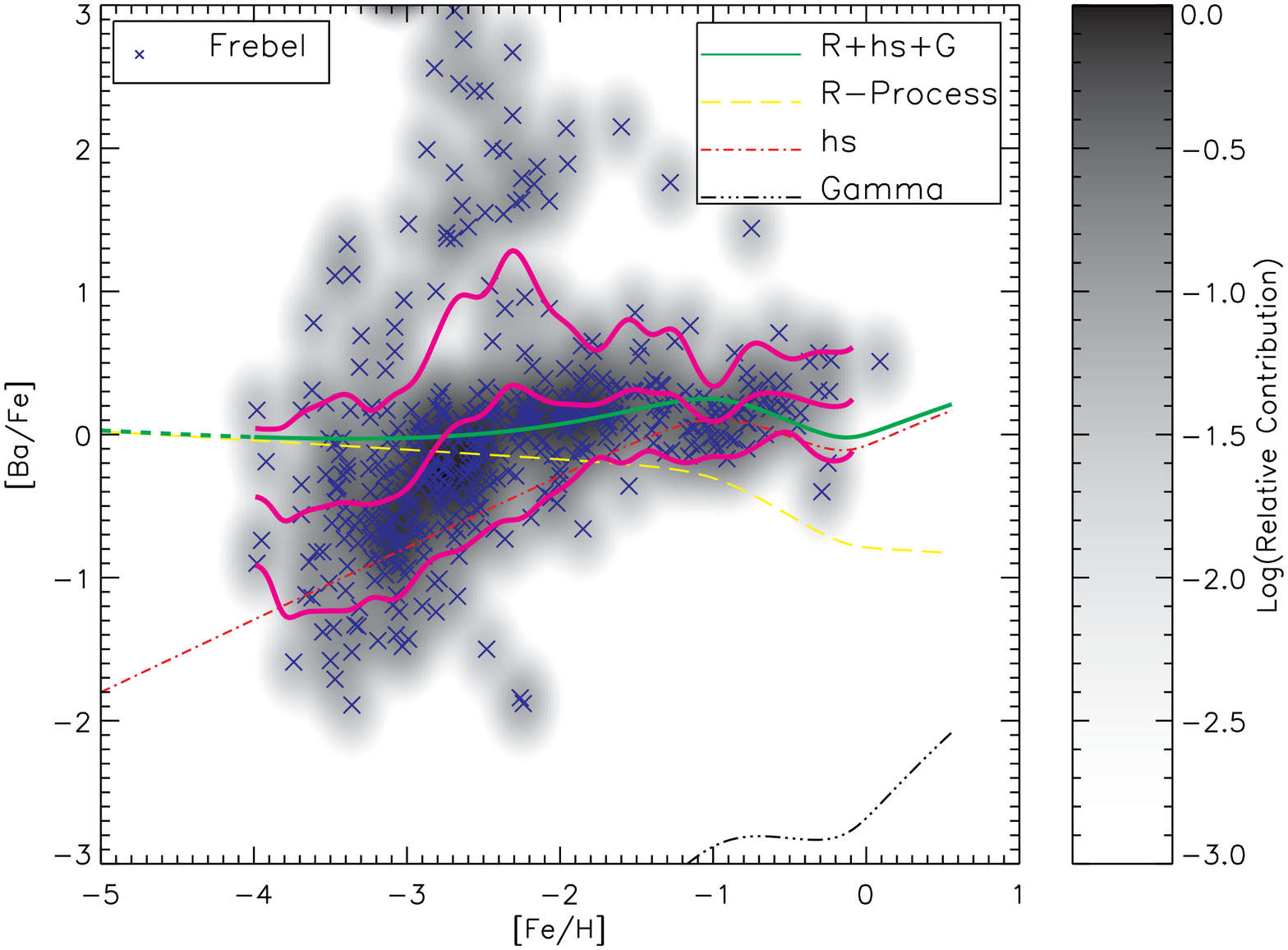}

\figcaption{\label{fig5}The resulting \x{model for} {[}Ba/Fe{]} found by parameter
fitting. The full \x{scaling} is shown in the solid thin (green)
line, with the\emph{ }heavy\emph{ s}-process, \emph{r}-process, and
$\gamma$-process components plotted in dot-dashed (red), long-dashed
(yellow), and dot-dot-dot-dashed (black) lines, respectively. \emph{\textsl{Blue x's}}:
Frebel (2010) data points. The rest of the figure follows the convention
of Fig.\,3. }

The minimized value of $\chi_{\mathrm{r}}^{2}=0.134$ was obtained
from the parameter value $h$ $\mathrm{=1.509}$. The trend of our
barium model at low ($\mathrm{\lesssim-2.0}$) metallicities implies
\x{abundances} that begins to track iron. At these low metallicities,
iron production is dominated by primary massive star contributions,
as is barium production dominated by the primary \emph{r}-process,
hence the {[}Ba/Fe{]} encounters a ``floor'' in its \x{abundances},
as is expected (e.g., \citealp{r61}).

Above $\mathrm{[Fe/H]=-1.793}$, secondary contributions from the
\emph{s}-process exceed \emph{r}-process contributions, and begin
to drive {[}Ba/Fe{]} upward to a local maximum, before Type Ia contributions
take the ratio back down to solar. This metallicity value of $\mathrm{[Fe/H]=-1.793}$
where heavy \emph{s}-process contributions equal \emph{r}-process
contributions is lower than the typical value of $[\mathrm{Fe/H]\approx-1.5}$
(found by, e.g., \citealp{r62,r61}), and is due to our main heavy
\emph{s}-process exponent being smaller than the theoretical value
of 2. Note that at all metallicities, the $\gamma$-process contributions
to elemental abundances are negligible.

\subsection{Weak S-Process and \emph{ls} Parameter}

The\emph{ }parameters constraining the weak \emph{s}-process and \emph{ls
}\x{scalings} are $w$ and $l$. Ideally one would wish to use {[}Ga/Fe{]}
or {[}Se/Fe{]} elemental data, both of which have significant weak
\emph{s}-process isotopic contributions to their elemental abundances
($\mathrm{\approx0.61}$ for Ga, $\approx0.21$ for Se), as well as
also lying on the first main \emph{s}-process peak. Unfortunately
data across a sufficient metallicity range for these elements is sparse,
and {[}Sr/Fe{]} data is used instead, which has $\approx0.09$ of
its elemental abundance due to the weak \emph{s}-process. In addition
to Frebel (2010) data, two additional sources were used that provide
observations of higher metallicity stars \citep{r32}. Sr has two
\emph{s}-only isotopes along the weak \emph{s}-process path, $^{86}$Sr
and $^{87}$Sr, along with one mixed isotope of \emph{r}- and \emph{s}-process
origin, $^{88}$Sr (although the \emph{r}-process component is negligible),
and one $\nu$\emph{p}-process isotope, $^{84}$Sr. Possible binary
contamination of the data was removed according to the same prescription
adopted for the {[}Eu/Fe{]} data. In Fig.\,\ref{fig6} we plot results
\x{of} the \x{model for} strontium with the best fit value for $w$
and $l$. The previously found best fit values for $a$, $b$, $f$,
and $h$ (from Section 4.1 and 4.2) were used for Fe and the main
\emph{s}-process contributions to Sr.

\includegraphics[scale=0.5,angle=90]{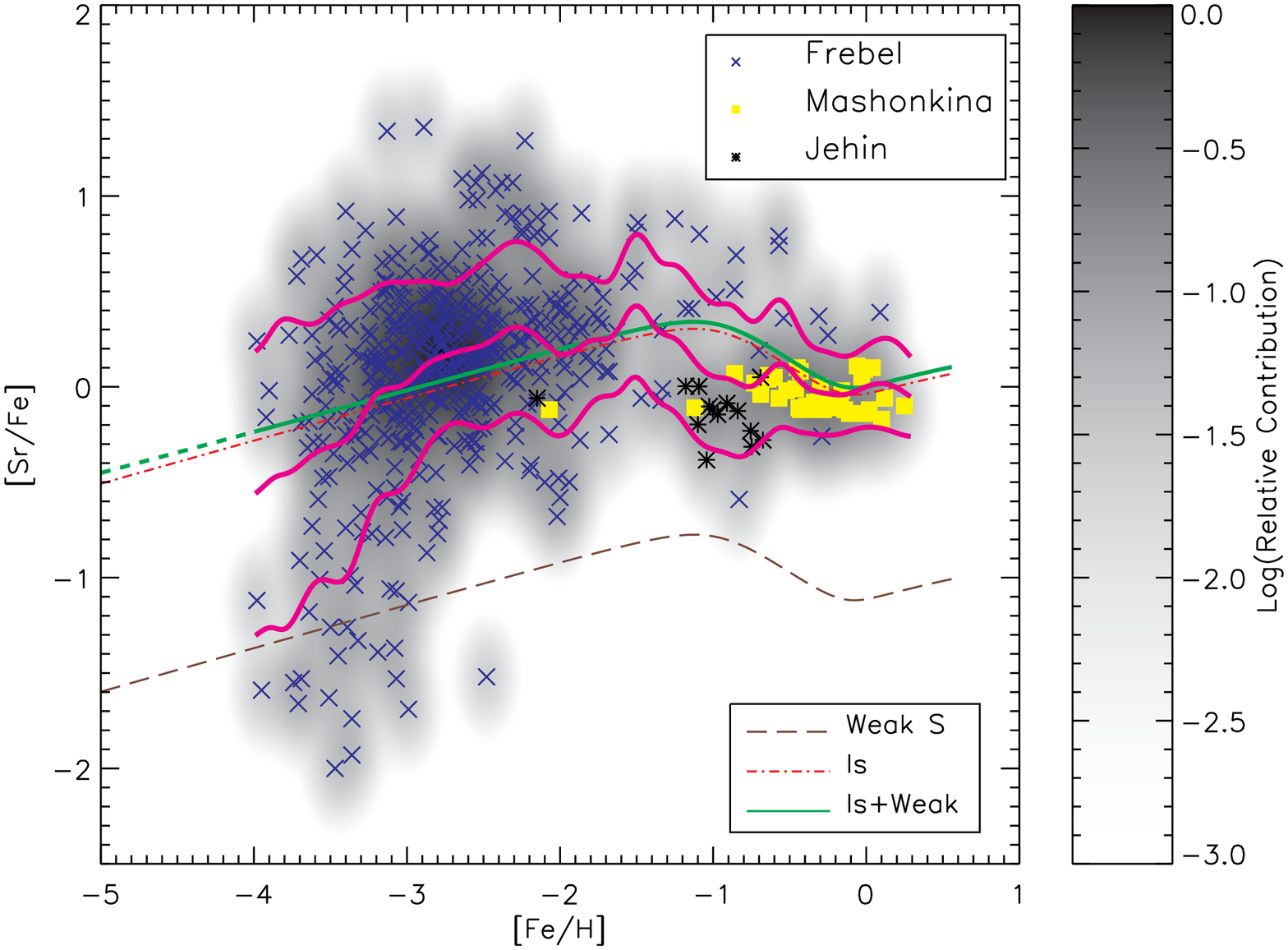}

\figcaption{\label{fig6}The resulting \x{model for} {[}Sr/Fe{]} found by parameter
fitting. The full \x{scaling} is shown in the solid thin (green)
line, with the light \emph{s}-process and weak \emph{s}-process components
plotted in dot-dashed (red) and long-dashed (brown) lines, respectively.
\emph{\textsl{Blue x's}}: Frebel (2010) data points. \emph{\textsl{Yellow crosses}}:
\citet{r23} data points. \emph{\textsl{Black asteriks}}: \citet{r13}
data points. The rest of the figure follows the convention of Fig.\,3. }

The value of $\mathrm{\chi_{r}^{2}=0.061}$ was obtained from the
parameter value $w$\emph{ }$=1.230$ and $l$ $=1.227$. At all metallicities,
contributions from the light \emph{s}-process exceed weak \emph{s}-process
contributions, and the $\nu$\emph{p}-process and \emph{r}-process\emph{
}contributions are negligible.

\subsection{``Strong'' S-Process Parameter }

The final free parameter constrains the third \emph{s}-process peak.
Pb data was taken from Frebel (2010), and binary stars were removed.
The remaining data set consists of four points only. Due to the paucity
of data, our usual standard of optimizing the elemental \x{scaling}
using a $\chi_{\mathrm{r}}^{2}$ analysis poorly constrains {[}Pb/Fe{]}.
Instead, the functional form given in Section 3.3 was used to fit
the free parameters \emph{$c$, $d$, }and\emph{ $g$ }by hand to
the four data points. The result is given in Fig.\,\ref{fig7}.

\includegraphics[scale=0.5,angle=90]{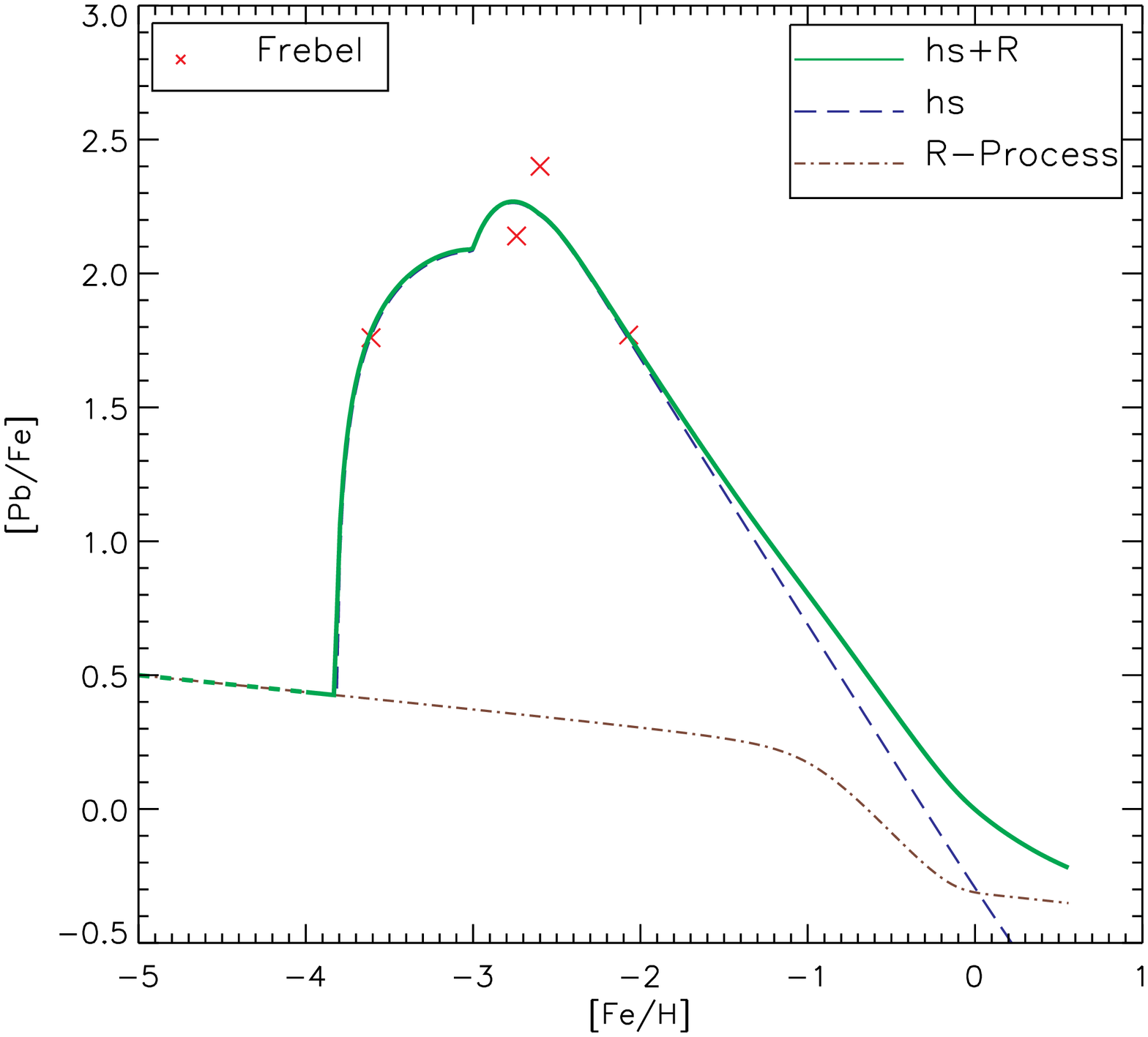}

\figcaption{\label{fig7}The resulting \x{model for} {[}Pb/Fe{]} found by parameter
fitting by hand. The full \x{scaling} is shown in the solid thin
(green) line, with the heavy \emph{s}-process and r-process components
plotted in long-dashed (blue) and dot-dashed (brown) lines, respectively.
\emph{\textsl{Red x's}}: Frebel (2010) data points. }

As shown in Fig.\,\ref{fig7}, {[}Pb/Fe{]} peaks at $\mathrm{[Fe/H]\thickapprox-2.5}$,
consistent with the AGB simulations by \citet{r86}. Below $\mathrm{[Fe/H]=-2.5}$,
{[}Pb/Fe{]} drops until the ``strong'' \emph{s}-process component
vanishes, and the elemental \x{scaling} is determined only by the
\emph{r}-process. The free parameters were found to be \emph{$c$}$=-2\cdot10^{-11}$\emph{,
$d$}$=200$\emph{, }and\emph{ $g$}$=-0.23$. It should be noted
that the few data points can only constrain the peak of the {[}Pb/Fe{]}
fit, and the drop of the abundance at $\mathrm{[Fe/H]\thickapprox-3.6}$
is not motivated by the data, but is an artifact of our chosen function
for ``strong'' \emph{s}-process evolution. It is clear that {[}Pb/Fe{]}
should indeed drop, as AGB stars do not produce isotopes at arbitrary
low metallicities, but the exact nature of the drop to the \emph{r}-process
may not be well represented by our model.

Furthermore, the ``kink'' at $\mathrm{[Fe/H]=-3}$ results from
the change in slope of the Fe \x{scaling}. As discussed in Section
3.1, massive star contributions are linearly interpolated between
their solar values and the values at $\mathrm{[Fe/H]=-3}$ given by
the massive star simulation. Below $\mathrm{[Fe/H]=-3}$, massive
star contributions are sent linearly to zero (in linear space). This
treatment changes the slope of {[}Fe{]} on either side of $\mathrm{[Fe/H]=-3}$,
which manifests as the observed ``kink'' in Fig.\,\ref{fig7}.
In reality we would expect a broad peak rather than the shown narrower
peak following a kink before descent to the \emph{r}-process floor.
This very rough treatment of the ``strong'' \emph{s}-process results
in larger uncertainties for the \emph{s}-process contributions to
Pb and Bi isotopes from our model at metallicities below $\mathrm{[Fe/H]<-2.5}$,
and further revision is left to future work.

\section{Results and Discussion\label{sec:Results-and-Discussion}}

The functional forms for all \x{scalings} are now fixed by adjusting
model to optimally fitting the observational data. A comparison is
given in Fig.\,\ref{fig8}.

\includegraphics[scale=0.5,angle=90]{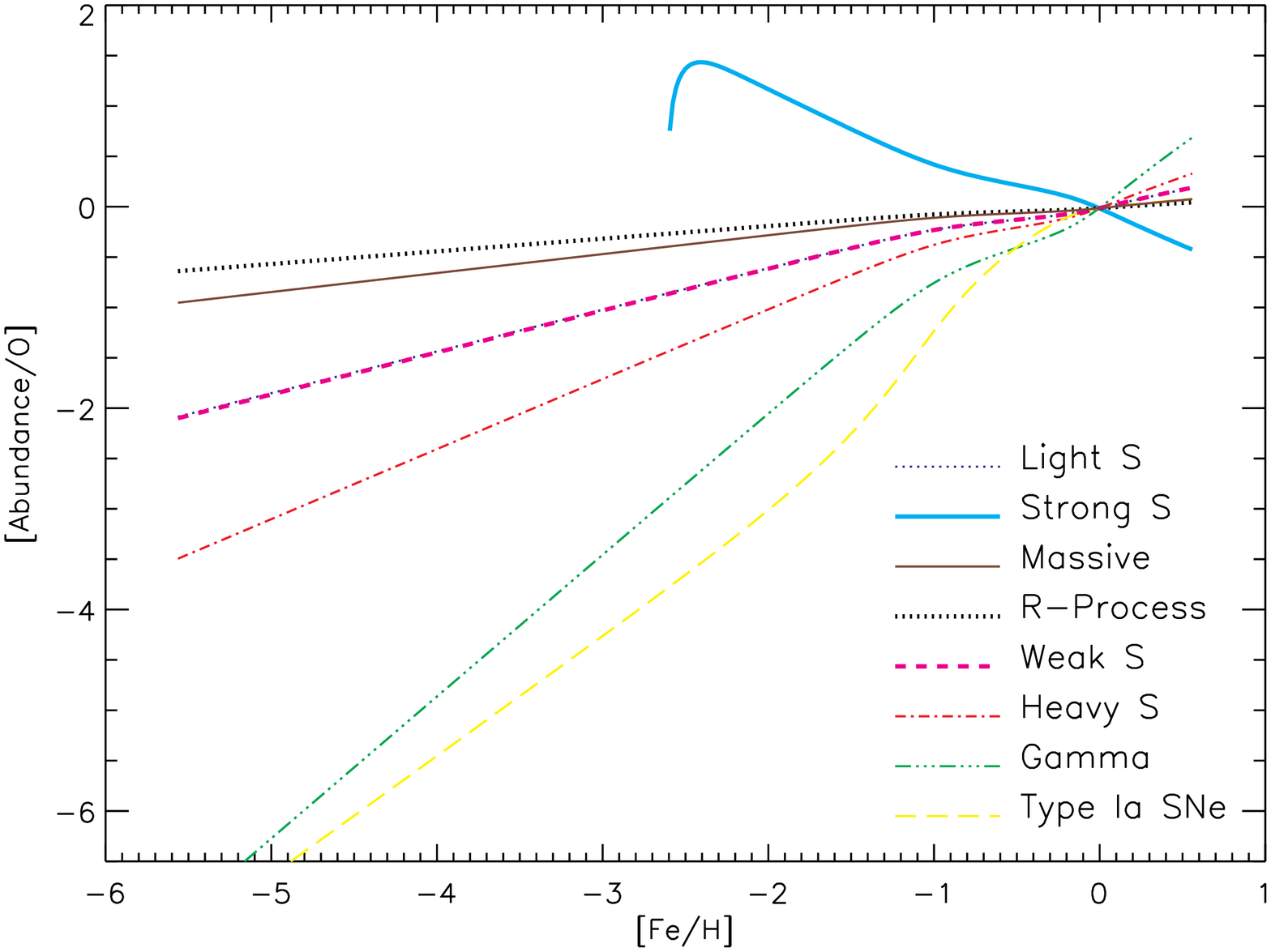}

\figcaption{\label{fig8} \x{The scaling functions} of the model contributions
relative to oxygen as functions of metallicity. The ``Massive''
line shows the scaling of the massive category's contribution to $^{56}\mathrm{Fe}$,
and is normalized to the solar contribution from this category only.}

Since each isotopic contribution in the ``massive'' category is
independently interpolated between solar and {[}Fe/H{]}, the ``massive''
(brown thin solid line) scaling shown in Fig.\,\ref{fig8} is an
example, and what is shown is the scaling for the massive contribution
to $^{56}\mathrm{Fe}$ only. Hence, this scaling gives $\mathrm{[{}^{56}Fe_{massive}/O]=log\left(^{56}Fe_{massive}/{}^{56}Fe_{massive,\odot}\right)-log\left(O/O_{\odot}\right)}$,
where $\mathrm{^{56}Fe_{massive,\odot}}$ is the massive contribution
to the solar abundance of $^{56}\mathrm{Fe}$. 

Type Ia SNe (yellow long-dashed line) contributions are negligible
($\mathrm{<1}$\,\% the solar value) below the Type Ia onset identified
by our model. After Type Ia onset at $\mathrm{[Fe/H]\approx-1.1}$,
the contributions climb smoothly to solar. It is concerning that the
Type Ia \x{scaling} begins to flatten at low metallicities until
its slope becomes less than the $\gamma$-process (green dot-dot-dot-dashed
line). Whereas contributions are negligible at these metallicities,
the $\mathrm{tanh(x)}$ function chosen for Type Ia fails to describe
the accurate physical picture, as the slope ideally should increase
sharply below onset to reflect Type Ia ``turning on.'' We accept
this behavior for the Type Ia \x{scaling} in part because contributions
are already negligible and would offer insignificant corrections to
the isotopic abundances if changed. Even though our Type Ia scaling
has a negligible impact on the abundances below the onset value, its
description in this range is not constrained at low metallicities.
In the present model it is only important for \textbf{$\mathrm{[Fe/H]>-1.1}$.}
Furthermore, the elemental data ends at $\mathrm{[Fe/H]\approx-4}$,
and hence the only constraint on Type Ia \x{scaling} below this metallicity
(where it flattens out) is that there is no contribution from the
BBN composition.

Both the massive star (brown solid line) and \emph{r}-process (black
thick dotted line) \x{scalings} show similar trends at all metallicities,
due to their shared primary nature. \x{Abundances} for the $\gamma$-process
and $\nu p$-process are \x{scaled} the same as the heavy \emph{s}-process
and \emph{r}-process, respectively. 

Both components of the \emph{s}-process do not show the typical behavior
of secondaries. The heavy component (red dot-dashed line) shows a
higher drop off at lower metallicities than the primary processes,
but with a slower exponent of 1.509 compared to the theoretical value
of 2. The light \emph{s}-process (blue dotted line) and weak \emph{s}-process
(thick pink dashed line) \x{scalings} behave similarly at all metallicities,
with an exponent intermediate between the heavy \emph{s}-process and
primary processes. The ``strong'' component (light-blue thick solid
line) displays supra-primary behavior above $\mathrm{[Fe/H]\thickapprox-2.5}$,
as is expected from its high abundance at low metallicities. Below
$\mathrm{[Fe/H]\thickapprox-2.5}$ it decreases at a rate larger than
any other process, until its contributions become zero and the \x{scaling}
is no longer plotted. In this range $\mathrm{[Fe/H]\lesssim-2.5}$
it is unlikely that our model correctly describes the ``strong''
component, as the \x{scaling} here is unconstrained by data.

The failure of our model to reproduce the theoretical secondary nature
is not new, and this discrepancy has been previously observed in the
data \citep{r63}. It may be possible to alleviate this discrepancy
by noting that at low metallicities rotating massive stars may have
an increased neutron exposure in two different ways. The first is
enhanced nitrogen production from CNO burning, which then burns into
neon to seed the neutron source \citep{r74,r73}. The second is an
earlier $^{\text{22}}$Ne ignition due to higher core temperatures
\citep{r73}. The effect of this larger neutron exposure is enhanced
weak \emph{s}-process production at low metallicities, driving an
increase in the abundance and resulting in a parameter value that
is closer to a primary rather than secondary process. 

Explaining the discrepancy for the heavy and light components is more
challenging. It may be plausible, however, that we are observing the
net result of a mixed isotopic history. That is, one could imagine
a composition with a given abundance of an isotope made from a primary
process in some astrophysical environment, that is then processed
in a different astrophysical environment at a later time. If another
abundance of this isotope is then made from a secondary process in
the new astrophysical environment, its history would change from primary
to secondary. Our model offers no way to track this effect. Indeed
the aggregate of these effects may be in fact what we observe, and
hence may be why our model gives an averaged main component parameter
that deviates from the theoretical value of 2.0. Of course, this only
applies to elements whose isotopic contributions can be made from
both primary and secondary sources. In Section\,\ref{sub:Free-Parameter-Values}
we summarize our optimized parameter values. The complete \x{scaling}
model for all elements is shown in Fig.\,\ref{fig9}.

\includegraphics[scale=0.5,angle=90]{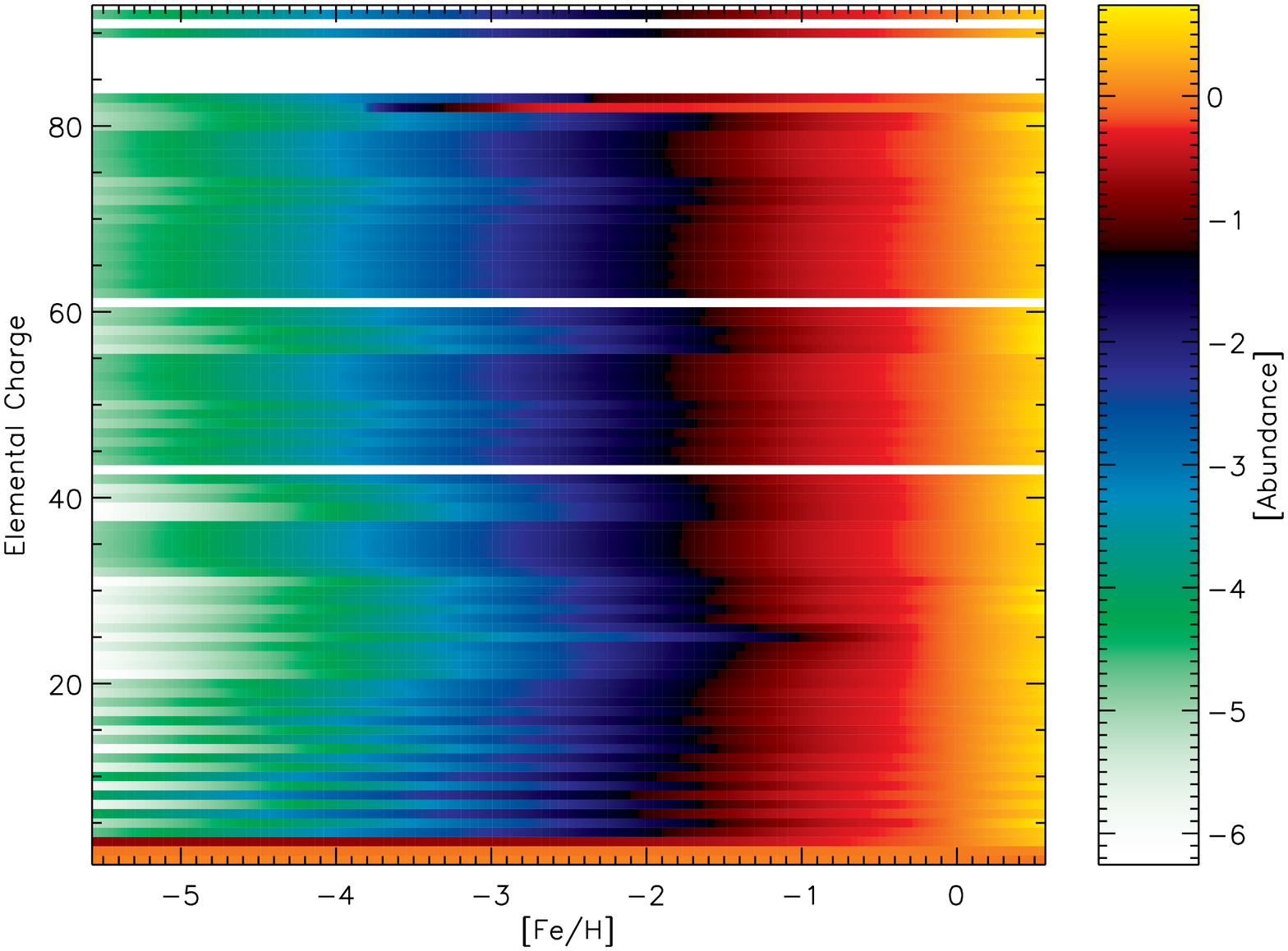}

\figcaption{\label{fig9}The complete elemental \x{scaling} of the model. The
abundances are given relative to their solar values.}

The traditional method of scaling solar abundances for inputs into
stellar simulations is equivalent to treating the Galaxy as though
all isotopic production is primary. A plot representing this approximation
would be similar to Fig.\,\ref{fig9}, but with all metals changing
their relative abundances at the same {[}Fe/H{]}, which would look
like a type of ``flag'' pattern, with all colors (representing relative
abundances) changing together. In contrast to this approximation,
Fig.\,\ref{fig9} shows the corrections offered by our model to the
traditional approximation due to the inclusion of Type Ia onset and
secondary processes. These corrections are identifiable as ``fingers''
which protrude in the horizontal axis, distorting the otherwise clean
``flag'' pattern, and occurring at elements that either lie on the
Fe-peak or that have strong secondary or other Type Ia contributions. 

The \x{scalings} that notably stand apart from the others are the
light elements: H, He, and Li. These begin with a much higher relative
abundance (compared to the metals) due to BBN, and so change less,
relative to their solar values compared to the other elements that
begin with zero BBN contributions. Additionally, Pb displays a relative
higher abundance at low metallicities due to the ``strong'' \emph{s}-process.
The sharp drop of Pb at $\mathrm{[Fe/H]\approx-3.8}$ is consistent
with Fig.\,\ref{fig7}, and is a very approximate treatment. 

In Section\,\ref{sub:Comparison-to-Linear}, we give the ratios of
the isotopic abundances from our scaling model over the abundances
generated from a linear interpolation between BBN and solar. The ratios
are computed for two sub-solar metallicities ($[Z]=-1$ and $[Z]=-3$),
and illustrate the corrections the model provides to the standard
approach.

Finally, our choice for the model parameter \emph{$\xi$} can now
be compared to the normalized metallicity Z/Z$_{\odot}$. The plot
is given in Fig.\,\ref{fig10}.

\includegraphics[scale=0.5,angle=90]{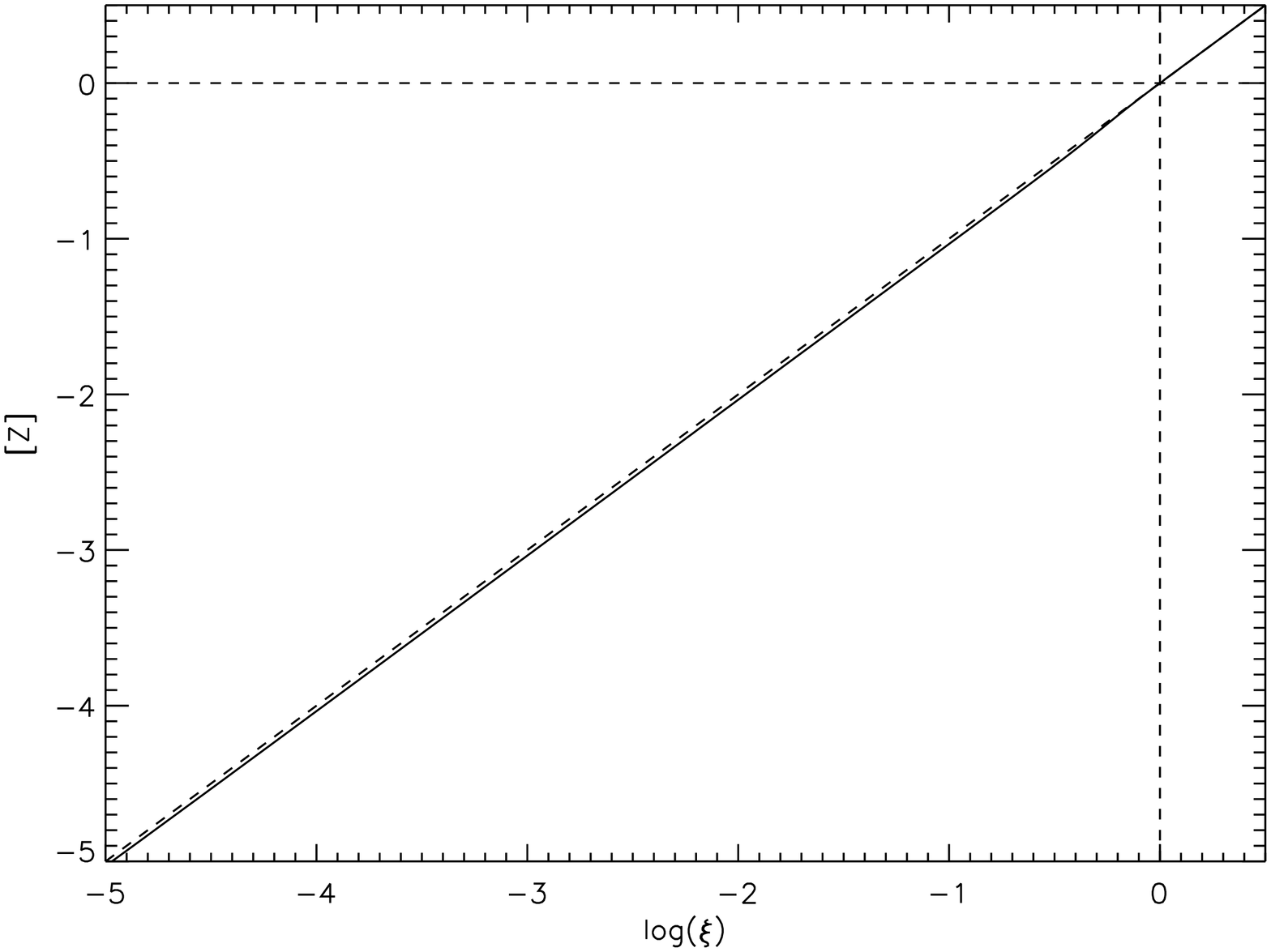}

\figcaption{\label{fig10}Comparison of our model parameter with Z/Z$_{\odot}$.
The behavior is linear to within 2\% for all values of $\log(\xi)$
and Z/Z$_{\odot}$. The dashed lines demarcate the origin, and display
the unity line for comparison.}

As shown in Fig.\,\ref{fig10}, our choice for the model parameter
indeed corresponds to Z/Z$_{\odot}$ to within 2\% for all values.
The deviation from true linear occurs close to solar metallicity,
when Type Ia SNe contributions become important and produce a small,
almost unidentifiable ``bump'' near $\mathrm{[Z]\thickapprox-0.2}$.
This comparison verifies our motivation for choosing functions for
primary and secondary processes as being proportional to a polynomial
of $\xi$.

\section{Conclusions\label{sec:Conclusions}}

\x{A metallicity-dependent Galactic isotopic decomposition} for all
stable isotopes has been constructed. The solar abundance pattern
was decomposed into several astrophysical processes responsible for
isotope synthesis. Parametric functions were chosen to scale the contributions
from each astrophysical process to give isotopic abundances, with
the solar abundance pattern and BBN used as boundary conditions. The
isotopic scalings were summed into elemental scalings and compared
with stellar data in the halo and disk to tune the fit parameters
of the model. The final scalings provide a complete isotopic abundance
pattern at any desired metallicity. The purpose of this work is to
provide \emph{isotopic} abundances that can be used as initial abundances
for stellar models in future work, \x{or other nucleosynthesis studies}.
Our model is a large improvement over approximating the input \emph{isotopic}
abundance pattern by simply scaling the solar abundances, and/or using
solar isotopic ratios. This is the first time this has been done in
a systematic way.

The interpretation of our offered solar abundance pattern decomposition
is approximate, as different assumptions (of varying reliability)
operate for different isotopes. The decompositions of the light isotopes
(below carbon) likely reflect the best current understanding, however,
the current understanding is admittedly an area of ongoing investigation.
The decomposition from carbon up to the Fe peak from massive stars
and Type Ia SNe are the result of scalings done on data from two models
that preserve the isotopic ratios across the two models, which distorts
the original abundance patterns taken from the simulations. Whereas
this preserves the salient features of these two processes, it should
be noted that the isotopes that are not CNO or Fe peak suffer a larger
uncertainty in their decompositions due to this scaling. Furthermore
we make no distinction between the several operative primary processes
that fall under the category of ``massive stars,'' which are dominated
by CCSNe yields but also necessarily include yields from stellar winds
of lower mass stars, $\nu$-process, and classical novae. 

The main \emph{s}-process yields used show good agreement with the
solar abundances, however, the weak \emph{s}-process computation only
reproduces the necessary abundances for three of the six \emph{s}-only
isotopes along the weak \emph{s}-process path. The method for computing
the weak \emph{s}-process should ideally rely upon stellar models,
unfortunately the results often display over-productions of several
isotopes above their solar values. Our more simple approach, whereas
less robust, has the advantage of only two isotopic over-productions,
and is calculated directly from the neutron capture cross-sections
and branching ratios. 

In addition to providing isotopic abundances for input into stellar
simulations, our methodology can also be applied to model other systems
such as dwarf galaxies, and tailoring the free parameters to relevant
data sets will then yield approximate chemical abundances for such
systems. Yet another application lies with stellar model fits. Currently
published stellar evolution models only give average yields for stars
over certain mass ranges. These averages must be fitted to available
observational data sets, as done for example by \citet{r10}. The
need for this fit is evident, since stellar models only address the
stellar nucleosynthesis part of GCE and neglect several processes
which will ultimately influence the subsequent ISM abundances over
many stellar lifetimes, processes such as infall and ISM mixing. The
age of the Galaxy (or timescale of GCE) is much greater than the typical
stellar lifetimes that contribute significantly to ISM enrichment;
hence this fitting is required to give stellar yields a more precise
physical meaning in a GCE context. Thus in addition to providing the
initial isotopic abundances for input into stellar simulations, the
abundances can also be used to fit the resulting stellar yields from
the model in a consistent fashion.

This work has laid out a basic method of isotopic decomposition as
a function of metallicity based on elemental observational data and
underlying nucleosynthesis processes for a complicated environment
like the Galaxy. Future work should have a more detailed look at specific
and less complicated environments like, e.g., dwarf galaxies that
have different contributions and would hence allow one to constrain
model parameters more uniquely (such as Type Ia onset contributions).
Another challenge will be to relate and identify the different nucleosynthesis
processes with the principle components found in observational work
like that of \citet{Ting2012}, and help to improve such principle
component analysis based on physical nucleosynthesis processes.

\acknowledgements{}

We would like to thank Yong-Zhong Qian for useful discussions on nucleosynthesis,
Nikos Prantzos for providing data from his own GCE model, and Anna
Frebel for helpful correspondence concerning uncertainties in stellar
abundances.

This research was supported by the US Department of Energy under grant
DE-GF02-87ER40328, by the DOE Program for Scientific Discovery through
Advanced Computing (SciDAC; DE-FC02-09ER41618), by NSF grant AST-1109394,
and by the Joint Institute for Nuclear Astrophysics (JINA; NSF grant
PHY02-16783). AH was supported by a future fellowship from the Australian
Research Council (ARC FT 120100363).

\section{Appendix}

\subsection{Fitting Procedure\label{sub:App-Fitting-Procedure}}

Here we explain the general algorithm for constructing elemental ratios
{[}X/Fe{]} from the isotopic scaling functions, and fitting to data.
A specific example will then be given.

A) For each isotope \emph{i} of element X, write the scaling function,
$\mathrm{\mathit{X}_{\mathit{i}}}\left(\xi\right)$, for all processes
that contribute to the solar abundance of $\mathrm{\mathit{X}}_{i}$,
using Equations\,4-13 and the solar abundance decomposition in Section\,\ref{sub:Solar-Abundance-Decomposition}.
This gives a scaling relation for the isotope as a function of $\xi$,
with one or more free parameters. 

B) Sum the isotopic scaling relations to give an elemental scaling
relation, 

\begin{equation}
\mathrm{\mathit{X}}\left(\xi\right)=\sum_{i}^{n}\mathrm{\mathit{X}_{\mathit{i}}}\left(\xi\right),
\end{equation}
where \emph{n} is the number of isotopes that comprise element X.

C) Repeat this process for the elements Fe and H.

D) The ratios {[}X/Fe{]} and {[}Fe/H{]} can then be evaluated, which
are functions of $\xi$ and one or more free parameters. For specific
free parameter values, a curve can be plotted in the {[}X/Fe{]} and
{[}Fe/H{]} plane. 

E) Observational data is then plotted on this plane, and each data
point is assigned a gaussian spread (Equation\,\ref{eq:gaussian}),
and the gaussian contributions in x- and y-axes are binned and averaged
(Section\,\ref{sec:Fitting-Scaling-Model}). 

F) A $\chi_{r}^{2}$ analysis is performed for the free parameter
values in {[}X/Fe{]} and {[}Fe/H{]}: for each free parameter value
a curve is defined for {[}X/Fe{]} vs. {[}Fe/H{]}, and the best-fit
parameter value is chosen that minimizes the $\chi_{r}^{2}$ between
the curve and the averaged data. This defines a unique curve for {[}X/Fe{]}
vs. {[}Fe/H{]}, and hence also gives unique functions for the elemental
\emph{and isotopic} scaling functions. 

We now give a specific example of the above steps A - C, for making
the elemental ratios {[}Au/Fe{]} and {[}Fe/H{]}. Note the above steps
D - F comprise Section\,\ref{sec:Fitting-Scaling-Model} for the
elements Mg, Eu, Ba, Sr, and Pb.

We first consider Au, which has only one stable isotope. The solar
abundance for $^{197}\mathrm{Au}$ has \emph{hs}-process and \emph{r}-process
components (Section\,\ref{sub:Solar-Abundance-Decomposition}), hence
these abundances scale as follows (using Equations\,4-13): 

\textbf{
\begin{equation}
^{197}\mathrm{Au}(\xi)={}^{197}\mathrm{Au_{\odot}^{s}}\cdot(\xi)^{h}+{}^{197}\mathrm{Au_{\odot}^{r}}\cdot(\xi)^{p},
\end{equation}
}where $^{197}\mathrm{Au_{\odot}^{s}}$ is the portion of the solar
abundance of $^{197}\mathrm{Au}$ from the \emph{hs}-process (according
to our solar abundance decomposition, see Section\,\ref{sub:Solar-Abundance-Decomposition}).
Similarly, $^{197}\mathrm{Au_{\odot}^{r}}$ is the portion of the
solar abundance of $^{197}\mathrm{Au}$ from the \emph{r}-process.
For illustration, we can evaluate $^{197}\mathrm{Au}(\xi)$ at $\xi=0,1$:

\textbf{
\begin{equation}
^{197}\mathrm{Au}(\xi=0)=0,\label{eq:16}
\end{equation}
}

\textbf{
\begin{equation}
^{197}\mathrm{Au}(\xi=1)={}^{197}\mathrm{Au_{\odot}^{s}}+{}^{197}\mathrm{Au_{\odot}^{r}}={}^{197}\mathrm{Au_{\odot}}.\label{eq:17}
\end{equation}
}Observe that Equation\,\ref{eq:16} gives the BBN abundance of \textbf{$^{197}\mathrm{Au}$},
whereas Equation\,\ref{eq:17} gives the solar abundance. When $\xi$
takes a value in between 0 and 1, we get a $^{197}\mathrm{Au}$ abundance
that is in between BBN and solar. This holds for all elemental scalings
(all scaling functions are monotonic). We now have a scaling relation
for $^{197}\mathrm{Au}$ using the continuous technical parameter
$\xi$. Note the function $^{197}\mathrm{Au(Z/Z_{\odot})}$ also must
go from BBN to solar for $Z/Z_{\odot}\in\left[0,1\right]$. Since
Au is a mono-isotopic element, the function of its isotope is also
the function of its element, $^{197}\mathrm{Au}(\xi)=\mathrm{Au}(\xi)$. 

The 4 stable isotopes of Fe have massive, Type Ia, and weak s-process
contributions (using Equations\,4-13), 

\textbf{
\begin{equation}
^{54}\mathrm{Fe}(\xi)=^{54}\mathrm{Fe_{\odot}^{Ia}}\cdot\mathit{\xi}\cdot[\tanh(\mathit{\mathit{a}\cdot\mathit{\xi-b}})+\tanh(\mathit{b})]/[\tanh(\mathit{a-b})+\tanh(\mathit{b})]+^{54}\mathrm{Fe_{\odot}^{*}}\cdot10^{m_{Fe54}\cdot\left(\mathrm{log}\left(\xi\right)-\mathrm{log}\left(\xi_{low}\right)\right)+\mathrm{log}\left(X_{i}^{\mathrm{sim}}\right)},
\end{equation}
}

\textbf{
\begin{equation}
^{56}\mathrm{Fe}(\xi)={}^{56}\mathrm{Fe_{\odot}^{Ia}}\cdot\mathit{\xi}\cdot[\tanh(\mathit{\mathit{a}\cdot\mathit{\xi-b}})+\tanh(\mathit{b})]/[\tanh(\mathit{a-b})+\tanh(\mathit{b})]+\mathrm{Fe_{\odot}^{*}}\cdot10^{m_{Fe56}\cdot\left(\mathrm{log}\left(\xi\right)-\mathrm{log}\left(\xi_{low}\right)\right)+\mathrm{log}\left(X_{Fe56}^{\mathrm{sim}}\right)},
\end{equation}
}

\textbf{
\begin{equation}
^{57}\mathrm{Fe}(\xi)=^{57}\mathrm{Fe_{\odot}^{ws}}\cdot\mathit{\xi}^{\mathit{w}}+^{57}\mathrm{Fe_{\odot}^{*}}\cdot10^{m_{Fe57}\cdot\left(\mathrm{log}\left(\xi\right)-\mathrm{log}\left(\xi_{low}\right)\right)+\mathrm{log}\left(X_{Fe57}^{\mathrm{sim}}\right)},
\end{equation}
}

\textbf{
\begin{equation}
^{58}\mathrm{Fe}(\xi)=^{58}\mathrm{Fe_{\odot}^{ws}}\cdot\mathit{\xi}^{\mathit{w}}+^{58}\mathrm{Fe_{\odot}^{*}}\cdot10^{m_{Fe58}\cdot\left(\mathrm{log}\left(\xi\right)-\mathrm{log}\left(\xi_{low}\right)\right)+\mathrm{log}\left(X_{Fe58}^{\mathrm{sim}}\right)}.
\end{equation}
}We then find the function for elemental Fe, $\mathrm{Fe}(\xi)=^{54}\mathrm{Fe}(\xi)+^{56}\mathrm{Fe}(\xi)+^{57}\mathrm{Fe}(\xi)+^{58}\mathrm{Fe}(\xi)$.
The ratio $[\mathrm{Au/Fe}]=\mathrm{log\left(\mathrm{Au}(\xi)/Au_{\odot}\right)-log\left(Fe(\xi)/Fe_{\odot}\right)}$
can then be constructed (which is a function of $\xi$). To find {[}Fe/H{]},
we consider the scaling functions for deuterium and $^{1}\mathrm{H}$:

\textbf{
\begin{equation}
\mathrm{D(\xi)=\mathrm{\mathit{\mathrm{D}}}_{\odot}\cdot\mathit{\xi^{p}}}+\mathrm{D}_{\mathrm{BBN}},
\end{equation}
}

\textbf{
\begin{equation}
^{1}\mathrm{H}(\xi)={}^{1}\mathrm{H}_{\odot}\cdot[1.0-\xi\cdot Z_{\odot}-\mathrm{Y}(\xi)-\mathrm{D}(\xi)],
\end{equation}
}

where $\mathrm{D}_{\mathrm{BBN}}$ is the BBN abundance of deuterium.
The helium function $\mathrm{Y}(\xi)$ is the sum of the isotopic
scalings of its two stable isotopes:

\textbf{
\begin{equation}
\mathrm{^{3}He(\xi)=^{3}He_{\odot}\cdot\mathit{\xi^{p}}}+\mathrm{^{3}He}_{\mathrm{BBN}},
\end{equation}
}

\textbf{
\begin{equation}
\mathrm{^{4}He(\xi)=^{4}He_{\odot}\cdot\mathit{\xi^{p}}}+^{4}\mathrm{He}_{\mathrm{BBN}}
\end{equation}
}

The elemental function for H is then, $\mathrm{H\left(\xi\right)={}^{1}\mathrm{H}(\xi)+D(\xi)}$,
and {[}Fe/H{]} can be found. The specific values used for the free
parameters \emph{h}, \emph{p}, \emph{a}, \emph{b}, and \emph{w} are
given in Section\,\ref{sub:Free-Parameter-Values}. This defines
a unique curve in the {[}Au/Fe{]} vs. {[}Fe/H{]} plane. All free parameter
values are determined by fitting the elemental functions {[}Mg/Fe{]},
{[}Eu/Fe{]}, {[}Ba/Fe{]}, {[}Sr/Fe{]}, and {[}Pb/Fe{]} to observational
data (Section\,\ref{sec:Fitting-Scaling-Model}). Using using Equations\,4-13,
Section\,\ref{sub:Solar-Abundance-Decomposition}, and Section\,\ref{sub:Free-Parameter-Values}
all elemental ratios and isotopic functions can be evaluated.

\subsection{Free Parameter Values\label{sub:Free-Parameter-Values}}

Table 1 summarizes the optimized parameter values found by fitting
to data.

\begin{deluxetable}{ccccccccccccc}  
\tablecolumns{3}  
\tablewidth{0pc}  
\tablecaption{Optimized Parameter Values}  
\tablehead{  
\colhead{Parameter} & \colhead{Best-fit Value} & \colhead{Description}}
\startdata
\emph{a} & 5.024 & Type Ia tanh Scaling Factor\\
\emph{b} & 2.722 & Type Ia tanh Shifting Factor\\
\emph{f} & 0.693 & Fraction of Solar \textsuperscript{56}Fe from Type Ia \\
\emph{p} & 0.938 & Primary Process Exponent\\
\emph{h} & 1.509 & \emph{hs}-process Exponent\\
\emph{l} & 1.227 & \emph{ls}-process Exponent\\
\emph{w} & 1.230 & Weak \emph{s}-process Exponent\\
\emph{c} & -2.e-11 & "Strong" tanh Coefficient\\
\emph{d} & 200 & "Strong" tanh Scaling Factor\\
\emph{g} & -0.23 & "Strong" tanh Shift Factor\\
\enddata  
\end{deluxetable}

\subsection{Solar Abundance Decomposition\label{sub:Solar-Abundance-Decomposition}}

Table 2 shows the solar abundance pattern decomposition for all stable
isotopes into the various astrophysical processes employed by the
model: Big Bang Nucleosynthesis, $\nu$-process/primary galactic cosmic
ray spallation/novae yields (together in a single category), secondary
galactic cosmic ray spallation, massive star yields (includes CCSNe,
stellar winds, $\nu$-process, and \emph{r}-process contributions
from carbon through zinc), Type Ia SNe yields, main \emph{s}-process
(which include all of ``strong'', \emph{ls}, and \emph{hs} components),
weak \emph{s}-process, $\nu$p-process, $\gamma$-process, and the
\emph{r}-process (from zinc through uranium). The solar abundances
in column 2 are from Lodders et al. \citeyearpar{r20} and are in
units of mole fractions. The various astrophysical processes in columns
3-12 show the fraction of the solar abundance attributed to each process,
and these fractions can be used to decompose any desired solar abundance
pattern. The fraction values are rounded to three significant figures.
Note that the remaining helium not made in BBN is from hydrogen burning,
which not explicitly shown in the the table.

\begin{deluxetable}{ccccccccccccc}  
\tablecolumns{12}  
\tablewidth{0pc}  
\rotate 
\tablecaption{Solar Abundance Decomposition}  
\tablehead{  
\colhead{Ion} & \colhead{Solar}   & \colhead{Main S}    & \colhead{Weak S} &  \colhead{R}    & \colhead{$\nu$P}   & \colhead{$\gamma$}    & \colhead{Ia} &
\colhead{Massive}      & \colhead{GCR} & \colhead{$\nu$/Novae/GCR}   &\colhead{BBN}}
\startdata
H1	&	7.0571E-01	&	\nodata	&	\nodata	&	\nodata	&	\nodata	&	\nodata	&	\nodata	&	\nodata	&	\nodata	&	\nodata	&	1.06	\\ H2	&	1.3691E-05	&	\nodata	&	\nodata	&	\nodata	&	\nodata	&	\nodata	&	\nodata	&	\nodata	&	\nodata	&	\nodata	&	1.57	\\ He3	&	1.1343E-05	&	\nodata	&	\nodata	&	\nodata	&	\nodata	&	\nodata	&	\nodata	&	\nodata	&	\nodata	&	\nodata	&	6.23E-01	\\ He4	&	6.8306E-02	&	\nodata	&	\nodata	&	\nodata	&	\nodata	&	\nodata	&	\nodata	&	\nodata	&	\nodata	&	\nodata	&	9.10E-01	\\ Li6	&	1.1479E-10	&	\nodata	&	\nodata	&	\nodata	&	\nodata	&	\nodata	&	\nodata	&	\nodata	&	3.00E-01	&	7.00E-01	&	\nodata	\\ Li7	&	1.3978E-09	&	\nodata	&	\nodata	&	\nodata	&	\nodata	&	\nodata	&	\nodata	&	\nodata	&	3.00E-01	&	5.05E-01	&	1.95E-01	\\ Be9	&	1.6640E-11	&	\nodata	&	\nodata	&	\nodata	&	\nodata	&	\nodata	&	\nodata	&	\nodata	&	2.50E-01	&	7.50E-01	&	\nodata	\\ B10	&	1.0146E-10	&	\nodata	&	\nodata	&	\nodata	&	\nodata	&	\nodata	&	\nodata	&	\nodata	&	2.50E-01	&	7.50E-01	&	\nodata	\\ B11	&	4.1043E-10	&	\nodata	&	\nodata	&	\nodata	&	\nodata	&	\nodata	&	\nodata	&	\nodata	&	2.50E-01	&	7.50E-01	&	\nodata	\\ C12	&	1.9355E-04	&	\nodata	&	\nodata	&	\nodata	&	\nodata	&	\nodata	&	1.12E-02	&	9.89E-01	&	\nodata	&	\nodata	&	\nodata	\\ C13	&	2.1747E-06	&	\nodata	&	\nodata	&	\nodata	&	\nodata	&	\nodata	&	2.19E-05	&	1.00	&	\nodata	&	\nodata	&	\nodata	\\ N14	&	5.7550E-05	&	\nodata	&	\nodata	&	\nodata	&	\nodata	&	\nodata	&	4.56E-06	&	1.00	&	\nodata	&	\nodata	&	\nodata	\\ N15	&	2.1158E-07	&	\nodata	&	\nodata	&	\nodata	&	\nodata	&	\nodata	&	3.38E-06	&	1.00	&	\nodata	&	\nodata	&	\nodata	\\ O16	&	4.2717E-04	&	\nodata	&	\nodata	&	\nodata	&	\nodata	&	\nodata	&	9.46E-03	&	9.91E-01	&	\nodata	&	\nodata	&	\nodata	\\ O17	&	1.5929E-07	&	\nodata	&	\nodata	&	\nodata	&	\nodata	&	\nodata	&	6.27E-05	&	1.00	&	\nodata	&	\nodata	&	\nodata	\\ O18	&	8.5638E-07	&	\nodata	&	\nodata	&	\nodata	&	\nodata	&	\nodata	&	8.13E-05	&	1.00	&	\nodata	&	\nodata	&	\nodata	\\ F19	&	2.1877E-08	&	\nodata	&	\nodata	&	\nodata	&	\nodata	&	\nodata	&	4.42E-06	&	1.00	&	\nodata	&	\nodata	&	\nodata	\\ Ne20	&	8.3148E-05	&	\nodata	&	\nodata	&	\nodata	&	\nodata	&	\nodata	&	8.22E-04	&	9.99E-01	&	\nodata	&	\nodata	&	\nodata	\\ Ne21	&	1.9932E-07	&	\nodata	&	\nodata	&	\nodata	&	\nodata	&	\nodata	&	3.65E-03	&	9.96E-01	&	\nodata	&	\nodata	&	\nodata	\\ Ne22	&	6.1139E-06	&	\nodata	&	\nodata	&	\nodata	&	\nodata	&	\nodata	&	3.81E-01	&	6.19E-01	&	\nodata	&	\nodata	&	\nodata	\\ Na23	&	1.5705E-06	&	\nodata	&	\nodata	&	\nodata	&	\nodata	&	\nodata	&	5.08E-03	&	9.95E-01	&	\nodata	&	\nodata	&	\nodata	\\ Mg24	&	2.2036E-05	&	\nodata	&	\nodata	&	\nodata	&	\nodata	&	\nodata	&	1.63E-02	&	9.84E-01	&	\nodata	&	\nodata	&	\nodata	\\ Mg25	&	2.7905E-06	&	\nodata	&	\nodata	&	\nodata	&	\nodata	&	\nodata	&	5.23E-03	&	9.95E-01	&	\nodata	&	\nodata	&	\nodata	\\ Mg26	&	3.0701E-06	&	\nodata	&	\nodata	&	\nodata	&	\nodata	&	\nodata	&	3.58E-03	&	9.96E-01	&	\nodata	&	\nodata	&	\nodata	\\ Al27	&	2.3014E-06	&	\nodata	&	\nodata	&	\nodata	&	\nodata	&	\nodata	&	7.35E-02	&	9.26E-01	&	\nodata	&	\nodata	&	\nodata	\\ Si28	&	2.5093E-05	&	\nodata	&	\nodata	&	\nodata	&	\nodata	&	\nodata	&	2.51E-01	&	7.49E-01	&	\nodata	&	\nodata	&	\nodata	\\ Si29	&	1.2741E-06	&	\nodata	&	\nodata	&	\nodata	&	\nodata	&	\nodata	&	2.65E-01	&	7.35E-01	&	\nodata	&	\nodata	&	\nodata	\\ Si30	&	8.3992E-07	&	\nodata	&	\nodata	&	\nodata	&	\nodata	&	\nodata	&	4.28E-01	&	5.72E-01	&	\nodata	&	\nodata	&	\nodata	\\ P31	&	2.2592E-07	&	\nodata	&	\nodata	&	\nodata	&	\nodata	&	\nodata	&	1.73E-01	&	8.27E-01	&	\nodata	&	\nodata	&	\nodata	\\ S32	&	1.0890E-05	&	\nodata	&	\nodata	&	\nodata	&	\nodata	&	\nodata	&	2.16E-01	&	7.84E-01	&	\nodata	&	\nodata	&	\nodata	\\ S33	&	8.5956E-08	&	\nodata	&	\nodata	&	\nodata	&	\nodata	&	\nodata	&	2.28E-01	&	7.72E-01	&	\nodata	&	\nodata	&	\nodata	\\ S34	&	4.8307E-07	&	\nodata	&	\nodata	&	\nodata	&	\nodata	&	\nodata	&	3.93E-01	&	6.07E-01	&	\nodata	&	\nodata	&	\nodata	\\ S36	&	1.9483E-09	&	\nodata	&	\nodata	&	\nodata	&	\nodata	&	\nodata	&	4.86E-01	&	5.14E-01	&	\nodata	&	\nodata	&	\nodata	\\ Cl35	&	1.0653E-07	&	\nodata	&	\nodata	&	\nodata	&	\nodata	&	\nodata	&	8.18E-02	&	9.18E-01	&	\nodata	&	\nodata	&	\nodata	\\ Cl37	&	3.4066E-08	&	\nodata	&	\nodata	&	\nodata	&	\nodata	&	\nodata	&	1.16E-01	&	8.84E-01	&	\nodata	&	\nodata	&	\nodata	\\ Ar36	&	2.1329E-06	&	\nodata	&	\nodata	&	\nodata	&	\nodata	&	\nodata	&	2.14E-01	&	7.86E-01	&	\nodata	&	\nodata	&	\nodata	\\ Ar38	&	3.8781E-07	&	\nodata	&	\nodata	&	\nodata	&	\nodata	&	\nodata	&	2.81E-01	&	7.19E-01	&	\nodata	&	\nodata	&	\nodata	\\ Ar40	&	6.0578E-10	&	\nodata	&	\nodata	&	\nodata	&	\nodata	&	\nodata	&	3.98E-01	&	6.02E-01	&	\nodata	&	\nodata	&	\nodata	\\ K39	&	9.5342E-08	&	\nodata	&	\nodata	&	\nodata	&	\nodata	&	\nodata	&	5.10E-02	&	9.49E-01	&	\nodata	&	\nodata	&	\nodata	\\ K40	&	1.5000E-10	&	\nodata	&	\nodata	&	\nodata	&	\nodata	&	\nodata	&	\nodata	&	1.00	&	\nodata	&	\nodata	&	\nodata	\\ K41	&	6.8806E-09	&	\nodata	&	\nodata	&	\nodata	&	\nodata	&	\nodata	&	6.07E-02	&	9.39E-01	&	\nodata	&	\nodata	&	\nodata	\\ Ca40	&	1.5925E-06	&	\nodata	&	\nodata	&	\nodata	&	\nodata	&	\nodata	&	2.97E-01	&	7.03E-01	&	\nodata	&	\nodata	&	\nodata	\\ Ca42	&	1.0629E-08	&	\nodata	&	\nodata	&	\nodata	&	\nodata	&	\nodata	&	1.93E-01	&	8.07E-01	&	\nodata	&	\nodata	&	\nodata	\\ Ca43	&	2.2179E-09	&	\nodata	&	\nodata	&	\nodata	&	\nodata	&	\nodata	&	1.71E-02	&	9.83E-01	&	\nodata	&	\nodata	&	\nodata	\\ Ca44	&	3.4270E-08	&	\nodata	&	\nodata	&	\nodata	&	\nodata	&	\nodata	&	5.79E-02	&	9.42E-01	&	\nodata	&	\nodata	&	\nodata	\\ Ca46	&	6.5714E-11	&	\nodata	&	\nodata	&	\nodata	&	\nodata	&	\nodata	&	9.58E-01	&	4.23E-02	&	\nodata	&	\nodata	&	\nodata	\\ Ca48	&	3.0721E-09	&	\nodata	&	\nodata	&	\nodata	&	\nodata	&	\nodata	&	3.16E-01	&	6.84E-01	&	\nodata	&	\nodata	&	\nodata	\\ Sc45	&	9.3722E-10	&	\nodata	&	\nodata	&	\nodata	&	\nodata	&	\nodata	&	3.26E-02	&	9.67E-01	&	\nodata	&	\nodata	&	\nodata	\\ Ti46	&	5.5503E-09	&	\nodata	&	\nodata	&	\nodata	&	\nodata	&	\nodata	&	3.51E-01	&	6.49E-01	&	\nodata	&	\nodata	&	\nodata	\\ Ti47	&	5.0040E-09	&	\nodata	&	\nodata	&	\nodata	&	\nodata	&	\nodata	&	5.00E-02	&	9.50E-01	&	\nodata	&	\nodata	&	\nodata	\\ Ti48	&	4.9602E-08	&	\nodata	&	\nodata	&	\nodata	&	\nodata	&	\nodata	&	2.65E-01	&	7.35E-01	&	\nodata	&	\nodata	&	\nodata	\\ Ti49	&	3.6394E-09	&	\nodata	&	\nodata	&	\nodata	&	\nodata	&	\nodata	&	3.53E-01	&	6.47E-01	&	\nodata	&	\nodata	&	\nodata	\\ Ti50	&	3.4887E-09	&	\nodata	&	\nodata	&	\nodata	&	\nodata	&	\nodata	&	1.00	&	8.27E-05	&	\nodata	&	\nodata	&	\nodata	\\ V50	&	1.9458E-11	&	\nodata	&	\nodata	&	\nodata	&	\nodata	&	\nodata	&	1.78E-01	&	8.22E-01	&	\nodata	&	\nodata	&	\nodata	\\ V51	&	7.7732E-09	&	\nodata	&	\nodata	&	\nodata	&	\nodata	&	\nodata	&	4.23E-01	&	5.77E-01	&	\nodata	&	\nodata	&	\nodata	\\ Cr50	&	1.5469E-08	&	\nodata	&	\nodata	&	\nodata	&	\nodata	&	\nodata	&	7.65E-01	&	2.35E-01	&	\nodata	&	\nodata	&	\nodata	\\ Cr52	&	2.9829E-07	&	\nodata	&	\nodata	&	\nodata	&	\nodata	&	\nodata	&	5.64E-01	&	4.36E-01	&	\nodata	&	\nodata	&	\nodata	\\ Cr53	&	3.3822E-08	&	\nodata	&	\nodata	&	\nodata	&	\nodata	&	\nodata	&	7.29E-01	&	2.71E-01	&	\nodata	&	\nodata	&	\nodata	\\ Cr54	&	8.4184E-09	&	\nodata	&	\nodata	&	\nodata	&	\nodata	&	\nodata	&	9.91E-01	&	8.88E-03	&	\nodata	&	\nodata	&	\nodata	\\ Mn55	&	2.5088E-07	&	\nodata	&	\nodata	&	\nodata	&	\nodata	&	\nodata	&	8.69E-01	&	1.31E-01	&	\nodata	&	\nodata	&	\nodata	\\ Fe54	&	1.3481E-06	&	\nodata	&	\nodata	&	\nodata	&	\nodata	&	\nodata	&	9.66E-01	&	3.35E-02	&	\nodata	&	\nodata	&	\nodata	\\ Fe56	&	2.1162E-05	&	\nodata	&	\nodata	&	\nodata	&	\nodata	&	\nodata	&	6.94E-01	&	3.07E-01	&	\nodata	&	\nodata	&	\nodata	\\ Fe57	&	4.8876E-07	&	\nodata	&	1.40E-04	&	\nodata	&	\nodata	&	\nodata	&	\nodata	&	1.00	&	\nodata	&	\nodata	&	\nodata	\\ Fe58	&	6.5018E-08	&	\nodata	&	1.79E-02	&	\nodata	&	\nodata	&	\nodata	&	\nodata	&	9.82E-01	&	\nodata	&	\nodata	&	\nodata	\\ Co59	&	6.3817E-08	&	\nodata	&	8.97E-03	&	\nodata	&	\nodata	&	\nodata	&	\nodata	&	9.91E-01	&	\nodata	&	\nodata	&	\nodata	\\ Ni58	&	9.0816E-07	&	\nodata	&	\nodata	&	\nodata	&	\nodata	&	\nodata	&	\nodata	&	1.00	&	\nodata	&	\nodata	&	\nodata	\\ Ni60	&	3.4988E-07	&	\nodata	&	3.80E-03	&	\nodata	&	\nodata	&	\nodata	&	\nodata	&	9.96E-01	&	\nodata	&	\nodata	&	\nodata	\\ Ni61	&	1.5206E-08	&	\nodata	&	3.74E-02	&	\nodata	&	\nodata	&	\nodata	&	\nodata	&	9.63E-01	&	\nodata	&	\nodata	&	\nodata	\\ Ni62	&	4.8485E-08	&	\nodata	&	7.33E-02	&	\nodata	&	\nodata	&	\nodata	&	\nodata	&	9.27E-01	&	\nodata	&	\nodata	&	\nodata	\\ Ni64	&	1.2348E-08	&	\nodata	&	2.81E-01	&	\nodata	&	\nodata	&	\nodata	&	\nodata	&	7.19E-01	&	\nodata	&	\nodata	&	\nodata	\\ Cu63	&	1.0184E-08	&	7.00E-03	&	1.61E-01	&	\nodata	&	\nodata	&	\nodata	&	\nodata	&	8.32E-01	&	\nodata	&	\nodata	&	\nodata	\\ Cu65	&	4.5381E-09	&	1.90E-02	&	7.27E-01	&	\nodata	&	\nodata	&	\nodata	&	\nodata	&	2.54E-01	&	\nodata	&	\nodata	&	\nodata	\\ Zn64	&	1.7153E-08	&	1.00E-03	&	6.67E-02	&	\nodata	&	\nodata	&	\nodata	&	\nodata	&	9.32E-01	&	\nodata	&	\nodata	&	\nodata	\\ Zn66	&	9.8409E-09	&	9.00E-03	&	2.70E-01	&	\nodata	&	\nodata	&	\nodata	&	\nodata	&	7.21E-01	&	\nodata	&	\nodata	&	\nodata	\\ Zn67	&	1.4461E-09	&	1.40E-02	&	4.08E-01	&	\nodata	&	\nodata	&	\nodata	&	\nodata	&	5.78E-01	&	\nodata	&	\nodata	&	\nodata	\\ Zn68	&	6.6135E-09	&	2.00E-02	&	5.19E-01	&	\nodata	&	\nodata	&	\nodata	&	\nodata	&	4.61E-01	&	\nodata	&	\nodata	&	\nodata	\\ Zn70	&	2.1869E-10	&	1.00E-03	&	\nodata	&	9.99E-01	&	\nodata	&	\nodata	&	\nodata	&	\nodata	&	\nodata	&	\nodata	&	\nodata	\\ Ga69	&	5.9779E-10	&	3.20E-02	&	7.41E-01	&	2.27E-01	&	\nodata	&	\nodata	&	\nodata	&	\nodata	&	\nodata	&	\nodata	&	\nodata	\\ Ga71	&	3.9674E-10	&	5.30E-02	&	9.47E-01	&	\nodata	&	\nodata	&	\nodata	&	\nodata	&	\nodata	&	\nodata	&	\nodata	&	\nodata	\\ Ge70	&	6.6233E-10	&	5.40E-02	&	9.46E-01	&	\nodata	&	\nodata	&	\nodata	&	\nodata	&	\nodata	&	\nodata	&	\nodata	&	\nodata	\\ Ge72	&	8.6283E-10	&	7.00E-02	&	6.80E-01	&	2.50E-01	&	\nodata	&	\nodata	&	\nodata	&	\nodata	&	\nodata	&	\nodata	&	\nodata	\\ Ge73	&	2.4071E-10	&	6.70E-02	&	6.88E-01	&	2.45E-01	&	\nodata	&	\nodata	&	\nodata	&	\nodata	&	\nodata	&	\nodata	&	\nodata	\\ Ge74	&	1.1211E-09	&	8.10E-02	&	6.73E-01	&	2.46E-01	&	\nodata	&	\nodata	&	\nodata	&	\nodata	&	\nodata	&	\nodata	&	\nodata	\\ Ge76	&	2.3219E-10	&	1.00E-03	&	\nodata	&	9.99E-01	&	\nodata	&	\nodata	&	\nodata	&	\nodata	&	\nodata	&	\nodata	&	\nodata	\\ As75	&	1.6585E-10	&	5.50E-02	&	4.47E-01	&	4.98E-01	&	\nodata	&	\nodata	&	\nodata	&	\nodata	&	\nodata	&	\nodata	&	\nodata	\\ Se74	&	1.6322E-11	&	\nodata	&	5.00E-01	&	\nodata	&	5.00E-01	&	\nodata	&	\nodata	&	\nodata	&	\nodata	&	\nodata	&	\nodata	\\ Se76	&	1.7196E-10	&	1.45E-01	&	8.55E-01	&	\nodata	&	\nodata	&	\nodata	&	\nodata	&	\nodata	&	\nodata	&	\nodata	&	\nodata	\\ Se77	&	1.4018E-10	&	6.80E-02	&	3.91E-01	&	5.41E-01	&	\nodata	&	\nodata	&	\nodata	&	\nodata	&	\nodata	&	\nodata	&	\nodata	\\ Se78	&	4.3645E-10	&	1.53E-01	&	6.58E-01	&	1.89E-01	&	\nodata	&	\nodata	&	\nodata	&	\nodata	&	\nodata	&	\nodata	&	\nodata	\\ Se80	&	9.1077E-10	&	8.90E-02	&	9.14E-03	&	9.02E-01	&	\nodata	&	\nodata	&	\nodata	&	\nodata	&	\nodata	&	\nodata	&	\nodata	\\ Se82	&	1.6030E-10	&	1.00E-03	&	\nodata	&	9.99E-01	&	\nodata	&	\nodata	&	\nodata	&	\nodata	&	\nodata	&	\nodata	&	\nodata	\\ Br79	&	1.4768E-10	&	7.40E-02	&	1.78E-01	&	7.48E-01	&	\nodata	&	\nodata	&	\nodata	&	\nodata	&	\nodata	&	\nodata	&	\nodata	\\ Br81	&	1.4368E-10	&	9.30E-02	&	3.99E-01	&	5.08E-01	&	\nodata	&	\nodata	&	\nodata	&	\nodata	&	\nodata	&	\nodata	&	\nodata	\\ Kr78	&	5.5028E-12	&	\nodata	&	5.00E-01	&	\nodata	&	5.00E-01	&	\nodata	&	\nodata	&	\nodata	&	\nodata	&	\nodata	&	\nodata	\\ Kr80	&	3.5341E-11	&	9.80E-02	&	7.52E-01	&	\nodata	&	1.50E-01	&	\nodata	&	\nodata	&	\nodata	&	\nodata	&	\nodata	&	\nodata	\\ Kr82	&	1.7707E-10	&	2.87E-01	&	6.83E-01	&	\nodata	&	3.00E-02	&	\nodata	&	\nodata	&	\nodata	&	\nodata	&	\nodata	&	\nodata	\\ Kr83	&	1.7542E-10	&	1.07E-01	&	2.56E-01	&	6.37E-01	&	\nodata	&	\nodata	&	\nodata	&	\nodata	&	\nodata	&	\nodata	&	\nodata	\\ Kr84	&	8.6451E-10	&	1.38E-01	&	1.71E-01	&	6.91E-01	&	\nodata	&	\nodata	&	\nodata	&	\nodata	&	\nodata	&	\nodata	&	\nodata	\\ Kr86	&	2.6143E-10	&	1.89E-01	&	\nodata	&	8.11E-01	&	\nodata	&	\nodata	&	\nodata	&	\nodata	&	\nodata	&	\nodata	&	\nodata	\\ Rb85	&	1.3932E-10	&	1.75E-01	&	3.16E-01	&	5.09E-01	&	\nodata	&	\nodata	&	\nodata	&	\nodata	&	\nodata	&	\nodata	&	\nodata	\\ Rb87	&	5.7339E-11	&	2.80E-01	&	\nodata	&	7.20E-01	&	\nodata	&	\nodata	&	\nodata	&	\nodata	&	\nodata	&	\nodata	&	\nodata	\\ Sr84	&	3.5322E-12	&	\nodata	&	5.00E-01	&	\nodata	&	5.00E-01	&	\nodata	&	\nodata	&	\nodata	&	\nodata	&	\nodata	&	\nodata	\\ Sr86	&	6.2461E-11	&	5.60E-01	&	4.40E-01	&	\nodata	&	\nodata	&	\nodata	&	\nodata	&	\nodata	&	\nodata	&	\nodata	&	\nodata	\\ Sr87	&	4.3651E-11	&	5.53E-01	&	4.47E-01	&	\nodata	&	\nodata	&	\nodata	&	\nodata	&	\nodata	&	\nodata	&	\nodata	&	\nodata	\\ Sr88	&	5.2333E-10	&	1.00	&	\nodata	&	\nodata	&	\nodata	&	\nodata	&	\nodata	&	\nodata	&	\nodata	&	\nodata	&	\nodata	\\ Y89	&	1.2610E-10	&	9.81E-01	&	1.90E-02	&	\nodata	&	\nodata	&	\nodata	&	\nodata	&	\nodata	&	\nodata	&	\nodata	&	\nodata	\\ Zr90	&	1.5089E-10	&	7.65E-01	&	1.49E-02	&	2.20E-01	&	\nodata	&	\nodata	&	\nodata	&	\nodata	&	\nodata	&	\nodata	&	\nodata	\\ Zr91	&	3.2913E-11	&	9.52E-01	&	1.27E-02	&	3.53E-02	&	\nodata	&	\nodata	&	\nodata	&	\nodata	&	\nodata	&	\nodata	&	\nodata	\\ Zr92	&	5.0282E-11	&	9.05E-01	&	6.81E-03	&	8.82E-02	&	\nodata	&	\nodata	&	\nodata	&	\nodata	&	\nodata	&	\nodata	&	\nodata	\\ Zr94	&	5.0969E-11	&	1.00	&	\nodata	&	\nodata	&	\nodata	&	\nodata	&	\nodata	&	\nodata	&	\nodata	&	\nodata	&	\nodata	\\ Zr96	&	8.2083E-12	&	4.63E-01	&	\nodata	&	5.37E-01	&	\nodata	&	\nodata	&	\nodata	&	\nodata	&	\nodata	&	\nodata	&	\nodata	\\ Nb93	&	2.1209E-11	&	9.14E-01	&	3.37E-03	&	8.26E-02	&	\nodata	&	\nodata	&	\nodata	&	\nodata	&	\nodata	&	\nodata	&	\nodata	\\ Mo92	&	1.0075E-11	&	\nodata	&	\nodata	&	\nodata	&	1.00	&	\nodata	&	\nodata	&	\nodata	&	\nodata	&	\nodata	&	\nodata	\\ Mo94	&	6.3478E-12	&	8.00E-03	&	\nodata	&	\nodata	&	9.92E-01	&	\nodata	&	\nodata	&	\nodata	&	\nodata	&	\nodata	&	\nodata	\\ Mo95	&	1.0986E-11	&	6.61E-01	&	3.59E-03	&	3.35E-01	&	\nodata	&	\nodata	&	\nodata	&	\nodata	&	\nodata	&	\nodata	&	\nodata	\\ Mo96	&	1.1564E-11	&	1.00	&	\nodata	&	\nodata	&	\nodata	&	\nodata	&	\nodata	&	\nodata	&	\nodata	&	\nodata	&	\nodata	\\ Mo97	&	6.6583E-12	&	6.11E-01	&	5.55E-03	&	3.83E-01	&	\nodata	&	\nodata	&	\nodata	&	\nodata	&	\nodata	&	\nodata	&	\nodata	\\ Mo98	&	1.6919E-11	&	7.94E-01	&	8.62E-03	&	1.97E-01	&	\nodata	&	\nodata	&	\nodata	&	\nodata	&	\nodata	&	\nodata	&	\nodata	\\ Mo100	&	6.8143E-12	&	4.30E-02	&	\nodata	&	9.57E-01	&	\nodata	&	\nodata	&	\nodata	&	\nodata	&	\nodata	&	\nodata	&	\nodata	\\ Ru96	&	2.6864E-12	&	\nodata	&	\nodata	&	\nodata	&	1.00	&	\nodata	&	\nodata	&	\nodata	&	\nodata	&	\nodata	&	\nodata	\\ Ru98	&	9.0586E-13	&	\nodata	&	\nodata	&	\nodata	&	1.00	&	\nodata	&	\nodata	&	\nodata	&	\nodata	&	\nodata	&	\nodata	\\ Ru99	&	6.1841E-12	&	3.04E-01	&	3.65E-03	&	6.92E-01	&	\nodata	&	\nodata	&	\nodata	&	\nodata	&	\nodata	&	\nodata	&	\nodata	\\ Ru100	&	6.1068E-12	&	1.00	&	\nodata	&	\nodata	&	\nodata	&	\nodata	&	\nodata	&	\nodata	&	\nodata	&	\nodata	&	\nodata	\\ Ru101	&	8.2694E-12	&	1.63E-01	&	2.31E-03	&	8.35E-01	&	\nodata	&	\nodata	&	\nodata	&	\nodata	&	\nodata	&	\nodata	&	\nodata	\\ Ru102	&	1.5294E-11	&	4.57E-01	&	\nodata	&	5.43E-01	&	\nodata	&	\nodata	&	\nodata	&	\nodata	&	\nodata	&	\nodata	&	\nodata	\\ Ru104	&	9.0261E-12	&	2.20E-02	&	\nodata	&	9.78E-01	&	\nodata	&	\nodata	&	\nodata	&	\nodata	&	\nodata	&	\nodata	&	\nodata	\\ Rh103	&	1.0078E-11	&	1.57E-01	&	\nodata	&	8.43E-01	&	\nodata	&	\nodata	&	\nodata	&	\nodata	&	\nodata	&	\nodata	&	\nodata	\\ Pd102	&	3.7684E-13	&	\nodata	&	\nodata	&	\nodata	&	2.50E-01	&	7.50E-01	&	\nodata	&	\nodata	&	\nodata	&	\nodata	&	\nodata	\\ Pd104	&	4.1157E-12	&	1.00	&	\nodata	&	\nodata	&	\nodata	&	\nodata	&	\nodata	&	\nodata	&	\nodata	&	\nodata	&	\nodata	\\ Pd105	&	8.2499E-12	&	1.46E-01	&	\nodata	&	8.54E-01	&	\nodata	&	\nodata	&	\nodata	&	\nodata	&	\nodata	&	\nodata	&	\nodata	\\ Pd106	&	1.0097E-11	&	5.45E-01	&	\nodata	&	4.55E-01	&	\nodata	&	\nodata	&	\nodata	&	\nodata	&	\nodata	&	\nodata	&	\nodata	\\ Pd108	&	9.7757E-12	&	6.95E-01	&	\nodata	&	3.05E-01	&	\nodata	&	\nodata	&	\nodata	&	\nodata	&	\nodata	&	\nodata	&	\nodata	\\ Pd110	&	4.3300E-12	&	2.80E-02	&	\nodata	&	9.72E-01	&	\nodata	&	\nodata	&	\nodata	&	\nodata	&	\nodata	&	\nodata	&	\nodata	\\ Ag107	&	6.8982E-12	&	1.59E-01	&	\nodata	&	8.41E-01	&	\nodata	&	\nodata	&	\nodata	&	\nodata	&	\nodata	&	\nodata	&	\nodata	\\ Ag109	&	6.4087E-12	&	2.71E-01	&	\nodata	&	7.29E-01	&	\nodata	&	\nodata	&	\nodata	&	\nodata	&	\nodata	&	\nodata	&	\nodata	\\ Cd106	&	5.3522E-13	&	\nodata	&	\nodata	&	\nodata	&	\nodata	&	1.00	&	\nodata	&	\nodata	&	\nodata	&	\nodata	&	\nodata	\\ Cd108	&	3.8108E-13	&	4.00E-03	&	\nodata	&	\nodata	&	\nodata	&	9.96E-01	&	\nodata	&	\nodata	&	\nodata	&	\nodata	&	\nodata	\\ Cd110	&	5.3479E-12	&	1.00	&	\nodata	&	\nodata	&	\nodata	&	\nodata	&	\nodata	&	\nodata	&	\nodata	&	\nodata	&	\nodata	\\ Cd111	&	5.4807E-12	&	3.54E-01	&	\nodata	&	6.46E-01	&	\nodata	&	\nodata	&	\nodata	&	\nodata	&	\nodata	&	\nodata	&	\nodata	\\ Cd112	&	1.0332E-11	&	7.00E-01	&	\nodata	&	3.00E-01	&	\nodata	&	\nodata	&	\nodata	&	\nodata	&	\nodata	&	\nodata	&	\nodata	\\ Cd113	&	5.2323E-12	&	4.02E-01	&	\nodata	&	5.98E-01	&	\nodata	&	\nodata	&	\nodata	&	\nodata	&	\nodata	&	\nodata	&	\nodata	\\ Cd114	&	1.2302E-11	&	8.37E-01	&	\nodata	&	1.63E-01	&	\nodata	&	\nodata	&	\nodata	&	\nodata	&	\nodata	&	\nodata	&	\nodata	\\ Cd116	&	3.2070E-12	&	1.58E-01	&	\nodata	&	8.42E-01	&	\nodata	&	\nodata	&	\nodata	&	\nodata	&	\nodata	&	\nodata	&	\nodata	\\ In113	&	2.0749E-13	&	\nodata	&	\nodata	&	\nodata	&	\nodata	&	1.00	&	\nodata	&	\nodata	&	\nodata	&	\nodata	&	\nodata	\\ In115	&	4.6314E-12	&	4.09E-01	&	\nodata	&	5.91E-01	&	\nodata	&	\nodata	&	\nodata	&	\nodata	&	\nodata	&	\nodata	&	\nodata	\\ Sn112	&	9.5212E-13	&	\nodata	&	\nodata	&	\nodata	&	\nodata	&	1.00	&	\nodata	&	\nodata	&	\nodata	&	\nodata	&	\nodata	\\ Sn114	&	6.4619E-13	&	\nodata	&	\nodata	&	\nodata	&	\nodata	&	1.00	&	\nodata	&	\nodata	&	\nodata	&	\nodata	&	\nodata	\\ Sn115	&	3.3241E-13	&	2.50E-02	&	\nodata	&	\nodata	&	\nodata	&	9.75E-01	&	\nodata	&	\nodata	&	\nodata	&	\nodata	&	\nodata	\\ Sn116	&	1.4253E-11	&	1.00	&	\nodata	&	\nodata	&	\nodata	&	\nodata	&	\nodata	&	\nodata	&	\nodata	&	\nodata	&	\nodata	\\ Sn117	&	7.5268E-12	&	5.14E-01	&	\nodata	&	4.86E-01	&	\nodata	&	\nodata	&	\nodata	&	\nodata	&	\nodata	&	\nodata	&	\nodata	\\ Sn118	&	2.3752E-11	&	7.27E-01	&	\nodata	&	2.73E-01	&	\nodata	&	\nodata	&	\nodata	&	\nodata	&	\nodata	&	\nodata	&	\nodata	\\ Sn119	&	8.4181E-12	&	5.88E-01	&	\nodata	&	4.12E-01	&	\nodata	&	\nodata	&	\nodata	&	\nodata	&	\nodata	&	\nodata	&	\nodata	\\ Sn120	&	3.1959E-11	&	7.62E-01	&	\nodata	&	2.38E-01	&	\nodata	&	\nodata	&	\nodata	&	\nodata	&	\nodata	&	\nodata	&	\nodata	\\ Sn122	&	4.5390E-12	&	4.20E-01	&	\nodata	&	5.80E-01	&	\nodata	&	\nodata	&	\nodata	&	\nodata	&	\nodata	&	\nodata	&	\nodata	\\ Sn124	&	5.6765E-12	&	\nodata	&	\nodata	&	1.00	&	\nodata	&	\nodata	&	\nodata	&	\nodata	&	\nodata	&	\nodata	&	\nodata	\\ Sb121	&	4.8656E-12	&	3.99E-01	&	\nodata	&	6.01E-01	&	\nodata	&	\nodata	&	\nodata	&	\nodata	&	\nodata	&	\nodata	&	\nodata	\\ Sb123	&	3.6387E-12	&	6.30E-02	&	\nodata	&	9.37E-01	&	\nodata	&	\nodata	&	\nodata	&	\nodata	&	\nodata	&	\nodata	&	\nodata	\\ Te120	&	1.2251E-13	&	\nodata	&	\nodata	&	\nodata	&	\nodata	&	1.00	&	\nodata	&	\nodata	&	\nodata	&	\nodata	&	\nodata	\\ Te122	&	3.3219E-12	&	1.00	&	\nodata	&	\nodata	&	\nodata	&	\nodata	&	\nodata	&	\nodata	&	\nodata	&	\nodata	&	\nodata	\\ Te123	&	1.1588E-12	&	1.00	&	\nodata	&	\nodata	&	\nodata	&	\nodata	&	\nodata	&	\nodata	&	\nodata	&	\nodata	&	\nodata	\\ Te124	&	6.1460E-12	&	1.00	&	\nodata	&	\nodata	&	\nodata	&	\nodata	&	\nodata	&	\nodata	&	\nodata	&	\nodata	&	\nodata	\\ Te125	&	9.1106E-12	&	2.08E-01	&	\nodata	&	7.92E-01	&	\nodata	&	\nodata	&	\nodata	&	\nodata	&	\nodata	&	\nodata	&	\nodata	\\ Te126	&	2.4186E-11	&	4.20E-01	&	\nodata	&	5.80E-01	&	\nodata	&	\nodata	&	\nodata	&	\nodata	&	\nodata	&	\nodata	&	\nodata	\\ Te128	&	4.0438E-11	&	3.60E-02	&	\nodata	&	9.64E-01	&	\nodata	&	\nodata	&	\nodata	&	\nodata	&	\nodata	&	\nodata	&	\nodata	\\ Te130	&	4.3133E-11	&	\nodata	&	\nodata	&	1.00	&	\nodata	&	\nodata	&	\nodata	&	\nodata	&	\nodata	&	\nodata	&	\nodata	\\ I127	&	2.9825E-11	&	5.50E-02	&	\nodata	&	9.45E-01	&	\nodata	&	\nodata	&	\nodata	&	\nodata	&	\nodata	&	\nodata	&	\nodata	\\ Xe124	&	1.9108E-13	&	\nodata	&	\nodata	&	\nodata	&	\nodata	&	1.00	&	\nodata	&	\nodata	&	\nodata	&	\nodata	&	\nodata	\\ Xe126	&	1.6569E-13	&	\nodata	&	\nodata	&	\nodata	&	\nodata	&	1.00	&	\nodata	&	\nodata	&	\nodata	&	\nodata	&	\nodata	\\ Xe128	&	3.3162E-12	&	1.00	&	\nodata	&	\nodata	&	\nodata	&	\nodata	&	\nodata	&	\nodata	&	\nodata	&	\nodata	&	\nodata	\\ Xe129	&	4.0773E-11	&	3.20E-02	&	\nodata	&	9.68E-01	&	\nodata	&	\nodata	&	\nodata	&	\nodata	&	\nodata	&	\nodata	&	\nodata	\\ Xe130	&	6.4998E-12	&	1.00	&	\nodata	&	\nodata	&	\nodata	&	\nodata	&	\nodata	&	\nodata	&	\nodata	&	\nodata	&	\nodata	\\ Xe131	&	3.2369E-11	&	7.40E-02	&	\nodata	&	9.26E-01	&	\nodata	&	\nodata	&	\nodata	&	\nodata	&	\nodata	&	\nodata	&	\nodata	\\ Xe132	&	3.9129E-11	&	3.00E-01	&	\nodata	&	7.00E-01	&	\nodata	&	\nodata	&	\nodata	&	\nodata	&	\nodata	&	\nodata	&	\nodata	\\ Xe134	&	1.4344E-11	&	4.00E-02	&	\nodata	&	9.60E-01	&	\nodata	&	\nodata	&	\nodata	&	\nodata	&	\nodata	&	\nodata	&	\nodata	\\ Xe136	&	1.1681E-11	&	\nodata	&	\nodata	&	1.00	&	\nodata	&	\nodata	&	\nodata	&	\nodata	&	\nodata	&	\nodata	&	\nodata	\\ Cs133	&	1.0102E-11	&	1.49E-01	&	\nodata	&	8.51E-01	&	\nodata	&	\nodata	&	\nodata	&	\nodata	&	\nodata	&	\nodata	&	\nodata	\\ Ba130	&	1.2868E-13	&	\nodata	&	\nodata	&	\nodata	&	\nodata	&	1.00	&	\nodata	&	\nodata	&	\nodata	&	\nodata	&	\nodata	\\ Ba132	&	1.2309E-13	&	\nodata	&	\nodata	&	\nodata	&	\nodata	&	1.00	&	\nodata	&	\nodata	&	\nodata	&	\nodata	&	\nodata	\\ Ba134	&	2.9398E-12	&	1.00	&	\nodata	&	\nodata	&	\nodata	&	\nodata	&	\nodata	&	\nodata	&	\nodata	&	\nodata	&	\nodata	\\ Ba135	&	8.0178E-12	&	2.87E-01	&	\nodata	&	7.13E-01	&	\nodata	&	\nodata	&	\nodata	&	\nodata	&	\nodata	&	\nodata	&	\nodata	\\ Ba136	&	9.5515E-12	&	1.00	&	\nodata	&	\nodata	&	\nodata	&	\nodata	&	\nodata	&	\nodata	&	\nodata	&	\nodata	&	\nodata	\\ Ba137	&	1.3661E-11	&	6.41E-01	&	\nodata	&	3.59E-01	&	\nodata	&	\nodata	&	\nodata	&	\nodata	&	\nodata	&	\nodata	&	\nodata	\\ Ba138	&	8.7207E-11	&	8.95E-01	&	\nodata	&	1.05E-01	&	\nodata	&	\nodata	&	\nodata	&	\nodata	&	\nodata	&	\nodata	&	\nodata	\\ La138	&	1.1371E-14	&	\nodata	&	\nodata	&	\nodata	&	\nodata	&	1.00	&	\nodata	&	\nodata	&	\nodata	&	\nodata	&	\nodata	\\ La139	&	1.2430E-11	&	6.96E-01	&	\nodata	&	3.04E-01	&	\nodata	&	\nodata	&	\nodata	&	\nodata	&	\nodata	&	\nodata	&	\nodata	\\ Ce136	&	5.9690E-14	&	\nodata	&	\nodata	&	\nodata	&	\nodata	&	1.00	&	\nodata	&	\nodata	&	\nodata	&	\nodata	&	\nodata	\\ Ce138	&	8.0227E-14	&	\nodata	&	\nodata	&	\nodata	&	\nodata	&	1.00	&	\nodata	&	\nodata	&	\nodata	&	\nodata	&	\nodata	\\ Ce140	&	2.8385E-11	&	8.87E-01	&	\nodata	&	1.13E-01	&	\nodata	&	\nodata	&	\nodata	&	\nodata	&	\nodata	&	\nodata	&	\nodata	\\ Ce142	&	3.5666E-12	&	1.92E-01	&	\nodata	&	8.08E-01	&	\nodata	&	\nodata	&	\nodata	&	\nodata	&	\nodata	&	\nodata	&	\nodata	\\ Pr141	&	4.6912E-12	&	5.08E-01	&	\nodata	&	4.92E-01	&	\nodata	&	\nodata	&	\nodata	&	\nodata	&	\nodata	&	\nodata	&	\nodata	\\ Nd142	&	6.2977E-12	&	1.00	&	\nodata	&	\nodata	&	\nodata	&	\nodata	&	\nodata	&	\nodata	&	\nodata	&	\nodata	&	\nodata	\\ Nd143	&	2.7997E-12	&	3.22E-01	&	\nodata	&	6.78E-01	&	\nodata	&	\nodata	&	\nodata	&	\nodata	&	\nodata	&	\nodata	&	\nodata	\\ Nd144	&	5.5255E-12	&	5.13E-01	&	\nodata	&	4.87E-01	&	\nodata	&	\nodata	&	\nodata	&	\nodata	&	\nodata	&	\nodata	&	\nodata	\\ Nd145	&	2.0265E-12	&	2.74E-01	&	\nodata	&	7.26E-01	&	\nodata	&	\nodata	&	\nodata	&	\nodata	&	\nodata	&	\nodata	&	\nodata	\\ Nd146	&	3.9348E-12	&	6.47E-01	&	\nodata	&	3.53E-01	&	\nodata	&	\nodata	&	\nodata	&	\nodata	&	\nodata	&	\nodata	&	\nodata	\\ Nd148	&	1.3132E-12	&	1.88E-01	&	\nodata	&	8.12E-01	&	\nodata	&	\nodata	&	\nodata	&	\nodata	&	\nodata	&	\nodata	&	\nodata	\\ Nd150	&	1.3482E-12	&	\nodata	&	\nodata	&	1.00	&	\nodata	&	\nodata	&	\nodata	&	\nodata	&	\nodata	&	\nodata	&	\nodata	\\ Sm144	&	2.2234E-13	&	\nodata	&	\nodata	&	\nodata	&	\nodata	&	1.00	&	\nodata	&	\nodata	&	\nodata	&	\nodata	&	\nodata	\\ Sm147	&	1.1171E-12	&	2.57E-01	&	\nodata	&	7.43E-01	&	\nodata	&	\nodata	&	\nodata	&	\nodata	&	\nodata	&	\nodata	&	\nodata	\\ Sm148	&	8.1212E-13	&	1.00	&	\nodata	&	\nodata	&	\nodata	&	\nodata	&	\nodata	&	\nodata	&	\nodata	&	\nodata	&	\nodata	\\ Sm149	&	9.9837E-13	&	1.25E-01	&	\nodata	&	8.75E-01	&	\nodata	&	\nodata	&	\nodata	&	\nodata	&	\nodata	&	\nodata	&	\nodata	\\ Sm150	&	5.3059E-13	&	1.00	&	\nodata	&	\nodata	&	\nodata	&	\nodata	&	\nodata	&	\nodata	&	\nodata	&	\nodata	&	\nodata	\\ Sm152	&	1.9303E-12	&	2.27E-01	&	\nodata	&	7.73E-01	&	\nodata	&	\nodata	&	\nodata	&	\nodata	&	\nodata	&	\nodata	&	\nodata	\\ Sm154	&	1.6408E-12	&	2.70E-02	&	\nodata	&	9.73E-01	&	\nodata	&	\nodata	&	\nodata	&	\nodata	&	\nodata	&	\nodata	&	\nodata	\\ Eu151	&	1.2804E-12	&	5.80E-02	&	\nodata	&	9.42E-01	&	\nodata	&	\nodata	&	\nodata	&	\nodata	&	\nodata	&	\nodata	&	\nodata	\\ Eu153	&	1.3977E-12	&	5.70E-02	&	\nodata	&	9.43E-01	&	\nodata	&	\nodata	&	\nodata	&	\nodata	&	\nodata	&	\nodata	&	\nodata	\\ Gd152	&	1.9865E-14	&	7.14E-01	&	\nodata	&	\nodata	&	\nodata	&	2.86E-01	&	\nodata	&	\nodata	&	\nodata	&	\nodata	&	\nodata	\\ Gd154	&	2.1352E-13	&	1.00	&	\nodata	&	\nodata	&	\nodata	&	\nodata	&	\nodata	&	\nodata	&	\nodata	&	\nodata	&	\nodata	\\ Gd155	&	1.4490E-12	&	5.90E-02	&	\nodata	&	9.41E-01	&	\nodata	&	\nodata	&	\nodata	&	\nodata	&	\nodata	&	\nodata	&	\nodata	\\ Gd156	&	2.0038E-12	&	1.75E-01	&	\nodata	&	8.25E-01	&	\nodata	&	\nodata	&	\nodata	&	\nodata	&	\nodata	&	\nodata	&	\nodata	\\ Gd157	&	1.5324E-12	&	1.11E-01	&	\nodata	&	8.89E-01	&	\nodata	&	\nodata	&	\nodata	&	\nodata	&	\nodata	&	\nodata	&	\nodata	\\ Gd158	&	2.4315E-12	&	2.82E-01	&	\nodata	&	7.18E-01	&	\nodata	&	\nodata	&	\nodata	&	\nodata	&	\nodata	&	\nodata	&	\nodata	\\ Gd160	&	2.1406E-12	&	6.00E-03	&	\nodata	&	9.94E-01	&	\nodata	&	\nodata	&	\nodata	&	\nodata	&	\nodata	&	\nodata	&	\nodata	\\ Tb159	&	1.7255E-12	&	8.50E-02	&	\nodata	&	9.15E-01	&	\nodata	&	\nodata	&	\nodata	&	\nodata	&	\nodata	&	\nodata	&	\nodata	\\ Dy156	&	6.1493E-15	&	\nodata	&	\nodata	&	\nodata	&	\nodata	&	1.00	&	\nodata	&	\nodata	&	\nodata	&	\nodata	&	\nodata	\\ Dy158	&	1.0432E-14	&	\nodata	&	\nodata	&	\nodata	&	\nodata	&	1.00	&	\nodata	&	\nodata	&	\nodata	&	\nodata	&	\nodata	\\ Dy160	&	2.5575E-13	&	1.00	&	\nodata	&	\nodata	&	\nodata	&	\nodata	&	\nodata	&	\nodata	&	\nodata	&	\nodata	&	\nodata	\\ Dy161	&	2.0742E-12	&	5.10E-02	&	\nodata	&	9.49E-01	&	\nodata	&	\nodata	&	\nodata	&	\nodata	&	\nodata	&	\nodata	&	\nodata	\\ Dy162	&	2.7974E-12	&	1.56E-01	&	\nodata	&	8.44E-01	&	\nodata	&	\nodata	&	\nodata	&	\nodata	&	\nodata	&	\nodata	&	\nodata	\\ Dy163	&	2.7338E-12	&	4.20E-02	&	\nodata	&	9.58E-01	&	\nodata	&	\nodata	&	\nodata	&	\nodata	&	\nodata	&	\nodata	&	\nodata	\\ Dy164	&	3.1032E-12	&	2.26E-01	&	\nodata	&	7.74E-01	&	\nodata	&	\nodata	&	\nodata	&	\nodata	&	\nodata	&	\nodata	&	\nodata	\\ Ho165	&	2.4767E-12	&	8.00E-02	&	\nodata	&	9.20E-01	&	\nodata	&	\nodata	&	\nodata	&	\nodata	&	\nodata	&	\nodata	&	\nodata	\\ Er162	&	9.8904E-15	&	\nodata	&	\nodata	&	\nodata	&	\nodata	&	1.00	&	\nodata	&	\nodata	&	\nodata	&	\nodata	&	\nodata	\\ Er164	&	1.1392E-13	&	7.40E-01	&	\nodata	&	\nodata	&	\nodata	&	2.60E-01	&	\nodata	&	\nodata	&	\nodata	&	\nodata	&	\nodata	\\ Er166	&	2.3839E-12	&	1.59E-01	&	\nodata	&	8.41E-01	&	\nodata	&	\nodata	&	\nodata	&	\nodata	&	\nodata	&	\nodata	&	\nodata	\\ Er167	&	1.6272E-12	&	9.10E-02	&	\nodata	&	9.09E-01	&	\nodata	&	\nodata	&	\nodata	&	\nodata	&	\nodata	&	\nodata	&	\nodata	\\ Er168	&	1.9196E-12	&	2.89E-01	&	\nodata	&	7.11E-01	&	\nodata	&	\nodata	&	\nodata	&	\nodata	&	\nodata	&	\nodata	&	\nodata	\\ Er170	&	1.0609E-12	&	1.23E-01	&	\nodata	&	8.77E-01	&	\nodata	&	\nodata	&	\nodata	&	\nodata	&	\nodata	&	\nodata	&	\nodata	\\ Tm169	&	1.1033E-12	&	1.25E-01	&	\nodata	&	8.75E-01	&	\nodata	&	\nodata	&	\nodata	&	\nodata	&	\nodata	&	\nodata	&	\nodata	\\ Yb168	&	8.5919E-15	&	\nodata	&	\nodata	&	\nodata	&	\nodata	&	1.00	&	\nodata	&	\nodata	&	\nodata	&	\nodata	&	\nodata	\\ Yb170	&	2.0796E-13	&	1.00	&	\nodata	&	\nodata	&	\nodata	&	\nodata	&	\nodata	&	\nodata	&	\nodata	&	\nodata	&	\nodata	\\ Yb171	&	9.8235E-13	&	2.03E-01	&	\nodata	&	7.97E-01	&	\nodata	&	\nodata	&	\nodata	&	\nodata	&	\nodata	&	\nodata	&	\nodata	\\ Yb172	&	1.5124E-12	&	4.29E-01	&	\nodata	&	5.71E-01	&	\nodata	&	\nodata	&	\nodata	&	\nodata	&	\nodata	&	\nodata	&	\nodata	\\ Yb173	&	1.1230E-12	&	2.65E-01	&	\nodata	&	7.35E-01	&	\nodata	&	\nodata	&	\nodata	&	\nodata	&	\nodata	&	\nodata	&	\nodata	\\ Yb174	&	2.2334E-12	&	6.02E-01	&	\nodata	&	3.98E-01	&	\nodata	&	\nodata	&	\nodata	&	\nodata	&	\nodata	&	\nodata	&	\nodata	\\ Yb176	&	9.0626E-13	&	8.30E-02	&	\nodata	&	9.17E-01	&	\nodata	&	\nodata	&	\nodata	&	\nodata	&	\nodata	&	\nodata	&	\nodata	\\ Lu175	&	1.0058E-12	&	1.77E-01	&	\nodata	&	8.23E-01	&	\nodata	&	\nodata	&	\nodata	&	\nodata	&	\nodata	&	\nodata	&	\nodata	\\ Lu176	&	2.9191E-14	&	1.00	&	\nodata	&	\nodata	&	\nodata	&	\nodata	&	\nodata	&	\nodata	&	\nodata	&	\nodata	&	\nodata	\\ Hf174	&	6.8707E-15	&	\nodata	&	\nodata	&	\nodata	&	\nodata	&	1.00	&	\nodata	&	\nodata	&	\nodata	&	\nodata	&	\nodata	\\ Hf176	&	2.2066E-13	&	1.00	&	\nodata	&	\nodata	&	\nodata	&	\nodata	&	\nodata	&	\nodata	&	\nodata	&	\nodata	&	\nodata	\\ Hf177	&	7.8866E-13	&	1.66E-01	&	\nodata	&	8.34E-01	&	\nodata	&	\nodata	&	\nodata	&	\nodata	&	\nodata	&	\nodata	&	\nodata	\\ Hf178	&	1.1570E-12	&	5.66E-01	&	\nodata	&	4.34E-01	&	\nodata	&	\nodata	&	\nodata	&	\nodata	&	\nodata	&	\nodata	&	\nodata	\\ Hf179	&	5.7769E-13	&	3.96E-01	&	\nodata	&	6.04E-01	&	\nodata	&	\nodata	&	\nodata	&	\nodata	&	\nodata	&	\nodata	&	\nodata	\\ Hf180	&	1.4878E-12	&	8.57E-01	&	\nodata	&	1.43E-01	&	\nodata	&	\nodata	&	\nodata	&	\nodata	&	\nodata	&	\nodata	&	\nodata	\\ Ta180	&	7.0387E-17	&	\nodata	&	\nodata	&	\nodata	&	\nodata	&	1.00	&	\nodata	&	\nodata	&	\nodata	&	\nodata	&	\nodata	\\ Ta181	&	5.7218E-13	&	4.51E-01	&	\nodata	&	5.49E-01	&	\nodata	&	\nodata	&	\nodata	&	\nodata	&	\nodata	&	\nodata	&	\nodata	\\ W180	&	4.4674E-15	&	\nodata	&	\nodata	&	\nodata	&	\nodata	&	1.00	&	\nodata	&	\nodata	&	\nodata	&	\nodata	&	\nodata	\\ W182	&	9.8815E-13	&	6.01E-01	&	\nodata	&	3.99E-01	&	\nodata	&	\nodata	&	\nodata	&	\nodata	&	\nodata	&	\nodata	&	\nodata	\\ W183	&	5.3377E-13	&	5.70E-01	&	\nodata	&	4.30E-01	&	\nodata	&	\nodata	&	\nodata	&	\nodata	&	\nodata	&	\nodata	&	\nodata	\\ W184	&	1.1427E-12	&	7.64E-01	&	\nodata	&	2.36E-01	&	\nodata	&	\nodata	&	\nodata	&	\nodata	&	\nodata	&	\nodata	&	\nodata	\\ W186	&	1.0600E-12	&	5.74E-01	&	\nodata	&	4.26E-01	&	\nodata	&	\nodata	&	\nodata	&	\nodata	&	\nodata	&	\nodata	&	\nodata	\\ Re185	&	5.6366E-13	&	2.81E-01	&	\nodata	&	7.19E-01	&	\nodata	&	\nodata	&	\nodata	&	\nodata	&	\nodata	&	\nodata	&	\nodata	\\ Re187	&	1.0169E-12	&	1.02E-01	&	\nodata	&	8.98E-01	&	\nodata	&	\nodata	&	\nodata	&	\nodata	&	\nodata	&	\nodata	&	\nodata	\\ Os184	&	3.6631E-15	&	\nodata	&	\nodata	&	\nodata	&	\nodata	&	1.00	&	\nodata	&	\nodata	&	\nodata	&	\nodata	&	\nodata	\\ Os186	&	2.9457E-13	&	1.00	&	\nodata	&	\nodata	&	\nodata	&	\nodata	&	\nodata	&	\nodata	&	\nodata	&	\nodata	&	\nodata	\\ Os187	&	2.3429E-13	&	4.06E-01	&	\nodata	&	\nodata	&	\nodata	&	5.94E-01	&	\nodata	&	\nodata	&	\nodata	&	\nodata	&	\nodata	\\ Os188	&	2.4584E-12	&	2.85E-01	&	\nodata	&	7.15E-01	&	\nodata	&	\nodata	&	\nodata	&	\nodata	&	\nodata	&	\nodata	&	\nodata	\\ Os189	&	2.9974E-12	&	4.30E-02	&	\nodata	&	9.57E-01	&	\nodata	&	\nodata	&	\nodata	&	\nodata	&	\nodata	&	\nodata	&	\nodata	\\ Os190	&	4.8745E-12	&	1.40E-01	&	\nodata	&	8.60E-01	&	\nodata	&	\nodata	&	\nodata	&	\nodata	&	\nodata	&	\nodata	&	\nodata	\\ Os192	&	7.5705E-12	&	3.40E-02	&	\nodata	&	9.66E-01	&	\nodata	&	\nodata	&	\nodata	&	\nodata	&	\nodata	&	\nodata	&	\nodata	\\ Ir191	&	6.8116E-12	&	1.90E-02	&	\nodata	&	9.81E-01	&	\nodata	&	\nodata	&	\nodata	&	\nodata	&	\nodata	&	\nodata	&	\nodata	\\ Ir193	&	1.1464E-11	&	1.30E-02	&	\nodata	&	9.87E-01	&	\nodata	&	\nodata	&	\nodata	&	\nodata	&	\nodata	&	\nodata	&	\nodata	\\ Pt190	&	4.7599E-15	&	\nodata	&	\nodata	&	\nodata	&	\nodata	&	1.00	&	\nodata	&	\nodata	&	\nodata	&	\nodata	&	\nodata	\\ Pt192	&	2.7132E-13	&	7.92E-01	&	\nodata	&	\nodata	&	\nodata	&	2.08E-01	&	\nodata	&	\nodata	&	\nodata	&	\nodata	&	\nodata	\\ Pt194	&	1.1429E-11	&	6.00E-02	&	\nodata	&	9.40E-01	&	\nodata	&	\nodata	&	\nodata	&	\nodata	&	\nodata	&	\nodata	&	\nodata	\\ Pt195	&	1.1728E-11	&	2.30E-02	&	\nodata	&	9.77E-01	&	\nodata	&	\nodata	&	\nodata	&	\nodata	&	\nodata	&	\nodata	&	\nodata	\\ Pt196	&	8.7505E-12	&	1.21E-01	&	\nodata	&	8.79E-01	&	\nodata	&	\nodata	&	\nodata	&	\nodata	&	\nodata	&	\nodata	&	\nodata	\\ Pt198	&	2.4834E-12	&	\nodata	&	\nodata	&	1.00	&	\nodata	&	\nodata	&	\nodata	&	\nodata	&	\nodata	&	\nodata	&	\nodata	\\ Au197	&	5.2934E-12	&	5.90E-02	&	\nodata	&	9.41E-01	&	\nodata	&	\nodata	&	\nodata	&	\nodata	&	\nodata	&	\nodata	&	\nodata	\\ Hg196	&	1.9119E-14	&	\nodata	&	\nodata	&	\nodata	&	\nodata	&	1.00	&	\nodata	&	\nodata	&	\nodata	&	\nodata	&	\nodata	\\ Hg198	&	1.2420E-12	&	1.00	&	\nodata	&	\nodata	&	\nodata	&	\nodata	&	\nodata	&	\nodata	&	\nodata	&	\nodata	&	\nodata	\\ Hg199	&	2.1024E-12	&	2.78E-01	&	\nodata	&	7.22E-01	&	\nodata	&	\nodata	&	\nodata	&	\nodata	&	\nodata	&	\nodata	&	\nodata	\\ Hg200	&	2.8778E-12	&	6.75E-01	&	\nodata	&	3.25E-01	&	\nodata	&	\nodata	&	\nodata	&	\nodata	&	\nodata	&	\nodata	&	\nodata	\\ Hg201	&	1.6423E-12	&	5.07E-01	&	\nodata	&	4.93E-01	&	\nodata	&	\nodata	&	\nodata	&	\nodata	&	\nodata	&	\nodata	&	\nodata	\\ Hg202	&	3.7209E-12	&	8.41E-01	&	\nodata	&	1.59E-01	&	\nodata	&	\nodata	&	\nodata	&	\nodata	&	\nodata	&	\nodata	&	\nodata	\\ Hg204	&	8.5538E-13	&	1.02E-01	&	\nodata	&	8.98E-01	&	\nodata	&	\nodata	&	\nodata	&	\nodata	&	\nodata	&	\nodata	&	\nodata	\\ Tl203	&	1.4648E-12	&	7.90E-01	&	\nodata	&	2.10E-01	&	\nodata	&	\nodata	&	\nodata	&	\nodata	&	\nodata	&	\nodata	&	\nodata	\\ Tl205	&	3.4967E-12	&	5.96E-01	&	\nodata	&	4.04E-01	&	\nodata	&	\nodata	&	\nodata	&	\nodata	&	\nodata	&	\nodata	&	\nodata	\\ Pb204	&	1.7961E-12	&	1.00E+00	&	\nodata	&	\nodata	&	\nodata	&	\nodata	&	\nodata	&	\nodata	&	\nodata	&	\nodata	&	\nodata	\\ Pb206	&	1.6714E-11	&	6.59E-01	&	\nodata	&	3.41E-01	&	\nodata	&	\nodata	&	\nodata	&	\nodata	&	\nodata	&	\nodata	&	\nodata	\\ Pb207	&	1.8496E-11	&	5.83E-01	&	\nodata	&	4.17E-01	&	\nodata	&	\nodata	&	\nodata	&	\nodata	&	\nodata	&	\nodata	&	\nodata	\\ Pb208	&	5.2941E-11	&	4.23E-01	&	\nodata	&	5.77E-01	&	\nodata	&	\nodata	&	\nodata	&	\nodata	&	\nodata	&	\nodata	&	\nodata	\\ Bi209	&	3.7589E-12	&	5.80E-02	&	\nodata	&	9.42E-01	&	\nodata	&	\nodata	&	\nodata	&	\nodata	&	\nodata	&	\nodata	&	\nodata	\\ Th232	&	1.1959E-12	&	\nodata	&	\nodata	&	1.00	&	\nodata	&	\nodata	&	\nodata	&	\nodata	&	\nodata	&	\nodata	&	\nodata	\\ U234	&	1.3317E-17	&	\nodata	&	\nodata	&	1.00	&	\nodata	&	\nodata	&	\nodata	&	\nodata	&	\nodata	&	\nodata	&	\nodata	\\ U235	&	1.5716E-13	&	\nodata	&	\nodata	&	1.00	&	\nodata	&	\nodata	&	\nodata	&	\nodata	&	\nodata	&	\nodata	&	\nodata	\\ U238	&	4.8994E-13	&	\nodata	&	\nodata	&	1.00	&	\nodata	&	\nodata	&	\nodata	&	\nodata	&	\nodata	&	\nodata	&	\nodata	\\ 
\enddata
\end{deluxetable}

\subsection{Comparison to Linear Interpolation\label{sub:Comparison-to-Linear}}

Table 3 shows the ratios of the isotopic abundances given by our scaling
model over a simple linear interpolation of abundances between BBN
and solar. Ratios for all isotopes are given at two different metallicities:
{[}Z{]}=-1, and {[}Z{]}=-3. 

\begin{deluxetable}{ccccccccccccc}  
\tablecolumns{9}  
\tablewidth{0pc}
\tablecaption{Ratios of Abundances (Scaling Model/Linear Interpolations) at different metallicities}  
\tablehead{  
\colhead{Isotope} & \colhead{[Z]=-1} & \colhead{[Z]=-3} & \colhead{Isotope} & \colhead{[Z]=-1} & \colhead{[Z]=-3} & \colhead{Isotope} & \colhead{[Z]=-1} & \colhead{[Z]=-3}}
\startdata
H1	&	0.9995	&	1.0000	&	   Mg24	&	1.0980	&	1.3675	&	   Ca42	&	0.6347	&	0.3899	\\      H2	&	0.9942	&	0.9998	&	   Mg25	&	0.4635	&	0.1006	&	   Ca43	&	0.3651	&	0.0503	\\     He3	&	1.0088	&	1.0003	&	   Mg26	&	0.4644	&	0.1009	&	   Ca44	&	0.4238	&	0.0855	\\     He4	&	1.0015	&	1.0001	&	   Al27	&	0.5634	&	0.2077	&	   Ca46	&	0.0097	&	0.0001	\\     Li6	&	0.9003	&	1.0831	&	   Si28	&	0.7892	&	0.8704	&	   Ca48	&	0.0038	&	0.0000	\\     Li7	&	0.9530	&	0.9999	&	   Si29	&	0.3133	&	0.0559	&	   Sc45	&	0.5466	&	0.1743	\\     Be9	&	0.9425	&	1.1584	&	   Si30	&	0.3129	&	0.0908	&	   Ti46	&	0.3777	&	0.1252	\\     B10	&	0.9425	&	1.1584	&	    P31	&	0.6071	&	0.3252	&	   Ti47	&	0.3253	&	0.0380	\\     B11	&	0.9425	&	1.1584	&	    S32	&	0.9211	&	1.2657	&	   Ti48	&	0.5270	&	0.2676	\\     C12	&	1.4204	&	2.9308	&	    S33	&	0.7507	&	0.7052	&	   Ti49	&	0.4319	&	0.1889	\\     C13	&	1.5575	&	3.7785	&	    S34	&	0.4206	&	0.1976	&	   Ti50	&	0.0075	&	0.0000	\\     N14	&	0.7076	&	0.3543	&	    S36	&	0.0920	&	0.0026	&	    V50	&	0.2747	&	0.0302	\\     N15	&	0.4852	&	0.1142	&	   Cl35	&	0.7423	&	0.4839	&	    V51	&	0.4066	&	0.1976	\\     O16	&	1.5271	&	3.6295	&	   Cl37	&	0.5896	&	0.2614	&	   Cr50	&	0.1986	&	0.1301	\\     O17	&	0.5994	&	0.2154	&	   Ar36	&	0.8476	&	0.9812	&	   Cr52	&	0.4000	&	0.3266	\\     O18	&	0.0598	&	0.0002	&	   Ar38	&	0.4697	&	0.1978	&	   Cr53	&	0.2532	&	0.2076	\\     F19	&	0.7012	&	0.3448	&	   Ar40	&	0.0504	&	0.0003	&	   Cr54	&	0.0145	&	0.0045	\\    Ne20	&	1.3117	&	2.2605	&	    K39	&	0.7738	&	0.5137	&	   Mn55	&	0.1331	&	0.1181	\\    Ne21	&	0.8830	&	0.6934	&	    K40	&	0.8825	&	0.6874	&	   Fe54	&	0.0512	&	0.0759	\\    Ne22	&	0.2099	&	0.0232	&	    K41	&	0.7554	&	0.4877	&	   Fe56	&	0.3112	&	0.3051	\\    Na23	&	0.7288	&	0.3911	&	   Ca40	&	0.6662	&	0.5927	&	   Fe57	&	0.7367	&	0.3999	\\    Fe58	&	0.0508	&	0.0037	&	   As75	&	0.8701	&	0.8665	&	    Y89	&	0.5928	&	0.2084	\\    Co59	&	0.6743	&	0.3067	&	   Se74	&	0.8711	&	0.8694	&	   Zr90	&	0.7162	&	0.5002	\\    Ni58	&	0.3176	&	0.0320	&	   Se76	&	0.5894	&	0.2048	&	   Zr91	&	0.6126	&	0.2552	\\    Ni60	&	0.8105	&	0.5328	&	   Se77	&	0.8948	&	0.9248	&	   Zr92	&	0.6423	&	0.3254	\\    Ni61	&	0.7482	&	0.4210	&	   Se78	&	0.6964	&	0.4568	&	   Zr94	&	0.5929	&	0.2084	\\    Ni62	&	0.7784	&	0.4777	&	   Se80	&	1.0984	&	1.4044	&	   Zr96	&	0.8939	&	0.9206	\\    Ni64	&	0.1666	&	0.0574	&	   Se82	&	1.1529	&	1.5333	&	   Nb93	&	0.6392	&	0.3180	\\    Cu63	&	0.7444	&	0.4228	&	   Br79	&	1.0115	&	1.1997	&	   Mo92	&	1.1535	&	1.5346	\\    Cu65	&	0.5393	&	0.1679	&	   Br81	&	0.8762	&	0.8808	&	   Mo94	&	1.1490	&	1.5240	\\    Zn64	&	0.7622	&	0.4474	&	   Kr78	&	0.8711	&	0.8694	&	   Mo95	&	0.7809	&	0.6532	\\    Zn66	&	0.4529	&	0.1032	&	   Kr80	&	0.6739	&	0.4042	&	   Mo96	&	0.5929	&	0.2084	\\    Zn67	&	0.3895	&	0.0946	&	   Kr82	&	0.6070	&	0.2453	&	   Mo97	&	0.8078	&	0.7169	\\    Zn68	&	0.3867	&	0.1117	&	   Kr83	&	0.9487	&	1.0517	&	   Mo98	&	0.7035	&	0.4702	\\    Zn70	&	1.1529	&	1.5333	&	   Kr84	&	0.9793	&	1.1234	&	  Mo100	&	1.1294	&	1.4776	\\    Ga69	&	0.7171	&	0.5062	&	   Kr86	&	1.0475	&	1.2840	&	   Ru96	&	1.1535	&	1.5346	\\    Ga71	&	0.5891	&	0.2044	&	   Rb85	&	0.8771	&	0.8825	&	   Ru98	&	1.1535	&	1.5346	\\    Ge70	&	0.5891	&	0.2044	&	   Rb87	&	0.9965	&	1.1633	&	   Ru99	&	0.9810	&	1.1266	\\    Ge72	&	0.7302	&	0.5370	&	   Sr84	&	0.8711	&	0.8694	&	  Ru100	&	0.5929	&	0.2084	\\    Ge73	&	0.7275	&	0.5305	&	   Sr86	&	0.5911	&	0.2066	&	  Ru101	&	1.0608	&	1.3154	\\    Ge74	&	0.7280	&	0.5315	&	   Sr87	&	0.5911	&	0.2065	&	  Ru102	&	0.8973	&	0.9286	\\    Ge76	&	1.1529	&	1.5333	&	   Sr88	&	0.5929	&	0.2084	&	  Ru104	&	1.1411	&	1.5054	\\   Rh103	&	1.0655	&	1.3264	&	  Sn115	&	0.5976	&	0.2134	&	  Xe128	&	0.5929	&	0.2084	\\   Pd102	&	0.7367	&	0.5438	&	  Sn116	&	0.5929	&	0.2084	&	  Xe129	&	1.1355	&	1.4922	\\   Pd104	&	0.5929	&	0.2084	&	  Sn117	&	0.8653	&	0.8530	&	  Xe130	&	0.5929	&	0.2084	\\   Pd105	&	1.0716	&	1.3410	&	  Sn118	&	0.7459	&	0.5705	&	  Xe131	&	1.1120	&	1.4365	\\   Pd106	&	0.8480	&	0.8119	&	  Sn119	&	0.8239	&	0.7548	&	  Xe132	&	0.9853	&	1.1368	\\   Pd108	&	0.7639	&	0.6129	&	  Sn120	&	0.7263	&	0.5241	&	  Xe134	&	1.1310	&	1.4816	\\   Pd110	&	1.1378	&	1.4975	&	  Sn122	&	0.9180	&	0.9776	&	  Xe136	&	1.1535	&	1.5346	\\   Ag107	&	1.0643	&	1.3238	&	  Sn124	&	1.1535	&	1.5346	&	  Cs133	&	1.0699	&	1.3370	\\   Ag109	&	1.0016	&	1.1752	&	  Sb121	&	0.9298	&	1.0055	&	  Ba130	&	0.5977	&	0.2136	\\   Cd106	&	0.5977	&	0.2136	&	  Sb123	&	1.1181	&	1.4511	&	  Ba132	&	0.5977	&	0.2136	\\   Cd108	&	0.5977	&	0.2135	&	  Te120	&	0.5977	&	0.2136	&	  Ba134	&	0.3097	&	0.0297	\\   Cd110	&	0.5929	&	0.2084	&	  Te122	&	0.5929	&	0.2084	&	  Ba135	&	0.9113	&	1.1027	\\   Cd111	&	0.9550	&	1.0652	&	  Te123	&	0.5929	&	0.2084	&	  Ba136	&	0.3097	&	0.0297	\\   Cd112	&	0.7611	&	0.6063	&	  Te124	&	0.5929	&	0.2084	&	  Ba137	&	0.6126	&	0.5700	\\   Cd113	&	0.9281	&	1.0015	&	  Te125	&	1.0369	&	1.2588	&	  Ba138	&	0.3983	&	0.1877	\\   Cd114	&	0.6843	&	0.4246	&	  Te126	&	0.9180	&	0.9776	&	  La138	&	0.5977	&	0.2136	\\   Cd116	&	1.0649	&	1.3251	&	  Te128	&	1.1333	&	1.4869	&	  La139	&	0.5662	&	0.4872	\\   In113	&	0.5977	&	0.2136	&	  Te130	&	1.1535	&	1.5346	&	  Ce136	&	0.5977	&	0.2136	\\   In115	&	0.9242	&	0.9922	&	   I127	&	1.1226	&	1.4617	&	  Ce138	&	0.5977	&	0.2136	\\   Sn112	&	0.5977	&	0.2136	&	  Xe124	&	0.5977	&	0.2136	&	  Ce140	&	0.4051	&	0.1998	\\   Sn114	&	0.5977	&	0.2136	&	  Xe126	&	0.5977	&	0.2136	&	  Ce142	&	0.9915	&	1.2457	\\   Pr141	&	0.7248	&	0.7701	&	  Gd157	&	1.0598	&	1.3676	&	  Yb171	&	0.9822	&	1.2291	\\   Nd142	&	0.3097	&	0.0297	&	  Gd158	&	0.9155	&	1.1102	&	  Yb172	&	0.7915	&	0.8890	\\   Nd143	&	0.8818	&	1.0500	&	  Gd160	&	1.1484	&	1.5256	&	  Yb173	&	0.9299	&	1.1358	\\   Nd144	&	0.7206	&	0.7626	&	  Tb159	&	1.0817	&	1.4067	&	  Yb174	&	0.6455	&	0.6287	\\   Nd145	&	0.9223	&	1.1223	&	  Dy156	&	0.5977	&	0.2136	&	  Yb176	&	1.0834	&	1.4097	\\   Nd146	&	0.6076	&	0.5609	&	  Dy158	&	0.5977	&	0.2136	&	  Lu175	&	1.0041	&	1.2682	\\   Nd148	&	0.9948	&	1.2517	&	  Dy160	&	0.3097	&	0.0297	&	  Lu176	&	0.3097	&	0.0297	\\   Nd150	&	1.1535	&	1.5346	&	  Dy161	&	1.1104	&	1.4579	&	  Hf174	&	0.5977	&	0.2136	\\   Sm144	&	0.5977	&	0.2136	&	  Dy162	&	1.0218	&	1.2999	&	  Hf176	&	0.3097	&	0.0297	\\   Sm147	&	0.9366	&	1.1479	&	  Dy163	&	1.1180	&	1.4714	&	  Hf177	&	1.0134	&	1.2848	\\   Sm148	&	0.3097	&	0.0297	&	  Dy164	&	0.9628	&	1.1945	&	  Hf178	&	0.6759	&	0.6828	\\   Sm149	&	1.0480	&	1.3465	&	  Ho165	&	1.0860	&	1.4142	&	  Hf179	&	0.8193	&	0.9387	\\   Sm150	&	0.3097	&	0.0297	&	  Er162	&	0.5977	&	0.2136	&	  Hf180	&	0.4304	&	0.2449	\\   Sm152	&	0.9619	&	1.1930	&	  Er164	&	0.3846	&	0.0775	&	  Ta180	&	0.5977	&	0.2136	\\   Sm154	&	1.1307	&	1.4940	&	  Er166	&	1.0193	&	1.2953	&	  Ta181	&	0.7729	&	0.8559	\\   Eu151	&	1.1045	&	1.4473	&	  Er167	&	1.0767	&	1.3977	&	   W180	&	0.5977	&	0.2136	\\   Eu153	&	1.1054	&	1.4488	&	  Er168	&	0.9096	&	1.0997	&	   W182	&	0.6464	&	0.6302	\\   Gd152	&	0.3921	&	0.0823	&	  Er170	&	1.0497	&	1.3495	&	   W183	&	0.6725	&	0.6768	\\   Gd154	&	0.3097	&	0.0297	&	  Tm169	&	1.0480	&	1.3465	&	   W184	&	0.5089	&	0.3849	\\   Gd155	&	1.1037	&	1.4458	&	  Yb168	&	0.5977	&	0.2136	&	   W186	&	0.6692	&	0.6708	\\   Gd156	&	1.0058	&	1.2713	&	  Yb170	&	0.3097	&	0.0297	&	  Re185	&	0.9164	&	1.1117	\\   Re187	&	1.0674	&	1.3811	&	  Pt194	&	1.1028	&	1.4443	&	  Tl203	&	0.4869	&	0.3457	\\   Os184	&	0.5977	&	0.2136	&	  Pt195	&	1.1340	&	1.5000	&	  Tl205	&	0.6506	&	0.6377	\\   Os186	&	0.3097	&	0.0297	&	  Pt196	&	1.0514	&	1.3525	&	  Pb204	&	10	&	1000	\\   Os187	&	0.4808	&	0.1389	&	  Pt198	&	1.1535	&	1.5346	&	  Pb206	&	6.9833	&	659.52	\\   Os188	&	0.9130	&	1.1057	&	  Au197	&	1.1037	&	1.4458	&	  Pb207	&	6.3110	&	583.64	\\   Os189	&	1.1172	&	1.4699	&	  Hg196	&	0.5977	&	0.2136	&	  Pb208	&	4.8955	&	423.89	\\   Os190	&	1.0353	&	1.3239	&	  Hg198	&	0.3097	&	0.0297	&	  Bi209	&	1.6666	&	59.446	\\   Os192	&	1.1248	&	1.4835	&	  Hg199	&	0.9189	&	1.1163	&	  Th232	&	1.1535	&	1.5346	\\   Ir191	&	1.1374	&	1.5060	&	  Hg200	&	0.5839	&	0.5188	&	   U234	&	1.1535	&	1.5346	\\   Ir193	&	1.1425	&	1.5151	&	  Hg201	&	0.7257	&	0.7716	&	   U235	&	1.1535	&	1.5346	\\   Pt190	&	0.5977	&	0.2136	&	  Hg202	&	0.4439	&	0.2690	&	   U238	&	1.1535	&	1.5346	\\   Pt192	&	0.3696	&	0.0680	&	  Hg204	&	1.0674	&	1.3811							 					 					 
\enddata  
\end{deluxetable} 

\clearpage{}
\end{document}